\documentclass[letterpaper]{article}
\usepackage{amsfonts,amssymb,amsmath,bm,bbm,euscript,mathrsfs,calligra}
\usepackage[usenames]{color}
\usepackage{color,soul}
\usepackage[utf8]{inputenc}
\usepackage[T1]{fontenc}
\usepackage{textcomp}
\usepackage{graphicx}
\usepackage[tight,sf]{subfigure}
\usepackage{listings}
\usepackage{hyperref}
\usepackage{bookmark}
\usepackage{authblk}

\DeclareMathAlphabet{\mathcalligra}{T1}{calligra}{m}{n}
\DeclareFontShape{T1}{calligra}{m}{n}{<->s*[2.2]callig15}{}

\begin{document}
\sf

\begin{center}
   \vskip 2em  {\huge 
Conformal symmetry breaking and self-similar spirals\\
}
\vskip 3em
{\large \sf  Jemal Guven }

\vskip 2em

\em{Instituto de Ciencias Nucleares\\
Universidad Nacional Aut\'onoma de M\'exico\\
Apdo. Postal 70-543, 04510 Cuidad de M\'exico, MEXICO}
\end{center}

 \vskip 1em
 
\begin{abstract}
\sf \small
Self-similar curves are a recurring motif in nature.   
The tension-free stationary states of conformally invariant energies describe the simplest 
curves of this form.
Planar logarithmic spirals, for example, are associated with  conformal arc-length; their unique properties reflect the symmetry and the manner of its breaking. Constructing their analogues in three-dimensions is not so simple.  The qualitative behavior of these states is controlled by two parameters, the conserved scaling current $S$ and the magnitude of the torque $M.$ Their conservation determines the curvature and the torsion. If the spiral apex is located at the origin, the conserved \textit{special} conformal current vanishes. Planar logarithmic spirals occur when $M$ and $S$ are tuned so that $4MS =1$.  More generally, the spiral exhibits internal structure, nutating between two fixed cones aligned along the torque axis. It expands monotonically as this pattern precesses about this axis. 
If  the spiral is supercritical ($4MS>1$) the cones are identical and oppositely oriented. 
The torsion changes sign where the projection along the torque  axis turns, the spiral twisting one way and then the other within each nutation.  
These elementary spirals  provide templates for understanding a broad range of self-similar spatial spiral patterns occurring in nature. In particular, supercritical trajectories  approximate rather well the nutating tip of the growing tendril in a climbing plant first described by Darwin. 
\end{abstract}



\vskip 3em

\section{Introduction}

\begin{figure}
 \begin{center}
\includegraphics[height=6cm]{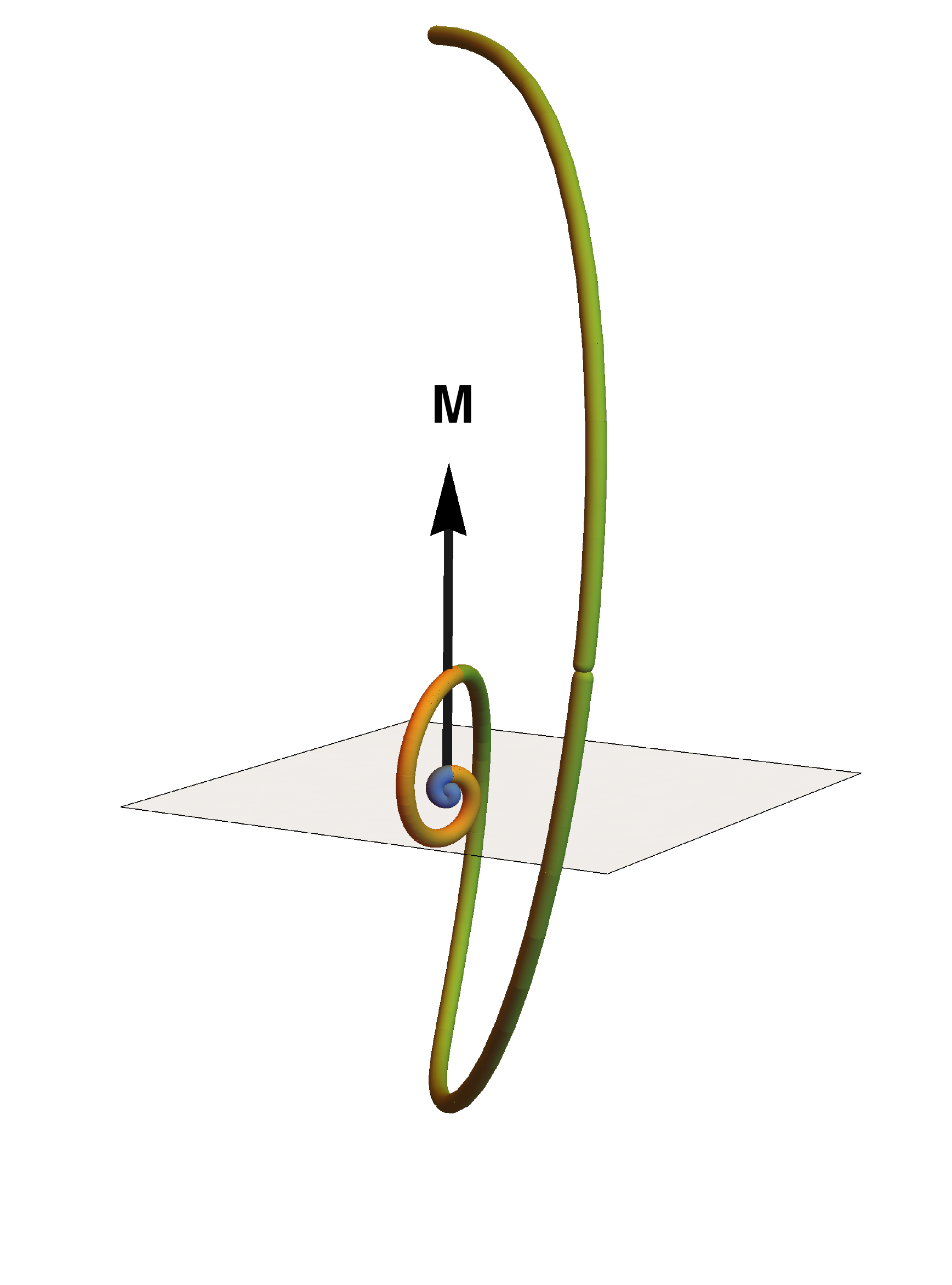}
\caption{\small \sf    
Three consecutive cycles (blue, yellow, green) of a supercritical self-similar spiral  with large $M$, described in detail in section \ref{TracingTrajectories}.  Within each cycle it nutates between vertical extrema, twisting one way on the way up, and another on the way down;  the torsion vanishes at the extrema. The torque direction $\mathbf{M}$ is indicated.  The spiral precession is captured by the projection of the trajectory onto a unit sphere, illustrated in Figure \ref{Fig:Sphere}.}
\label{Fig:Template}.
\end{center}
\end{figure}

Perhaps the simplest self-similar curves are the one-parameter family of planar logarithmic spirals. They
are characterized completely by their rate of radial growth. There is no internal structure \cite{Paper1}.  
 Logarithmic spirals are found in nature on any scale in which one can meaningfully speak of a curve. They famously form the most prominent morphological feature of spiral galaxies and weather patterns, the shells of mollusks  and the seed heads of plants. The cochlea in the inner ear and the pattern of nerve cells in the cornea also display this unmistakable pattern. 
More accurately,  this pattern is an approximation. 
\\\\
Self-similar spirals themselves need not be planar and in general they will exhibit internal structure: any of the examples that come to mind are, at best, planar only in approximation. 
 In his groundbreaking treatise \textit{On Growth and Form} \cite{Thompson},  written over a hundred years ago,  D'Arcy Thompson contemplated three-dimensional self-similar spiral geometries
 in the context of biological growth. 
More recently,  the \textit{warped} spiral geometry of our own Milky Way has been  attributed  to torques  generated within its inner disk \cite{WarpGalaxy}. 
It has also been argued that the spiral geometry of the Cochlea does not appear to fit a logarithmic  template, overturning existing assumptions \cite{Cochlea}. Significantly, the natural self-similar analogues of logarithmic spirals free to explore three-dimensional space exhibit internal structure.  Even their planar projections do not fit a logarithmic template.
\\\\
A natural place to search for self-similar curves is among the equilibrium states of scale-invariant energies along curves. Among such energies there will be conformal invariants, invariant under transformations preserving angles. The additional symmetry is the invariance under inversion 
in spheres, which implies translational invariance in the inverted space, dual in the Lie algebra to translations in the Euclidean space.  Just as the tension $\mathbf{F}$ is the 
Noether current associated with translational invariance, there is a 
additional conserved current $\mathbf{G}$, associated with the invariance under the composition of  an inversion, a translation, followed by another inversion that undoes the first one. 
In quantum field theory, physically consistent scale invariant theories invariably turn out to exhibit the larger symmetry, as discussed, for example, in reference \cite{NakayamaReview}.  Classically, there is no corresponding inevitability; although 
D'Arcy Thompson, and others before him, appreciated  the role of conformal invariance in biological growth processes, where resources need to be optimized.  The conformal invariant energy along a curve may not be the obvious simplest scale-invariant choice. The additional symmetry does, however, imply an additional conservation law which constrains its equilibrium states significantly.
At third order in derivatives, the simplest conformally invariant energy along curves is the conformal arc-length, given by  \cite{Sharpe1994,Musso1994}, 
\begin{equation}
H= \int ds \, \big( {\kappa'}^2 + \kappa^2 \tau^2  \big)^{1/4}\,,
\label{eq:Hdef}
 \end{equation}
Here $\kappa$ and $\tau$ are the Frenet curvature and torsion, $s$ is arc-length and prime denotes a derivative with respect to $s$.  This energy is the unique conformal invariant at this order in derivatives.
\\\\
 In reference \cite{Paper2},  
a variational framework was developed to construct the equilibrium states of the three-dimensional conformal arc-length.  Here,  the details of this construction will be completed for those states that exhibit self-similarity.  These states are necessarily tension-free. The reason is simple; tension introduces a length scale.  Because the tension is not conformally invariant,  this symmetry is broken when it vanishes.  
\\\\
In \cite{Paper2},  the four conserved currents associated with conformal invariance, the tension $\mathbf{F}$, the torque $\mathbf{M}$, the scaling current $S$ and the special conformal current $\mathbf{G}$ were identified. 
\\\\
The two independent parameters characterizing tension-free curves are  the magnitude of the torque $M$ and the scaling current $S$. This independence contrasts sharply with the planar reduction where  the two are constrained to satisfy $4MS=1$ \cite{Paper1}.  
Setting the  tangential tension to zero, the conserved scaling current  $S$ completely determines the Frenet torsion in terms of the curvature. It also bounds its magnitude, while placing a lower bound on the monotonic falloff of the Frenet curvature along a spiral. 
\\\\
Modulo the vanishing  tension, the 
conservation of torque provides a quadrature for the dimensionless variable $\Sigma=-\kappa'/\kappa^2$ with a potential controlled by the parameters $M$ and $S$ describing periodic self-similar behavior.  Integrating first the quadrature to determine $\Sigma$, a second integration determines $\kappa$ (and $\tau$) as a function of arc-length;  so in principle, one could stop here and use the fundamental theorem of  curves to reconstruct the trajectory in space \cite{doCarmo}. One would, however, be at a loss to correlate the periodic behavior described by the quadrature with the self-similar structure that reveal themselves when the trajectories are traced. This is where the special conformal current $\mathbf{G}$ associated with the additional symmetry comes into play.  In a tension-free equilibrium state, this current is not translationally invariant, but it does transform by a constant vector. If the origin is located at the spiral apex, $\mathbf{G}$ vanishes. This is consistent with  
the duality between $\mathbf{F}$ and $\mathbf{G}$  evident in the Lie algebra of the conformal group.  
\\\\
Modulo the torque quadrature, 
the vector identity $\mathbf{G}=0$ reveals the internal structure within the spiral geometry in a polar chart adapted to the spiral apex and the torque axis. 
One discovers that an irreducible cycle is associated with each spiral describing its nutation between two fixed circular cones  as it advances about the torque direction. Scaled and rotated appropriately,  the complete spiral is generated by iterating a single cycle.
The spiral grows monotonically out from its apex as the cycle precesses about the torque axis.
This structure  is not evident in the Frenet data.  
\\\\
In parameter space Bernoulli's planar logarithmic spirals sit where $M$ and $S$ are fine-tuned to their planar values: $4M S=1$.  Subcritical ($4MS<1$) and supercritical ($4SM>1$) spirals differ qualitatively in significant ways.
\\\\
If $4MS>1$,  the spiral nutates between two identical oppositely oriented cones as it expands, the precession of these cycles forming an expanding \textit{rosette} bounded by the two cones.  The torsion changes sign within each nutation cycle.
This sign change is correlated with the reversal of the 
projection along the torque axis. The spiral twists one way on its way up along the axis and untwists on the way down.
Just as a growing planar logarithmic spiral  intersects all lines on the plane, supercritical spirals will intersect every plane as they expand.  
\\\\
While every supercritical spiral can be reached by a continuous deformation of a planar logarithmic spiral,
if $S$ is fixed, however, there will be a lower bound: if $2S>1$, the conical opening tends smoothly towards $\pi/2$ as $4MS\downarrow1$, coinciding with a planar logarithmic spiral; if $2S<1$, however, the polar angle defining the conical opening tends in this limit towards an angle, strictly less than $\pi/2$, so that the limiting spiral is neither planar nor logarithmic: it exhibits internal structure but does not nutate. If $M$ is large, on the other hand, the distinction between large and small $S$ dissolves: the limiting cones close onto the poles so that the spiral nutates from one pole to the other, free to range throughout three-dimensional space as this expanding pattern precesses about the torque axis. Three consecutive cycles of such a spiral  are illustrated in Figure \ref{Fig:Template}. 
\\\\
The simplest analogues of logarithmic spirals---expanding helices on a cone---familiar
in the computer graphics literature (eg. \cite{Harary2011})---do not occur as small deformations of logarithmic spirals.  In general, spirals with $4MS <1$ nutate between two coaxial cones oriented alike, rising monotonically along the torque axis. As $4MS\uparrow1$,  the outer cone splays out into a plane; whereas its inner counterpart does not; the asymptotic behavior of the limiting spiral is logarithmic but its behavior near its apex is not. It will be shown that,  while the approach to the limit $4MS =1$ from above and below differ, the two limiting geometries $4MS\to1$ are identical.
\\\\
The behavior of supercritical spirals may provide a clue 
to understanding an intriguing process in plant biology:
the spiraling motion of the tip of the growing tendril of a cucumber or any other climbing plant, described by Charles Darwin, and dubbed circumnutation by him \cite{Darwin1880}.  A relatively qualitatively recent 
review from a plant biologist's perspective  is provided in reference \cite{Stolarz2009}.  It is also quite instructive   
to look at one  of the numerous time-lapse videos posted on youtube illustrating 
the elaborate exploration of its spatial environment made by the tendril as it grows \cite{youtube}. The trajectory of its tip approximates at least qualitatively over several revolutions 
one of the tension-free supercritical trajectories constructed in this paper.  It can be argued that this is probably not a coincidence: gravity may select the torque axis, but in a first approximation, 
the tendril does not have any external yardstick against which to measure its progress through its environment: this ignorance would be expected to be reflected in a scale invariant trajectory.  
This trajectory is not a simple conical corkscrew typically adopted by growing seedlings,
even though it could be mistaken for one in a  single cycle. 
Tendril growth and seeding growth serve different purposes. The tendril is  searching for a support it can attach to;
if it fails to find one it will wither and die.  Evolutionary pressure is likely to have selected an optimal search strategy to find this support without the benefit of a marker to assess its progress.  The projected \textit{rosette} patterns, described and illustrated qualitatively in \cite{Stolarz2009},  can be fitted to  a supercritical spiral with a large $4MS$ ratio.  In this regime, the growth rate is the lowest.\footnote{\sf  
One should not, of course, conflate the dynamical trajectory (the motion of the tendril tip) with the conformation of the tendril itself.}  
First, of course, it would be useful to trace tendril trajectories accurately and compare them to self-similar templates.  The focus of  recent work in this direction 
\cite{Stolarz2014} has been on the upward motion of seedlings.  Unlike tendril growth, however, there is no reason to expect these trajectory to be self-similar.  A challenge for the future is to construct a  model predicting the tendril trajectory. In this context,  it is noteworthy that a recent mathematical model of the motion of growing seedlings does predict conical helices  \cite{Meroz2016}.  
\\\\
Supercritical spirals also provide a pointer towards a solution of a not-unrelated three-dimensional analog of a two-dimensional problem addressed in the computer science literature \cite{Finch2}. A  hapless swimmer is lost at sea and cannot see the shore (a line), nor do they know how far away it is. What course should they navigate to improve their odds of reaching this  shore? It has been conjectured that it is  a logarithmic spiral.  Finch et al. also contemplated  \textit{the three-dimensional analog, for which shores are planes in space},  which let them to ask \textit{if an appropriate extension of spiral has ever been
examined in the past}\cite{Finch2}. It appears that natural selection has been nudging growing plant tendrils towards a solution of this scale-free three-dimensional problem for the past 200 or so million years. 

\section{ The Conserved Currents }

Consider an arc-length parametrized curve $s\to \mathbf{X}(s)$ in three-dimensional Euclidean space with the inner product between two vectors denoted by a centerdot separating them. 
Let prime denote a derivative with respect to arc-length  so that $\mathbf{t}=\mathbf{X}'$ is the unit  tangent vector to the curve. Also
let $\{\mathbf{t}, \mathbf{N}, \mathbf{B}\}$ denote 
the Frenet frame adapted to this curve, satisfying $\mathbf{t}' = -\kappa \mathbf{N}$, 
$\mathbf{N}'= \kappa\,\mathbf{t} - \tau \,\mathbf{B}$,
and $\mathbf{B}'=\tau\,\mathbf{N}$, where $\kappa$ and $\tau$ are the Frenet curvature and torsion respectively. 
\\\\
In \cite{Paper2},  the Euler Lagrange equations and conserved currents 
associated with the conformally invariant energy (\ref{eq:Hdef})  were derived.  The most general 
conformal transformation induced on the curve is of the form, 
\begin{equation}
\label{eq:delX}
\delta \mathbf{X} = \mathbf{a} + \mathbf{b}\times \mathbf{X} + \lambda \,\mathbf{X} +  \delta_\mathbf{c} \mathbf{X}\,,
 \end{equation}
consisting of a translation $\mathbf{a}$, a rotation characterized by its axial vector $\mathbf{b}$, rescaling by $\lambda$
and a special conformal transformation $\delta_\mathbf{c} \mathbf{X}$  (the composition of an inversion in a sphere with a translation followed by a second inversion, linearized in the intermediate translation $\mathbf{c}$), given by 
$\delta_\mathbf{c} \mathbf{X} = |\mathbf{x}|^ 2 \, \mathrm{R}_\mathbf{X} \,  \mathbf{c}$,
where $\mathrm{R}_\mathbf{X}$ is the linear operator describing a reflection in the plane orthogonal to $\mathbf{X}$,
$\mathrm{R}_\mathbf{X} =
{\sf 1} - 2 \hat{ \mathbf{X}} \otimes \hat{\mathbf{X}}$ (the hat denotes a unit vector).
If the energy is conformally invariant, then
\begin{equation}
\label{eq:delHdef}
\int ds\, \mathbf{F}' \cdot \delta\mathbf{X} +  \int ds\, \mathcal{J}' =0\,,
 \end{equation}
where 
\begin{equation}
\label{eq:Jdef} 
\mathcal{J}= \mathbf{a}\cdot \mathbf{F} + \mathbf{b}\cdot \mathbf{M} + \lambda \,S  + 
 \mathbf{c}\cdot\mathbf{G}\,.
 \end{equation}
Here $\mathbf{F}$ is the tension, $\mathbf{M}$ is the torque, $S$ and $\mathbf{G}$
are  the scaling and special  conformal currents respectively.
In equilibrium, $\mathbf{F}'=0$ and each of the currents is conserved so that $\mathbf{F}$, $\mathbf{M}$, $S$, and $\mathbf{G}$ are constant along the curve.
\\\\
When $H$ is  the conformal arc-length (\ref{eq:Hdef}), 
the torque is given by  \cite{Paper2}
\begin{equation}
\label{eq:Mdef}
\mathbf{M}= \mathbf {X} \times \mathbf{F} - H_1  \mathbf{N}  + H_2 \mathbf{B}
- 2\mathcal{S}_{12} \mathbf{t}\,,
\end{equation}
where 
\begin{subequations}
\label{eq:H12S12}
\begin{eqnarray}
2H_1 
&=& 
- (\mu^3  \kappa')'+\mu^3 \kappa \tau^2 \,;\\
2 H_2 
&=& 
 - (\mu^3\kappa^2\tau)'/\kappa\,;\\
2\mathcal{S}_{12} 
&=&  
- \mu^3 \kappa^2 \tau/2\,,
\end{eqnarray}
\end{subequations}
and the shorthand 
\begin{equation}
\label{eq:mudef}
\mu=\big( {\kappa'}^2 + \kappa^2 \tau^2  \big)^{-1/4}
\end{equation}
is introduced. 
The scaling current is given by 
\begin{equation}
\label{eq:Scaledef}
 S=  -\mathbf{F}\cdot 
\mathbf{X} 
 + S_D\,,\end{equation}
where 
\begin{equation}
\label{eq:SDcal}
S_D := - \kappa\,{\partial \mathcal{H}}/{\partial \kappa' }  = -\mu^3\kappa \kappa'/2\,.
\end{equation}   
Finally the special conformal current is given by 
\begin{eqnarray}
\mathbf{G}
= 2 \mathbf{X}\times (\mathbf{M} - \mathbf{X}\times\mathbf{F}) 
-   |\mathbf{X}|^2 \mathbf{F}-  2 S\,  \mathbf{X}
-\, \mu^3 (\kappa' \, \mathbf{N} +  \kappa \tau\, \mathbf{B}) 
\,,\label{eq:GdefMS}
\end{eqnarray}
involving the three conserved currents $\mathbf{F}$, $\mathbf{M}$ and $S$.
\\\\
Notice that $S$ and $\mathbf{M}$ possess the same dimensions as the energy, $H$. As such they are dimensionless.
The tension has dimensions of inverse length; the conformal current $\mathbf{G}$ has the dimension of length. An explicit expression for $\mathbf{F}$ has not been written down: this is because it is never used explicitly in this paper. Its construction, like that of the other currents, is presented in reference \cite{Paper2}.

\section{Tension-free States}

First one looks at the implications of the 
conserved scaling current  in tension-free states.

\subsection{Conserved scaling current}

If $\mathbf{F}=0$,  Eq.(\ref{eq:Scaledef}) implies that
$S_D$ itself will be conserved. Using the identity Eq.(\ref{eq:SDcal}), the conservation law   
can be cast as the statement 
\begin{equation}
 \mu^3 \kappa \kappa'/2 = -S\,,
\label{eq:Fparzero}
 \end{equation}
where $S$ is the constant value of the scaling current. If $S>0$, then $\kappa'<0$.  Introducing the 
 dimensionless ratios $ \Sigma$ and $\Gamma$, defined by
\begin{equation} \Sigma = (-) 
\frac{\kappa'}{\kappa^2}\,;\quad \Gamma = \frac{\tau}{\kappa}\,,
\label{eq:SigGamdef}
\end{equation}
Eq.(\ref{eq:Fparzero}) can be recast as an algebraic constraint connecting these two variables:
\begin{equation}
\label{eq:GamSig0}
\Big[
 \Sigma^2 + \Gamma^2 \Big]^3
=  \frac{1}{16S^4}\, \Sigma^4
\,,\end{equation}
which can be recast as the identity  
\begin{equation}
\label{eq:GamSig}
\Gamma^2
= \Sigma^2 \left(\frac{1}{(4S^2 \Sigma)^{2/3}}-1\right)
\,.\end{equation}
The torsion at a point along the curve is completely determined by the value of the curvature and its first derivative. 
The awkward fractional powers appearing in Eq.(\ref{eq:GamSig}) can be avoided by introducing   
a change of curvature variable and parameter, 
$\gamma$: 
\begin{subequations}
\label{eq:defsZgamma}
\begin{eqnarray}
Z &=& \Sigma^{2/3}>0\,;\\
\gamma &=& (2S)^{-4/3}\,.
\end{eqnarray}
\end{subequations}
With respect to these variables Eq.(\ref{eq:GamSig}) assumes the simple quadratic form 
\begin{equation}
\label{eq:GamZ}
\Gamma^2 = Z^2 (\gamma - Z)\,.
\end{equation}
Now examine the consequences of this constraint.

\subsubsection{ Vanishing torsion and a  lower bound on $\kappa$}
\label{kappabound}

Eq.(\ref{eq:GamZ}) places an upper bound on $Z$: $Z\le \gamma$,  and this bound is
saturated when $\tau=0$ and only then.\footnote{\sf 
In section \ref{Torquequadrature}, it will be seen that $Z'=0$ when $Z=\gamma$.  As shown there, while $\tau=0$ when $Z=\gamma$,  $\tau'$ doe not vanish.}
 In terms of $\Sigma$, 
\begin{equation}
|\Sigma| \le \frac{1}{(2S)^2}\,;
\label{eq:Sigmabound}
\end{equation}
if  $Z=\gamma$ everywhere along the curve, it will necessarily be planar with $\kappa=  (2S)^2/s$, 
which describes a logarithmic spiral.  Details  are provided in reference \cite{Paper1}.
\\\\
Suppose that  $\kappa'$ is negative somewhere. 
In section \ref{MotioninPotential}, in the context of the conservation of torque,
it will be seen that $Z=0$ is  never accessible, unless trivially everywhere along the curve.   
This implies that 
$\kappa'$ must be negative everywhere. Thus the curvature decreases monotonically with $s$. Integrating across the inequality (\ref{eq:Sigmabound}) one then sees that 
\begin{equation}
\kappa  s \ge 4S^2\,,
\label{eq:kappabound}
\end{equation} 
so that the scaling current $S$ places a lower bound on the falloff of $\kappa$.  This justifies  
the interpretation of $S$ as a dimensionless measure of curvature.  
\\\\
According to Eq.(\ref{eq:GamZ}),
the torsion vanishes when either $Z=0$ or $Z=\gamma$. 
In section \ref{MotioninPotential}, it will be seen that $Z=\gamma$ is not always accessible; and as pointed out, $Z=0$ is  never accessible.

\subsubsection{An upper bound on $\tau$ by $\kappa$}
\label{torsionbound}

It is evident from Eq.(\ref{eq:GamZ}) that
$\Gamma$ has a maximum with respect to $Z$ when $Z_{\Gamma\,{\sf max}}=2\gamma/3$ if this value of $Z$ is accessible.  Accessibility will be addressed in \ref{tau/kappabound}. There is now a bound on $\tau$ by $\kappa$, given by 
\begin{equation}
\label{eq:Gammabound}
\tau^2 \le  4\gamma^3 \kappa^2 /27 = \kappa^2 /(108 S^4) \,.
\end{equation} 
Thus scale invariance bounds the magnitude of the torsion by the curvature. 
Significantly, the bounds (\ref{eq:kappabound}) and (\ref{eq:Gammabound}) are independent of the magnitude of the torque $M.$
\\\\
Eq.(\ref{eq:GamSig0}) suggests the existence of  non-planar  equilibrium states with constant values of both $\Sigma$ and $\Gamma$. 
The conservation of torque will, however, imply that $\Sigma$ cannot be constant unless  the magnitude of the torque is fine-tuned, and this tuning is only possible if $S$ itself is bounded: $2S<1$.

\subsection{Torque conservation provides a quadrature for $\Sigma$}
\label{Torquequadrature}

\begin{figure}[htb]
\begin{center}
\includegraphics[height=6cm]{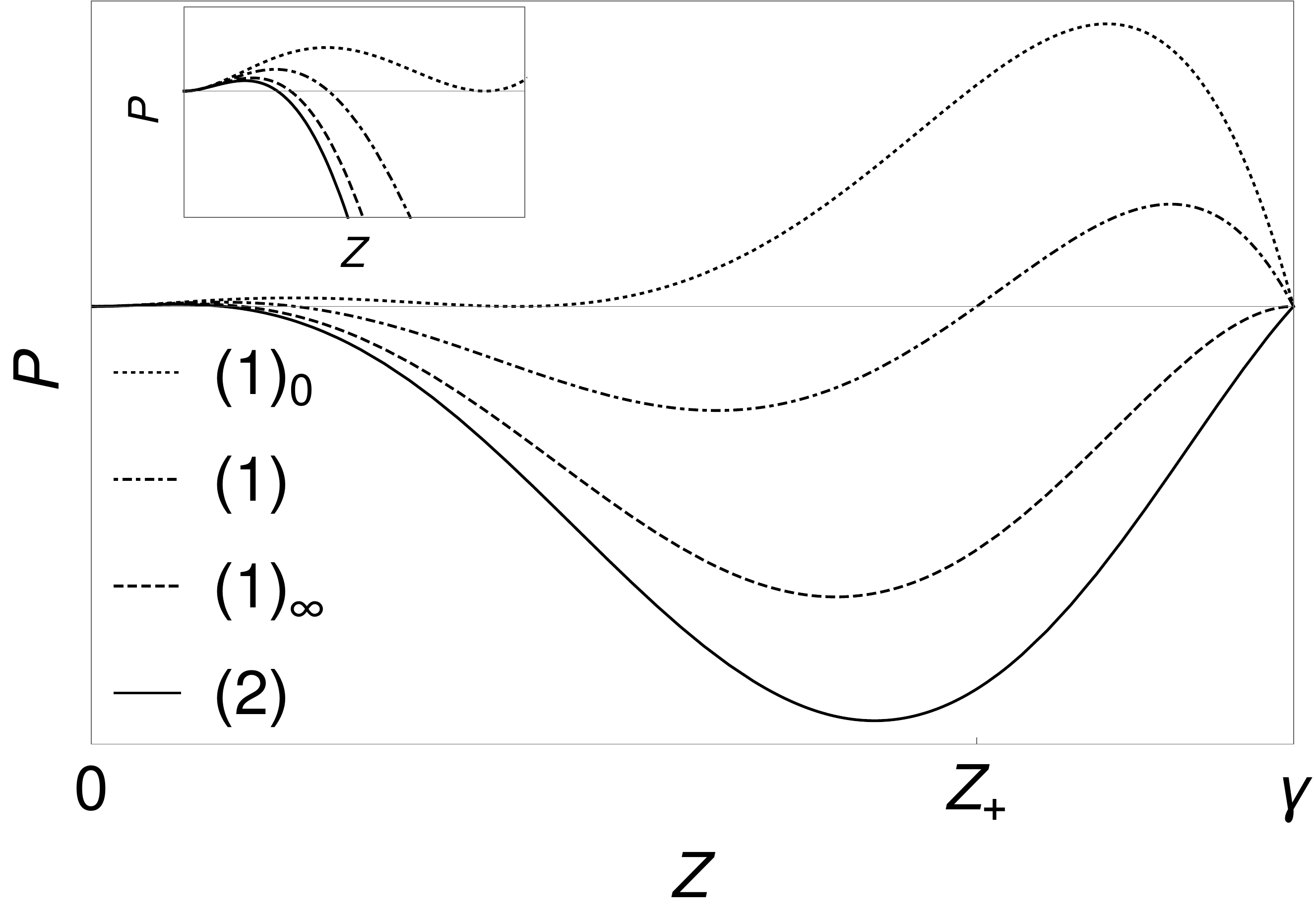}
\end{center}
\caption{\small \sf The potential for fixed $\gamma=2$ and four distinct increasing values of $m$. $(1)_0$ $m=2.16$ (the two 
roots of the quadratic coincide within the interval $[0,2]$); $(1)$ $m=2.5$ (the two roots lie within the interval $[0,2)$); $(1)_\infty$ $m=2.828$ ($Z_+=2$);
$(2)$ $m=3$ ($Z_+>2$).  The position of $Z_+$ is indicated in case $(1)$; in general it moves from left to right as $m$ increases. The qualitative behavior of the state will depend sensitively on the value of $m$  (cf. section \ref{MotioninPotential}).}\label{Potential1}
\end{figure}

The consequences of torque conservation in tension-free states is now explored. 
\\\\
In a tension-free state, the constant torque vector defined by Eq.(\ref{eq:Mdef}), like $S$, becomes translationally invariant; it will define the spiral axis. 
Its squared magnitude, $\mathbf{M}^2$, like $S$ is Euclidean invariant.  Its conservation provides a quadrature for the curvature variable 
$\Sigma$, modulo the 
conservation of $S$ as captured by  Eq.(\ref{eq:GamSig})) which
allows $\Gamma=\tau/\kappa$ to be eliminated in favor of $\Sigma$. 
\\\\
To see this, note that, along tension-free curves,  the expressions for  $H_1$, $H_2$ and $S_{12}$  appearing in Eq.(\ref{eq:Mdef}), given by 
Eqs.(\ref{eq:H12S12}),  can be 
cast in terms of $Z$, ${Z^\bullet}$,  and ${\sf sign}\,(\tau)$: 
\begin{subequations}
\label{eq:HISIJ}
\begin{eqnarray}
H_1 
&=&  S\,[\Sigma^2 + \Gamma^2] /\Sigma =  S \gamma Z^{1/2}  \,;
\\
 H_2 
&=&-S [ \Gamma/\Sigma]'/\kappa 
= {\sf sign}(\tau) S \gamma Z^\bullet/ (2  Z^{3/2} (\gamma  -Z)^{1/2})\,;
\\
2\mathcal{S}_{12} 
&=&
 S  \Gamma /\Sigma  =  {\sf sign}(\tau) S (\gamma -Z)^{1/2}/Z^{1/2}
\,,\end{eqnarray}
\end{subequations}
where the identities 
\begin{subequations}
\label{eq:GamZGambul}
\begin{eqnarray}
(\Gamma/\Sigma)^2 &=& \gamma Z^{-1} -1\,;\\
(\Gamma/ \Sigma)^\bullet &=& - {\sf sign}\,(\tau) \gamma Z^\bullet/ (2  Z^{3/2} (\gamma  -Z)^{1/2})\,,
\end{eqnarray}
\end{subequations}
following from
Eq.(\ref{eq:GamSig}) and the redefinitions following it, have been used. In Eq.(\ref{eq:HISIJ}b) and (\ref{eq:GamZGambul}b) the variable $\Theta$ is introduced, defined by $d\Theta = \kappa ds$; the dot (or bullet) from here on signifies a derivative with respect to $\Theta$ (not conformal arc-length as used in reference \cite{Paper2}). 
\\\\
Using 
Eq.(\ref{eq:Mdef}) with $\mathbf{F}=0$, and
the identities collected in Eqs.(\ref{eq:HISIJ}) the constant magnitude of the torque $M^2=\mathbf{M}^2$ is now given by 
\begin{equation}
\label{eq:Msquared}
\mathbf{M}^2 
=
S^2 \, \left( \gamma^2 Z  
+  \gamma^2 \,\frac{{{Z^\bullet}}^2}{4 Z^3 (\gamma -Z)} + \gamma/Z -1  \right)\,.
\end{equation}
The distinction between positive and negative torsion does not 
feature yet but it will further on. Eq.(\ref{eq:Msquared}) can be rearranged to provide the quadrature for $Z$: 
\begin{equation}
\label{eq:Quadrature}
\frac{\gamma^2}{4}  {Z^\bullet}^2 + P(Z; \gamma,m) =0\,,
\end{equation}
where the potential $P$ is a fifth order polynomial in $Z$ given by
\begin{equation}
\label{eq:Pdef}
P=  Z^2 (\gamma -Z) \Big[\gamma^2 Z^2  - (m^2 +1) Z  + \gamma \Big]\,.
\end{equation} 
Here the ratio of the torque to the scaling constant $m$, defined by
$m^2 = M^2/S^2$, 
has been introduced.
The quadrature Eq.(\ref{eq:Quadrature}),  unlike Eq.(\ref{eq:GamZ}) (or, equivalently, Eq.(\ref{eq:GamSig})), 
involves the torque $M$ (through the ratio $m$) as well as the scaling constant $\gamma$.  Each admissible set of values of $\gamma$ and $m$  will parametrize a unique tension-free state. The immediate task is to identify the 
admissible values of these parameters. The potential is plotted  with a fixed value $\gamma=2$, for various significant values of $m$ in Figure \ref{Potential1}. 

\subsection{Motion of a particle in a potential}
\label{MotioninPotential}

\begin{figure}[htb]
\begin{center}
\subfigure[$(\gamma,m)$ Plane]{\includegraphics[height=6cm]{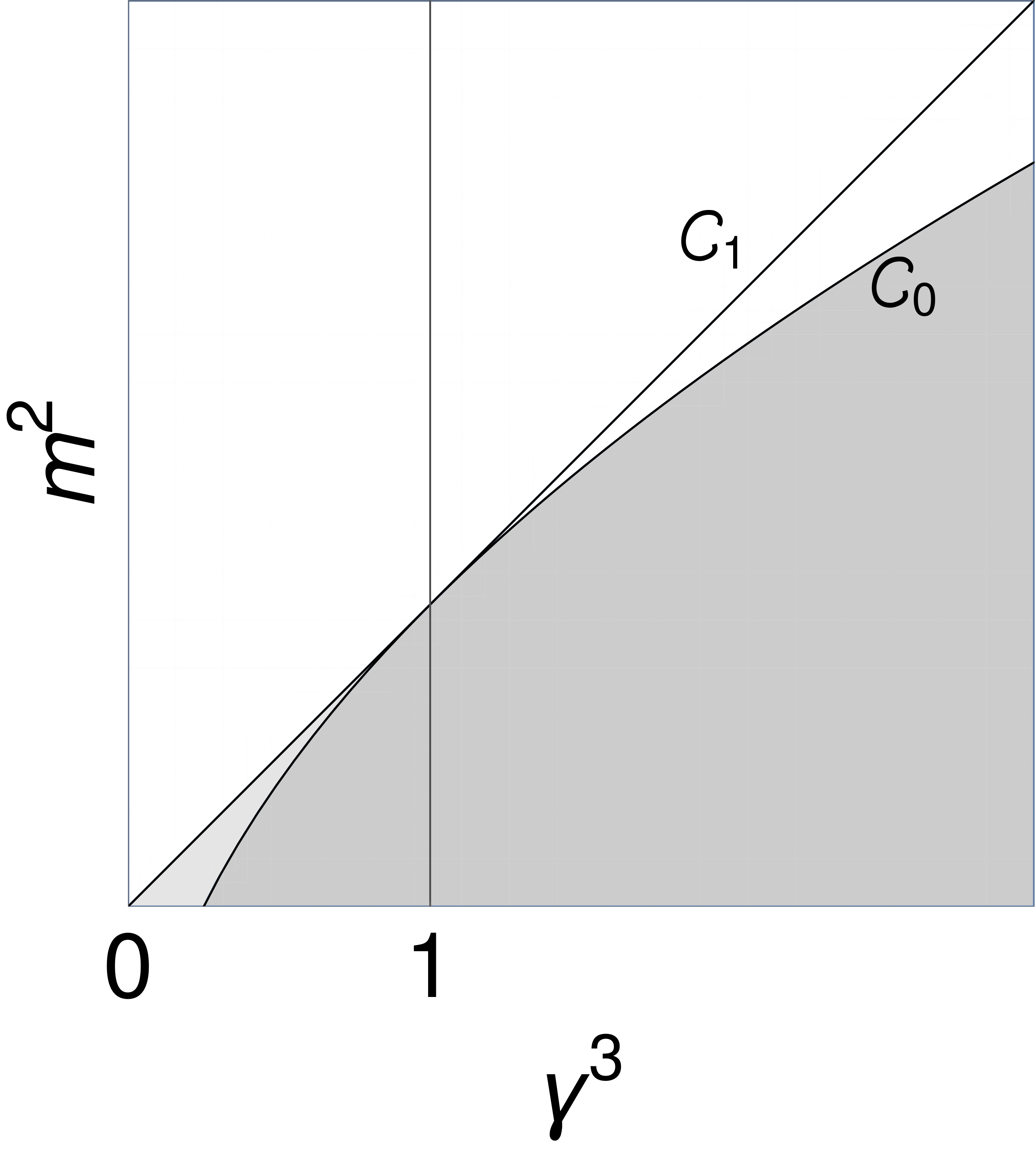}}
\hfil
\subfigure[$(S,M)$ Plane]{\includegraphics[height=6cm]{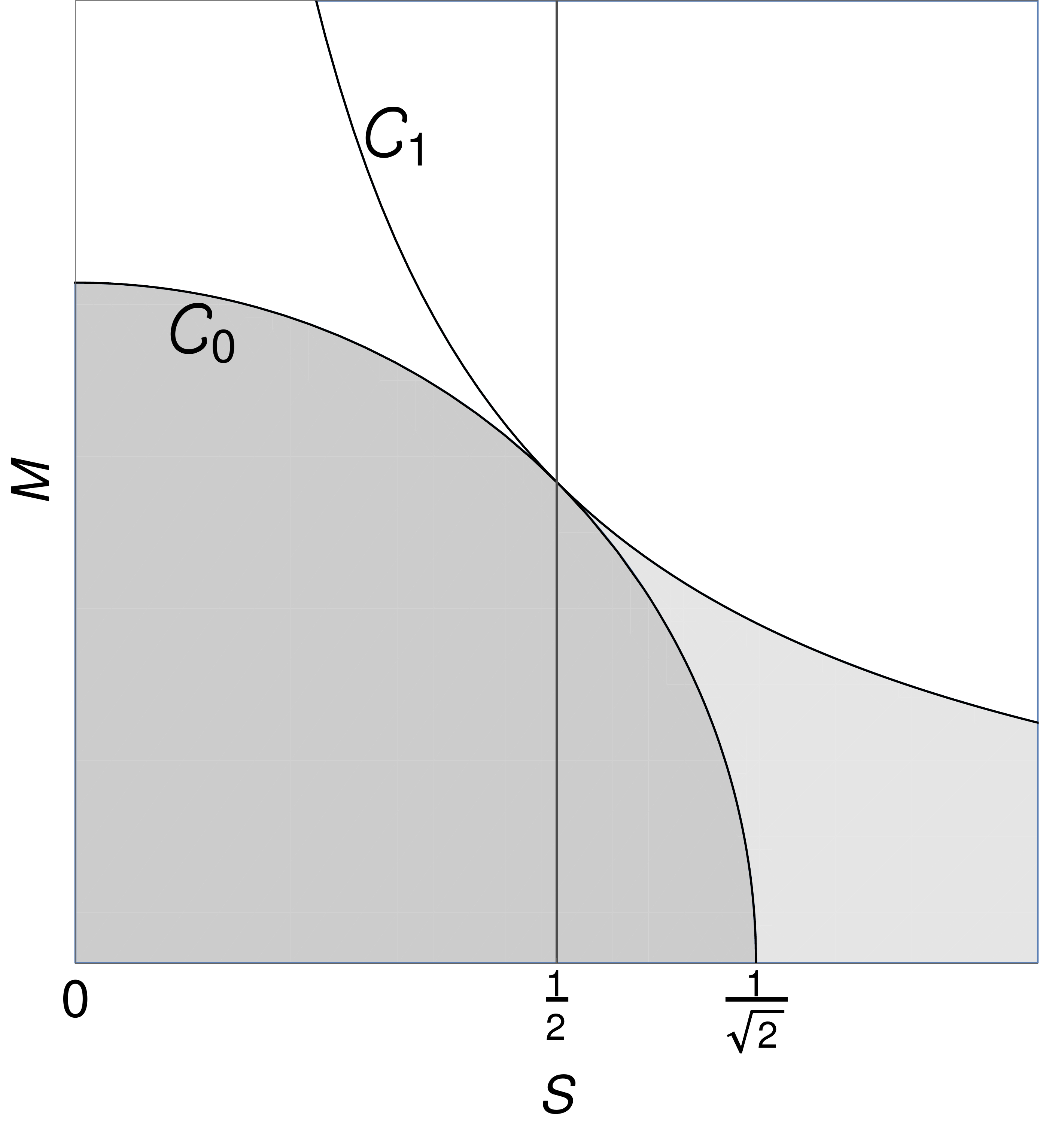}}
\caption{\small \sf
Parameter Space: The curves  $\mathcal{C}_0$ and $\mathcal{C}_1$, defined  respectively by $(m^2+1)^2 - 4\gamma^3=0$ and $m^2 = \gamma^3$, are displayed in  Figure \ref{Fig:Phases}a. Their representation 
with respect to the original parameters $M$ and $S$ are displayed in (b): $\mathcal{C}_0$ is defined there by 
 $M^2 + S^2=1/2$;  $\mathcal{C}_1$ is defined by
$4 M S = 1$. $\mathcal{C}_0$ and $\mathcal{C}_1$ touch with a common tangent at a single point 
$\gamma=1=m$, or equivalently $S=1/2=M$.
 (i)  
Each point along the phase boundary $\mathcal{C}_1$ represents a logarithmic spiral.  
If  $\gamma<1$  the limit towards $\mathcal{C}_1$ from above coincides with this logarithmic spiral; if $\gamma>1$, however,  the limit is neither logarithmic nor planar  (cf. \ref{Limits}).  
(ii) The (gray) region below $\mathcal{C}_0$, as well as the (light gray) region
bounded between $\mathcal{C}_0$ and $\mathcal{C}_1$ when $\gamma<1$ ($S>1/2$), are not accessed in tension-free states. This breaks the (unexpected) symmetry between $M$ and $S$ in Figure \ref{Fig:Phases}(b). 
When $\gamma>1$ the limit towards $\mathcal{C}_1$ from below 
coincides with its non-trivial counterpart in the limit from above (cf. \ref{Limits}). 
 (iii)  Each point along $\mathcal{C}_0$ with $\gamma>1$ represents  a conical helix (cf. \ref{conehelix}).(iv) 
Each point between $\mathcal{C}_0$ and $\mathcal{C}_1$ with $\gamma>1$ represents a rising spiral nutating between two nested coaxial cones; above $\mathcal{C}_1$,  the spiral nutates between opposite oriented cones (cf. section \ref{Coscycle}).
This structure, not evident in the quadrature,
will emerge as a consequence of the conserved special conformal current. 
} \label{Fig:Phases}
\end{center}
\end{figure}

The analysis of the quadrature is facilitated by interpreting it in terms of the motion of a particle in a potential. For the moment, ignore  the 
fact that $Z$ (or indeed $\Sigma$) is a composite variable, involving derivatives of the Frenet curvature, or indeed that the conservation law for $\mathbf{G}$, defined by  Eq.(\ref{eq:GdefMS}), has yet to be examined.
\\\\
The potential $P$ possesses a double root at $Z=0$ and another at $Z=\gamma$. The variable $Z$, by definition, is positive 
and Eq.(\ref{eq:Sigmabound}) bounds $Z$ by $\gamma$.  The two roots of the quadratic are given by \footnote{\sf
If these roots are not real, the potential will be positive everywhere within the interval $[0,\gamma]$ and, as a consequence the only  accessible values of $Z$ are $Z=0$ and $Z=\gamma$, corresponding respectively to a circle and a planar logarithmic spiral.}
\begin{equation}
\label{eq:roots}
Z_\pm = \frac{(m^2 +1)}{2\gamma^2} \pm \frac{1}{2\gamma^2} \sqrt{ (m^2+1)^2 -4\gamma^3}\,.\end{equation}
{\bf The origin $Z=0$ is isolated}: 
Note that the smaller root $Z_-$ is strictly positive for all finite $m$; this implies that the origin is always isolated: as a consequence there are no tension-free deformations of a circle and no non-trivial trajectory reaches it. Whereas Eq.(\ref{eq:GamZ}) did not rule out $\tau$ vanishing at $Z=0$, the quadrature does in any tension-free trajectory: thus, as anticipated in section \ref{kappabound}, $\tau=0$ if and only if $Z=\gamma$.
\\\\
{\bf A lower bound on the torque}:
The reality of the roots in \ref{eq:roots}) places a lower bound on 
$m$,
\begin{equation}
\label{eq:m22lessgam3}
(m^2+1)^2 \ge 4 \gamma^3\,.
\end{equation} 
There is thus a minimum torque $M$ for each legitimate value of $S$
in tension-free states.  
\\\\
The two roots of the quadratic coincide when $m$ and $\gamma$ are tuned such that the bound (\ref{eq:m22lessgam3}) is saturated, or 
\begin{equation}
\label{eq:calC0}
(m^2+1)^2 = 4 \gamma^3\,.
\end{equation}
The curve in parameter space, described by Eq.(\ref{eq:calC0})  is indicated $\mathcal{C}_0$ in Figure \ref{Fig:Phases}. 
The common root is given by $Z_0= {(m^2 +1)}/{2\gamma^2}= \gamma^{-1/2}$, or equivalently, $\Sigma_0 = 2S$. 
\\\\
{\bf  Conical Helices:  $(m^2+1)^2 = 4 \gamma^3$, $\gamma\ge1$}
\\\\
To be accessible, the coincident root must also lie below $Z=\gamma$, the bound 
derived in section \ref{kappabound}.  This requires $\gamma\ge1$ (or equivalently $2S\le 1$).  
The potential  is plotted in Figure \ref{Potential1} for $\gamma=2$, 
where it is labeled $(1)_0$.
The corresponding geometry is  a rising spiral helix with a fixed pitch wound upon a circular cone.  These states are constructed explicitly in \ref{conehelix}.  
\\\\
{\bf  Deepening the well}:
When Eq.(\ref{eq:m22lessgam3}) is satisfied with strict inequality,
the quadratic possesses two distinct positive roots.  Three possibilities can be distinguished which depend on the location of $Z_-$ and $Z_+$ with respect to $Z=\gamma$:  
both $Z_-,Z_+ <\gamma$ (subcritical); $Z_-<\gamma<Z_+$ (supercritical) and 
both $Z_+,Z_->\gamma$. One of two roots of the quadratic must also lie below  $\gamma$ for otherwise the potential would not be accessible. This rules out the third possibility just as it did with coincident roots. The condition $Z_+=\gamma$, partitioning the accessible region of parameter space into supercritical and subcritical regions, is given by
\begin{equation}
\label{eq:calC1}
m^2 = \gamma^3\,,
\end{equation}
is represented by the curve $\mathcal{C}_1$ in Figure \ref{Fig:Phases}.  The curves $\mathcal{C}_0$ and $\mathcal{C}_1$ touch with a common tangent at the point $(1,1)$ on the $(\gamma,m)$ plane. 

\vskip1pc\noindent {\bf Subcritical Torque: $2\gamma^{3/2} -1 < m^2 < \gamma^3$; $\gamma\ge1$} 
\\\\
This is the region of parameter space bounded between $\mathcal{C}_0$ and $\mathcal{C}_1$, with $\gamma>1$. $Z_+$ is bounded from above and $Z_-$ from below:
$Z_+ < [\gamma^3 +1 + |\gamma^3-1|]/2\gamma^2$, 
 $Z_-> [\gamma^3 +1 - |\gamma^3-1|]/2\gamma^2$.
\\\\ If $\gamma>1$, then $Z_+ <\gamma$.  In Figure \ref{Potential1}, the potential, labeled $(1)$ is plotted for parameter values $\gamma=2$, $m=2.5$. $Z$ oscillates  
within the band $[Z_-,Z_+]$, which does not include the point $Z=\gamma$.  As such, the torsion never vanishes.
Such states correspond to deformed helices bounded, as explained in section \ref{Coscycle},
between two cones. The two cones coalesce when the roots coincide: $Z_-=Z_+$.
\\\\
If $\gamma<1$,  then $Z_- > 1 > \gamma$, so that  the well is inaccessible. Thus the region of parameter space below $\mathcal{C}_1$ is inaccessible when $\gamma<1$.  This forbidden region is indicated
in light gray in Figure \ref{Fig:Phases}.
\\\\
{\bf Supercritical Torque: $m^2 > \gamma^3$}
\\\\\
Now $Z_+>\gamma$ for all $\gamma$, and $Z_-<\gamma$.
The upper turning point in the potential well occurs always at $Z=\gamma$ and  $Z$ oscillates  
within the band $[Z_-, \gamma]$.  Such a potential, labeled  $(2)$, is plotted in Figure \ref{Potential1} for the parameter values, $\gamma=2$, $m=3$. According to Eq.(\ref{eq:GamZ}), the torsion $\tau$ vanishes at the upper turning point $Z=\gamma$.  
\\\\
One can also see that $\tau^\bullet\ne 0$ when $\tau=0$ so that
$\tau$ changes sign at the upper turning point, $Z=\gamma$. 
To demonstrate this, it is sufficient to show that $\Gamma^\bullet\ne0$ at $Z=\gamma$.
The identity Eq.(\ref{eq:GamZ}) and the quadrature (\ref{eq:Quadrature}) together imply that 
\begin{equation}
\label{eq:Gammadot}
(\Gamma^\bullet)^2  = \frac{4}{\gamma^2} Z^{4} (Z - Z_-)(Z_+ - Z) \,,
\end{equation} 
when $Z=\gamma$, 
where $Z_-$ and $Z_+$ are the roots of the quadratic appearing in the quadrature given by (\ref{eq:roots}). Evidently $\Gamma^\bullet$ does not vanish unless $Z_-=\gamma$ or $Z_+=\gamma$, two cases that will be discussed in \ref{Limits}.
In section \ref{Coscycle},  it will be seen that the sign of $\tau$ correlates directly with the rise and fall of the spiral along the torque axis,
with $\tau=0$ at the the extremal values of the projection of the position vector along this axis.
 \\\\
 The behavior of the potential describing limiting subcritical ($m^2\uparrow \gamma^3$) and supercritical ($m^2\downarrow \gamma^3$) spirals is described in \ref{Limits}.

\subsection{Expressing $\kappa$ as a function of $s$}
\label{kappas}

For  each value of $S$ and $M,$ the quadrature  determines $Z$ as a function of $\Theta=\int \kappa ds$, given implicitly by
\begin{equation}
\label{eq:ThetaZ}
\Theta (Z) = - \frac{1}{2}
\, \int_Z^{{\rm min}\, (\gamma,Z_+)}\, \frac{dZ}{Z\sqrt{(\gamma-Z) (Z_+-Z) (Z-Z_-)}} \,.
\end{equation}
The right hand side can be expressed in terms of an Elliptic function of the third kind.  The choice was made not to dwell on this exact solution of the quadrature, suggesting---misleadingly in the author's view---that this exact form is essential for an understanding of the equilibrium geometry. 
\\\\
So long as $Z_+\ne \gamma$, the \textit{motion} in the potential is periodic in $\Theta$, with period $\Theta_0=2\Theta(Z_-)$. This is the relevant period if the spiral is subcritical; if it is supercritical, however, the sign change in $\tau$ at $Z=\gamma$ means that its period is doubled to $2\Theta_0$, corresponding not to one but to two complete oscillations in the well, 
\\\\
To determine $\kappa$ as a function of $s$,  use $\Sigma=-\kappa'/\kappa^2= -  \kappa^\bullet/\kappa$ to first express $\kappa$ as a function of $\Theta$,
\begin{equation}
\label{kapthet}
\kappa (\Theta)/ \kappa(0) =  e^{-\int_0^\Theta d\Theta_1\, Z^{3/2} (\Theta_1)}\,,\end{equation}
where $\kappa(0)$ is the value of $\kappa$ at the zero of $\Theta$.  It is now clear that 
$\kappa$ decreases exponentially with $\Theta$, modulated by the oscillation in the well with period $\Theta_0$: over one period
\begin{equation}
\label{kapthet0}
\kappa_0 = \kappa(\Theta_0)/\kappa(0) =  e^{- \int_0^{\Theta_0} d\Theta_1\, Z^{3/2} (\Theta_1)}\,.
 \end{equation}
Simple bounds can now be placed on $\kappa_0$ in terms of the turning points and the period:
$ e^{- Z^{3/2}_{\sf max} \Theta_0 } \le \kappa_0 \le  e^{- Z^{3/2}_- \Theta_0 } $,
where $Z_{\sf max}={\sf Min}(Z_+,\gamma)$. 
The behavior of $\kappa_0$ as a function of $\gamma$ and $m$ will be taken up in section \ref{SelfSim} in the context of scaling and self-similarity.  It is evident that $\kappa_0$ ranges from $0$ to $1$. The  factor $\kappa_0$ will depend sensitively on the parameters. 
\\\\
Arc length $s$ is related to $\Theta$ by $s^\bullet= \kappa^{-1}$, so that
\begin{equation}
\label{eq:sThet}
s (\Theta) - s(0) = \kappa_0^{-1} \int_0^\Theta d\Theta_1\,
e^{\int_0^{\Theta_1} d\Theta_2\, \Sigma (\Theta_2)}\,.
\end{equation}
Arc-length increases \textit{exponentially} with $\Theta$.  The constant $s(0)$  is the arc-length along the spiral from its apex to the position marking $\Theta=0$.
\\\\
Inverting Eq.(\ref{eq:sThet}) and substituting into (\ref{kapthet}), provides the functional dependence of $\kappa$ on $s$. 
Note that 
\begin{equation}
\kappa' = \frac{ \kappa^\bullet}{s^\bullet}= \Sigma (\Theta) e^{-2\int_0^\Theta d\Theta_1\, \Sigma (\Theta_1)}\,.
\end{equation}
The dependence of the torsion on $s$ is now determined by Eq.(\ref{eq:GamSig}). 
Whereas $\kappa$ decreases monotonically, $\tau$ generally does not.  
Having determined  $\kappa$ and $\tau,$ 
it is possible to appeal to the fundamental theorem of curves to reconstruct the space curve.   It is possible, however, to say much more  
by examining the  conserved special conformal current.  

\section{Conformal current and spatial trajectories}
\label{reconstruction}

In tension-free states, it was seen that the 
scaling current and torque are translationally invariant. The constant conformal current 
(\ref{eq:GdefMS}), given by  
\begin{equation}
\mathbf{G}_0/2=    \mathbf{X}\times \mathbf{M} 
- S\,\left[ \mathbf{X} - 
\left( \frac{1}{\kappa}\right)  \, 
\mathbf{N}
- \left(\frac{\tau}{\kappa'}\right)  \, 
\mathbf{B} \right]\,,
\label{eq:Gground2}
\end{equation}
is not.  In equilibrium, however, it transforms by a constant vector under translation; as a consequence,  with a judicious choice of origin, it is always possible to set $\mathbf{G}_0=0$.
This privileged point, perhaps not surprisingly, is the spiral apex.  

\subsection{The projection  along the torque axis $X_\|$}
\label{Xparpr0}

Setting  $\mathbf{G}_0=0$ in Eq.(\ref{eq:Gground2}),  projecting on $\mathbf{M}$, and using Eq.
(\ref{eq:Mdef}) for the projections of $\mathbf{M}$ onto the Frener normal vectors, one obtains the identity 
\begin{equation}
\label{eq:XonM}
X_\|= \mathbf{X}\cdot\hat{\mathbf{M}} =
 \frac{1}{M\kappa}\,\left(  H_2
-  \left(\frac{\kappa \tau}{\kappa'}\right)  H_1 \right)
\end{equation}
for the projection of the position vector along $\mathbf{M}$  in tension-free states.
Using Eqs. (\ref{eq:HISIJ}a) and (b), 
the term in brackets appearing on the right hand side of Eq.(\ref{eq:XonM})  can be cast in terms of 
$Z$ and ${Z^\bullet}$ and ${\sf sign}(\tau)$:
\begin{eqnarray}
\label{bracket}
 H_2 +
\left(\frac{\Gamma}{\Sigma}\right)  H_1
&=&
{\sf sign}(\tau) \frac{S}{Z^{3/2} (\gamma-Z)^{1/2}}\left(\frac{\gamma}{2} {Z^\bullet} +\gamma Z^{3/2} (\gamma-Z)\right)\,.
\end{eqnarray}
The quadrature (\ref{eq:Quadrature}) is now used to eliminate ${Z^\bullet}$, modulo its sign, in favor of the potential so that
\begin{equation}
\label{eq:XonM2}
\kappa  X_\| = {\sf sign} (\tau)  \frac{ \gamma }{m}
\frac{1}{ Z^{1/2} } W_\pm(Z)\,,
\end{equation}
where  the shorthands
\begin{equation}
\label{eq:Wdef}
W_\pm ={\sf sign} ({Z^\bullet})  \, Q+ Z^{1/2} 
(\gamma -Z)^{1/2}\,,
\end{equation}
as well as
\begin{equation}
\label{eq:Qdef}
Q(Z) =(Z-Z_-)^{1/2}(Z_+- Z)^{1/2} \ge 0\,,
\end{equation}
are introduced. Here $Z_\pm$ are the two roots of the quadratic, given by (\ref{eq:roots}). 
In general $W_+>  0$, and $W_+\ge W_-$, with equality at  $Z_\pm$. 
In a supercritical spiral, $W_-=-W_+$ at $Z=\gamma$; thus $W_-$ necessarily vanishes somewhere between $Z_-$ and $Z=\gamma$. Mid-plane crossing with $X_\|=0$ will be examined in section \ref{Coscycle}.
\\\\
In a supercritical spiral, Eq.(\ref{eq:XonM2}) describes four dimensionless functions of $Z$, 
one for each of the four possible sign pairings of $\tau$ and $Z^\bullet$: $(-,-), (-,+), (+,-)$  and $(+,+)$.  

\subsection{The polar radius with respect to the spiral apex}
\label{distance}

With $\mathbf{G}_0=0$, Eq.(\ref{eq:Gground2}) implies  
\begin{equation}
\label{eq:XcrossM}
M^2 X_\perp^2 + S^2 \rho^2  = \frac{S^2}{\kappa^2} \left(1 + \frac{\Gamma^2}{\Sigma^2}\right)=  \frac{S^2 \gamma}{\kappa^2 Z}\,,\end{equation}
where 
$\rho=|\mathbf{X}|$ is the polar radius, and $X_\perp= |\mathbf{X}\times \hat{\mathbf{M}} |$ is the distance to the torque axis.  The scaling identity has been used again to obtain the final expression. 
Eliminating $X_\perp$ in Eq.(\ref{eq:XcrossM}) in favor of $\rho$ and $X_\|$, and using Eq.(\ref{eq:XonM2}) for $\kappa X_\|$, the 
identity 
\begin{equation}
\label{eq:kapparho}
(m^2  + 1)\kappa^2  \rho^2 =
\frac{\gamma}{Z }\, \left[ \gamma W_\pm(Z)^2 
+1 \right]
\end{equation}
follows.
Eq.(\ref{eq:kapparho}) defines two functions of $Z$, corresponding to the ${\sf sign} (Z^\bullet)$. 
\\\\
In section \ref{rhoprimebound}, it will be confirmed that $\rho'\le 1$, and that $\rho$ is a monotonically increasing function of $s$.
\\\\
If one instead eliminates $X_\|$ in Eq.(\ref{eq:XcrossM}),  the identity 
\begin{equation}
\label{eq:Xperp}
m^2 X_\perp^2 = \frac{\gamma}{\kappa^2 Z} -  \rho^2 
\end{equation}
follows. Positivity of the left hand side places an upper bound on $\kappa \rho$ in terms of $Z$.  
The variables 
$\kappa \rho$ and $\kappa X_\perp$, unlike $\kappa X_\|$,  are independent of ${\sf sign}(\tau)$.   

\section{Invariant cones}
\label{Coscycle}

\subsection{The Cosine cycle}
\label{Coscycle1}

\begin{figure}[htb]
\begin{center}
\subfigure[]{\includegraphics[height=5cm]{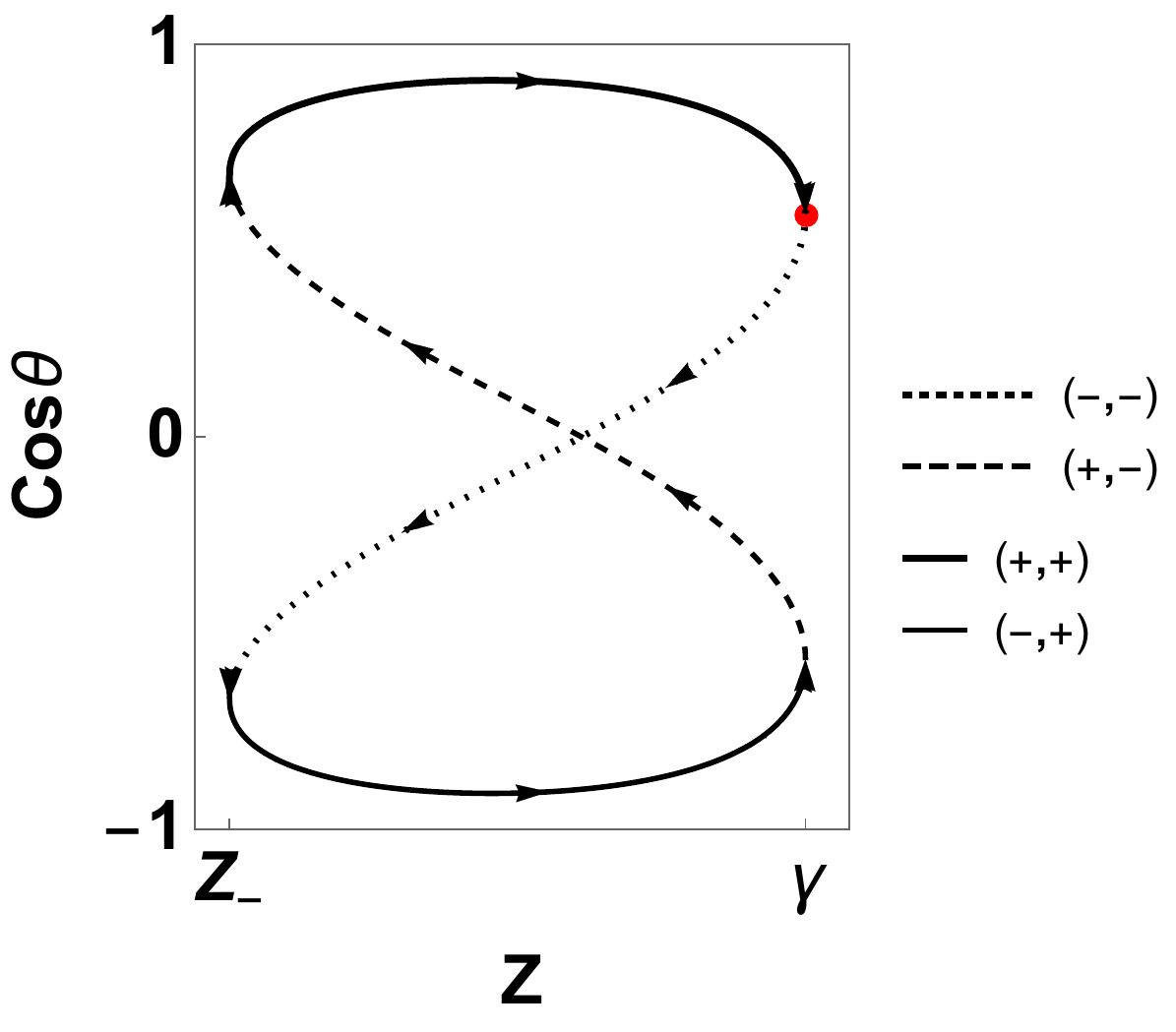}}
\hskip.5cm
\subfigure[]{\includegraphics[height=5cm]{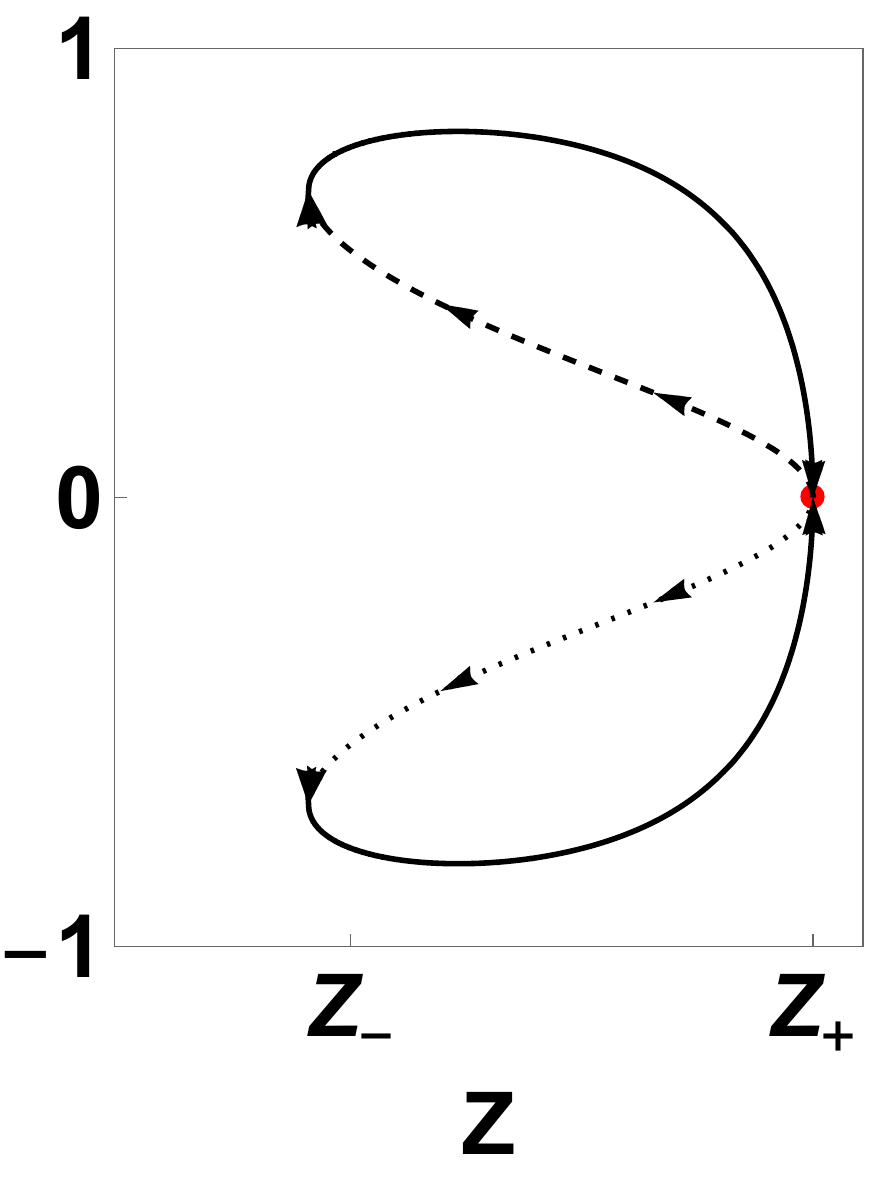}}
\hskip.5cm
\subfigure[]{\includegraphics[height=5cm]{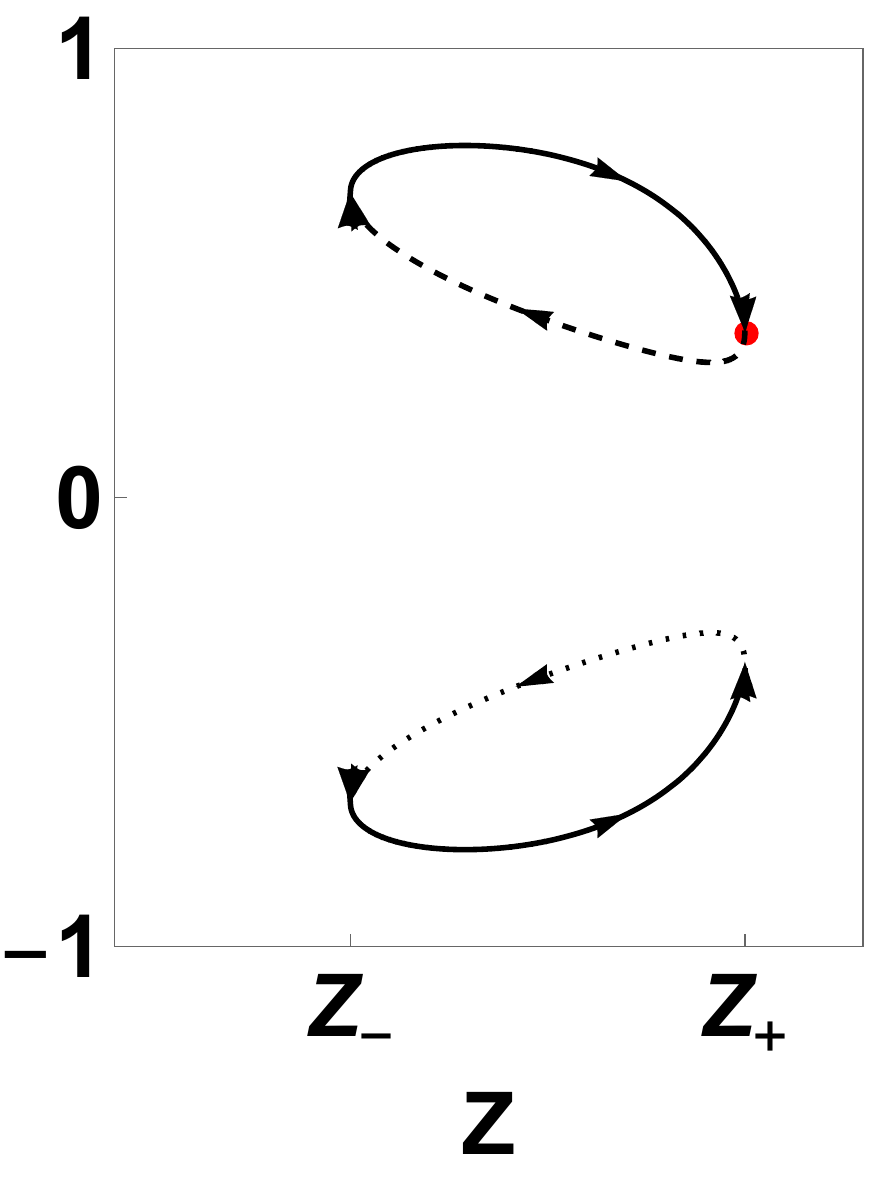}}
\caption{\small \sf (a)
Supercritical $\cos\theta$ cycle ($\gamma=1.2, m=\sqrt{2}$); (b) Critical Cycle ($\gamma=1.2, m=1.3$);
(c) Subcritical cycle ($\gamma=1.2, m=1.25$) and its
reflected twin.  The critical cycle occurs when $m^2=\gamma^3$ lying on the interface $\mathcal{C}_1$  in parameter space (cf. Figure \ref{Fig:Phases}).  } \label{Fig:Coscycle}
\end{center}
\end{figure}

The spherical polar angle in the adapted chart is the angle 
that the position vector makes with the torque axis.
Using Eqs.(\ref{eq:XonM2}) and (\ref{eq:kapparho}) to identify the ratio $ X_\|/\rho$, 
$\cos\theta$ can be   
expressed in the form
\begin{equation}
\label{eq:cosvsZ}
\cos\theta_{(\pm,\pm)}(Z) = {\sf sign} (\tau)\,
\frac{\sqrt{m^2 +1}}{m}\, \frac{
W_\pm(Z) }
{\sqrt{W_\pm(Z)^2
+ \gamma^{-1} }}
\,.
\end{equation}
Like the expression Eq.(\ref{eq:XonM2}) for $X_\|$, in a supercritical spiral, Eq.(\ref{eq:cosvsZ}) describes  four functions of $Z$.  Notice that the explicit dependence on $\kappa$ in the expressions for $X_\|$ and $\rho$ cancel.  Modulo ${\sf sign} (\tau)$ and ${\sf sign} (Z^\bullet)$, 
the polar angle depends only on $Z$.
\\\\
Together the graphs of the four functions  $\cos\theta_{(\pm,\pm)}(Z)$ describe a closed figure of eight in $(Z,\cos \theta)$ space.  This \textit{cycle} is illustrated in Figure \ref{Fig:Coscycle}a for the values $\gamma=1.2$ and $m=\sqrt{2}$. Inspection of this cycle reveals significant features of the  
internal structure.
 \\\\
The cycle is completed in  two complete oscillations within the potential well, one for each sign of $\tau$. 
Using the functional form $Z(\Theta)$ determined by  the integration of the quadrature given by Eq.(\ref{eq:ThetaZ}), with $Z(0)=\gamma$, the cycle in turn determines the polar angle as a function of the rotation angle $\Theta$,
\begin{equation} 
\cos\theta  
 = \begin{cases}
        \cos\theta_{(-,-)} \,, & 0\leq \Theta\leq \Theta_0/2\\
       \cos\theta_{(-,+)}  \,, & \Theta_0/2\leq \Theta \leq \Theta_0\\
       \cos\theta_{(+,-)}  \,,  &  \Theta_0\leq \Theta\leq 3\Theta_0/2\\
       \cos\theta_{(+,+)}  \,,  &  3\Theta_0/2\leq \Theta\leq  2\Theta_0  \,,     
          \end{cases}
 \end{equation}
with period $2\Theta_0$.
\noindent The polar angle ranges between a minimum  $\theta_{\sf Min}$ (a  maximum of $\cos\theta$) and a maximum $\pi-\theta_{\sf Min}$. The minimum is always realized
along the $(+,+)$ segment,  the maximum along $(-,+)$.   Each value of $\theta$ describes a circular cone  in Euclidean space, its apex coincident with the spiral apex, its axis aligned along the torque direction. 
The complete cycle thus describes spiral nutation between two identical oppositely-oriented cones with opening angles $\theta_{\sf Min}$ and $\pi - \theta_{\sf Min}$ 
bounding the trajectory away from the axis: the trajectory never enters the region within the cones.
Various relevant properties of $\theta_{\sf Min}$ will be presented in \ref{BoundingCones}.
\\\\
The spiral always grazes the bounding cones tangentially. These do \textit{not} occur at the turning points of the potential, except in the limit $m\to \infty$.  
Indeed the derivative $d\cos\theta/ dZ $ diverges at $Z= Z_-$ and ${\sf Min} (Z_+,\gamma)$;
$(\cos\theta)^\bullet$  remains finite ($\ne 0$) 
 when $Z$ coincides with  a turning point because $Z^\bullet$ 
vanishes with the same power.
 \\\\
Supercritical Cosine cycles are symmetric under reflection in the mid-plane: $\theta\to\pi-\theta$. However,
the spiral trajectory itself is not. This is because it both precesses and expands as it nutates. In particular,
the beginning and endpoints of the cycle never represent the same point in Euclidean space.

\subsection{
Mid-plane crossings in supercritical spirals}
\label{mid}

The mid plane $X_\|=0$ (or $\cos\theta=0$) is crossed 
twice per cycle, once on the way down along the $(-,-)$ cycle segment and again on the way up along the $(+,-)$ segment (evident in Figure \ref{Fig:Coscycle}a). Along both segments, $Z^\bullet<0$, $\tau$ positive in one, negative in the other.  The crossing occurs when $Z$ satisfies the linear equation $W_-(Z)=0$. Using Eq.(\ref{eq:Wdef}), 
as well as Eq.(\ref{eq:roots}) to express $Z_-$ and $Z_+$ in terms of $\gamma$ and $m$,  
crossings occurs when $Z=Z_{\theta=\pi/2}$, where $Z_{\theta=\pi/2}$ is given by
\begin{equation}
\label{eq:ztheta0}
Z_{\theta=\pi/2} =  \gamma/ [m^2 +1 - \gamma^3] \,.
\end{equation}
It is simple to confirm that $Z_{\theta=\pi/2}$ lies within the interval $[Z_-,\gamma]$ if
$m^2 \ge  \gamma^3$.  it is also observed that $Z_{\theta=\pi/2} \to Z_- \to 0$ in the limit $m\to\infty$. Where $Z_{\theta=\pi/2}$ lies with respect to the turning points as well as the value of $Z$ where $\theta$ assumes its minimum is illustrated graphically in Figure 16a
in \ref{BoundingCones}.

\subsection{$X_\|$ oscillates about the mid-plane in supercritical spirals}
\label{Xparosc}
\begin{figure}
\begin{center}
\subfigure[]{\includegraphics[scale=0.40]{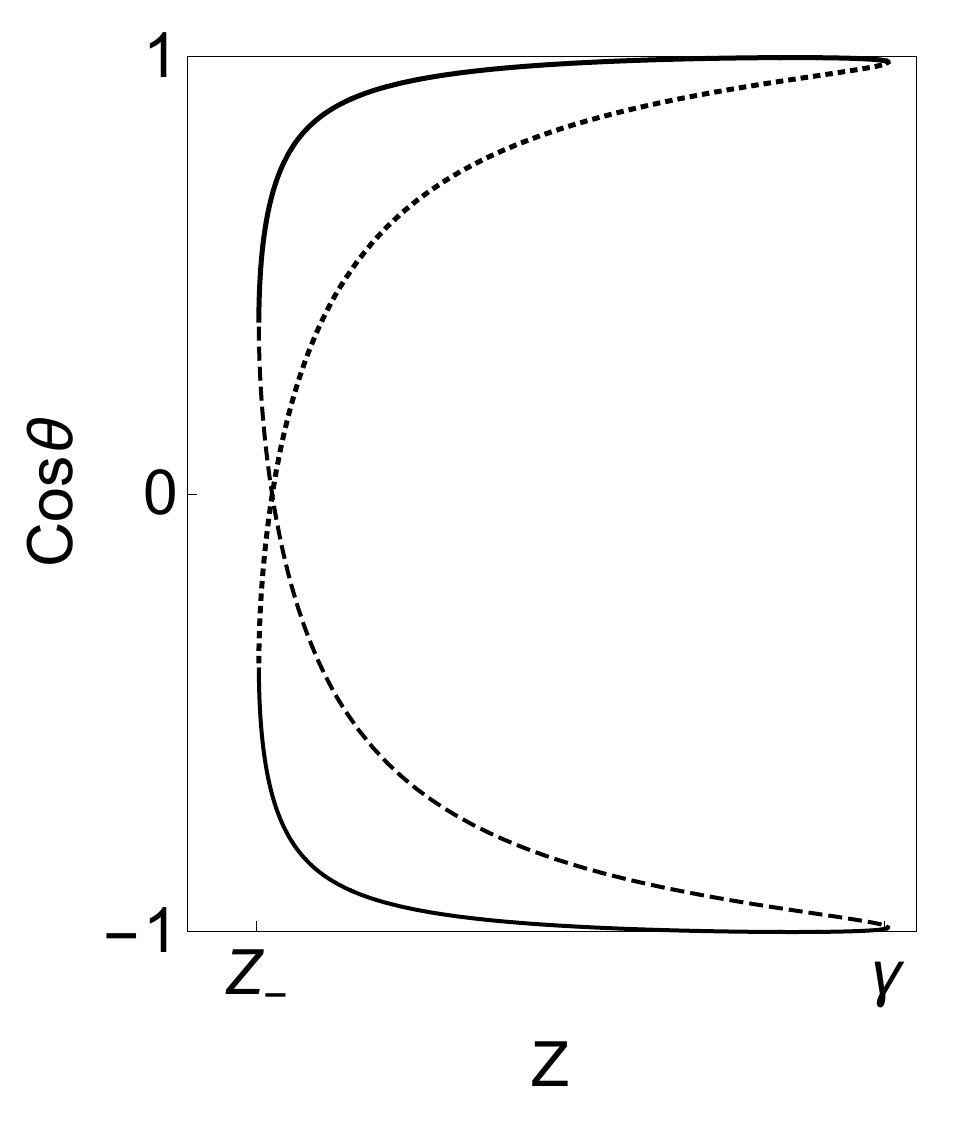}}
\hskip.5cm
\subfigure[]{\includegraphics[scale=0.40]{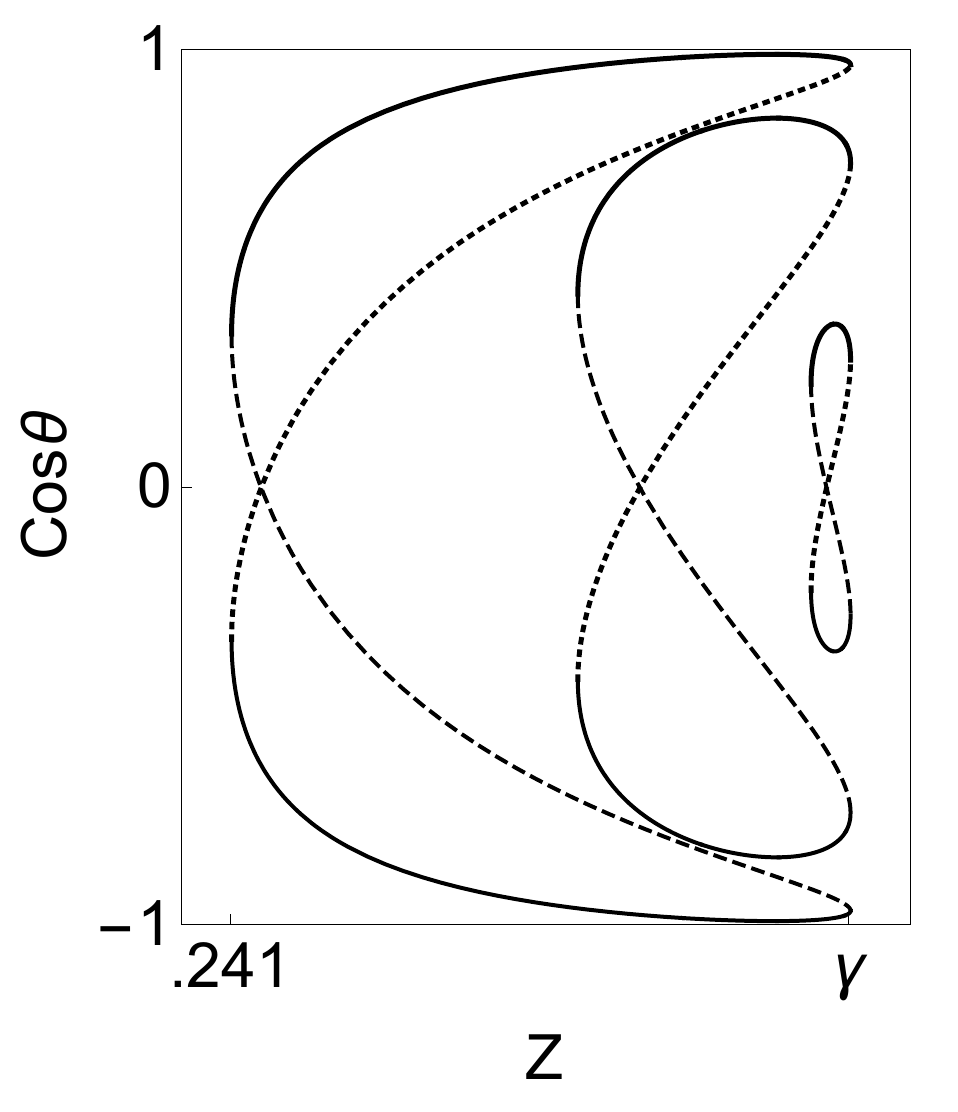}}
\hskip.5cm
\subfigure[]{\includegraphics[scale=0.40]{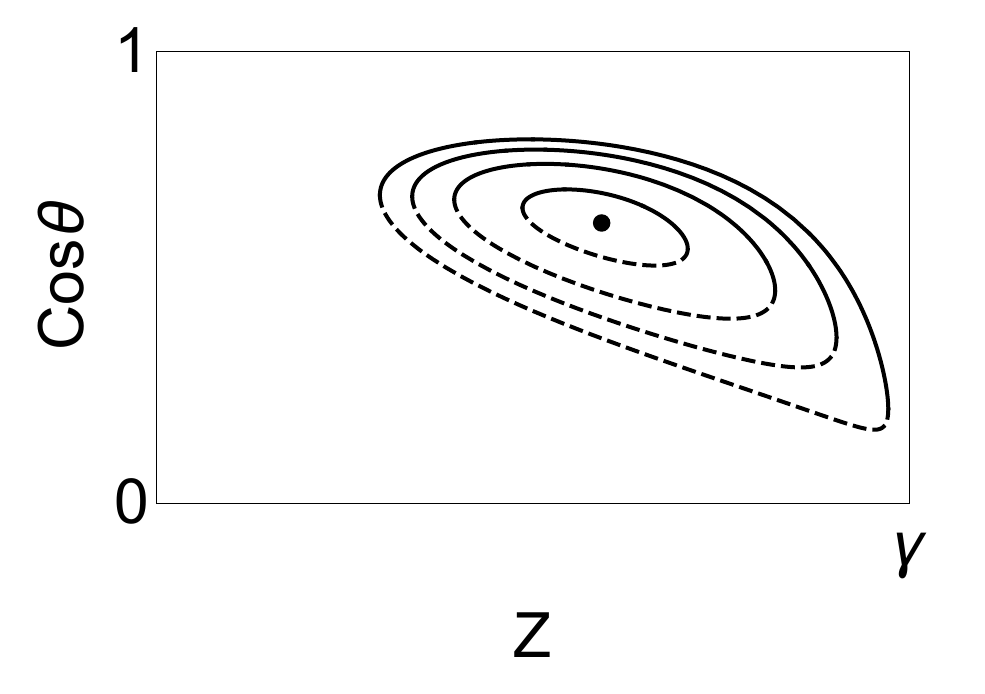}}
\caption{\small \sf (a)
Extremely asymmetric supercritical cycle ($\gamma=1.2, m=3$); (b) $\gamma<1$ Sequence of supercritical cycles: ($\gamma=0.75, m=1.5,0.85,0.67$).  As $m$ is lowered towards $\gamma^3$, the cycle vanishes; this contrasts with the behavior when $\gamma>1$. As $m$ increases the cycle grows both along $Z$ and $\theta_{\sf Min}\to 0$, in a manner qualitatively identical to (a);
(c) Nested subcritical cycles ($\gamma=1.2, m=1.31,1.3,1.29,1.28 $)  As $m$ tends towards its lower bound ($\mathcal{C}_0$), the cycles degenerate to a single point, representing  a helix wound on a single  cone with  fixed $Z$ and $\theta$, described in  \ref{conehelix}. 
}
\end{center}\label{Morecycles}
\end{figure}

In all supercritical spirals, $X_\|$ oscillates with increasing amplitude along the torque axis.  The extrema of $X_\|$ occur at the upper turning point in the potential,  $Z=\gamma$, thus correlating with the changing sign of $\tau$. 
To see this, begin with the simple identity,
\begin{equation}
\label{eq:XparptdotM}
 X_\|' = \mathbf{t}\cdot\hat{\mathbf{M}}\,,
 \end{equation}
connecting the velocity along the torque axis to the cosine of the angle that 
the tangent vector along the curve makes with that direction. Now 
using the definition (\ref{eq:Mdef}) of $\mathbf{M}$, with $\mathbf{F}=0$,
as well as the identity (\ref{eq:HISIJ}c),  it is straightforward to express the right hand side of Eq.(\ref{eq:XparptdotM}) as a function of $Z$, modulo ${\sf sign}(\tau)$,
 so that
 \begin{equation}
\label{eq:XparprimeZ}
X_\|'=
 {\sf sign} \, (\tau)\,\frac{1}{m}  \frac{1}{Z^{1/2}}  (\gamma  - Z)^{1/2}\,.
\end{equation}
It is now evident  that $X_\|'$ vanishes if and only if $Z=\gamma$. 
It is also simple to confirm that the magnitude of $X_\|'$ is  
bounded by $1$ for all accessible values of $\gamma$ and $m$.
\\\\
Suppose that $X_\|$ is a local maximum at $Z=\gamma$ with $\tau=0$. Let $\Theta=0$. 
As $\Theta$ increases $X_\|$ falls, twisting  increasingly ($\tau<0$ say) as it does; it crosses the mid-plane with a finite negative $\tau$, reaching a local minimum  on completing a complete oscillation in the well when $\Theta=\Theta_0$ and $Z$ returns to $\gamma$, and $\tau$ returns to zero.\footnote{\sf Note that the maximum value of $\tau$ does not occur where the mid-plane is crossed.}
 As $\Theta$ is increased, the motion along the axis is reversed and $\tau$ changes sign, the spiral now twisting in the opposite direction; $X_\|$ returns to a new local maximum when $\Theta=2\Theta_0$.  Whereas $\kappa X_\|$ is periodic in $\Theta$, with period $2\Theta_0$, 
it is clear from Eq.(\ref{eq:XonM2}) that the amplitude of oscillation about the torque axis grows as $\kappa^{-1}$. This dependence will be examined further in section \ref{SelfSim}. 
\\\\
{\bf  Turning points of $X_\|$ in supercritical spirals lie on  invariant cones} 
\\\\
Eq.(\ref{eq:cosvsZ}) implies that the opening angle 
$\cos\theta$ assume the values  
\begin{equation}\label{eq:costhetaXparmax}
\cos \theta_\gamma = \pm 
\, \sqrt{\frac{m^2+1}{m^2}}
 \sqrt{1 -\frac{\gamma^3}{m^2} }\Big/
\sqrt{ 1 -\frac{\gamma^3}{m^2 +1}}
\end{equation}
at the extrema of $X_\|$ which occur when $Z=\gamma$.
Whereas the trajectory touches the cones bounding it away from the torque direction tangentially, 
it intersects the two turning cones where $X_\|'=0$ 
with a non-vanishing  angle of incidence ($\theta'\ne 0$). 
\\\\
Evidently $\theta_\gamma<\theta_{\sf Min}$ (cf. Figure \ref{Fig:Coscycle}a or 18b).  
\\\\
 The polar angle $\theta$ assumes the value $\theta_\gamma$ at two different values of $Z$ along each cycle, once at $Z=\gamma$ 
but also again at a second lower value of $Z$, $\gamma_1$ say (cf. Figure \ref{Fig:Coscycle}a).  Clearly $\gamma_1$ is not a local maximum of $X_\|$, and nor does the torsion vanish there. In an expanding spiral, the $\theta_\gamma$ cone is always entered when $Z=\gamma_1$, and exited with $Z=\gamma$ (the red point in Figure \ref{Fig:Coscycle}a). Along which cycle segment it enters will depend on the relative values of the cosine at $Z_-$ and at $\gamma_1$, which will depend on the value of $m$
(in the sequence illustrated in Figure \ref{Morecycles}b it always occurs along the $(+,+)$ segment). 
Whatever the case may be,  the spiral geometry is asymmetric with respect to exit and re-entry of 
the $\theta_\gamma$ cone.  Between entry and exit $X_\|$ increases (if positive), reaching a local maximum upon exit where the torsion vanishes and changes sign; on its sojourn within this cone the trajectory grazes the interior bounding cone with $\theta=\theta_{\sf Min}$ once.
\\\\
It is also evident from inspection of Eq.(\ref{eq:XparprimeZ}) that  
the fastest ascent always occurs at  $Z_-$ (in subcritical trajectories the slowest ascent occurs at $Z_+$).  It may at first seem surprising that this does not occur where the trajectory crosses the mid plane, but not if one recalls that the trajectory is not symmetric with respect to reflection in this plane.
The angle $\Psi$  that the tangent makes with $\mathbf{M}$ (cf. Eq.(\ref{eq:XparptdotM})) is minimized  at this value of $Z$ with $\cos \Psi_{\sf max}= \pm ( 2\gamma^3/ [m^2+1 - \sqrt{ (m^2+1)^2 - 4 \gamma^3}]- 1)^{1/2}/m$.  
As $m\to\infty$,  for fixed $\gamma$,  $\cos \Psi_{\sf max} \to 1$ independent of $\gamma$. 
Yet there is no finite critical value above which $\mathbf{t}$ is aligned or anti-aligned with $\mathbf{M}$.
\\\\
If $m$ is increased for fixed $\gamma$, the cycle becomes increasingly asymmetric. In the limit $m\to \infty$,  $\theta_\gamma \downarrow 0$, and with it so also goes $\theta_{\sf Min}$. This behavior is displayed in Figure \ref{Morecycles}a for $\gamma=1.2$, and again in Figure  \ref{Morecycles}b for $\gamma<1$.  Notice, however, that $\theta_{\sf Min}=0$, for all practical purposes, at even modest values of $m$.  A consequence is that the region of Euclidean space excluded by the bounding cones disappears and the trajectory is freed to range  through three-dimensional space.  
\\\\
It is also evident in Figure  \ref{Morecycles}b that the  \textit{area}  enclosed by the large $m$ limiting cycles tends to zero. In the absence of any obvious metric, it is not clear, however, 
 if this area possesses geometrical significance.
 \\\\
The distinction in the behavior of limiting supercritical spirals with $\gamma<1$ and $\gamma>1$ is discussed in \ref{Limits}. How this distinction is reflected in the bounding cones is  examined in \ref{LimitSuper} (contrast Figure \ref{Fig:Coscycle}b and the leftmost and smallest figure of eight in Figure \ref{Morecycles}b).
\\\\
If $\gamma>1$, 
the supercritical figure of eight morphs discontinuously on reducing $m$ into two disconnected, non-self-intersecting subcritical cycles as  $\mathcal{C}_1$ is crossed (cf. Figure \ref{Fig:Phases}).  This behavior  is displayed in the sequence (a-c) in Figure \ref{Fig:Coscycle}.  The  critical cycle with $m^2=\gamma^3$,  illustrated in the central panel, has a divergent period associated wth the negative curvature of the potential at $Z=\gamma$.  In this limit, $Z=\gamma$ is only reached  asymptotically as described in section \ref{Limits}. Mid-plane crossing does not occur. Indeed,  $Z_{\theta=\pi/2}$, given by
Eq.(\ref{eq:ztheta0}), coincides with $\gamma$. 
 
\subsection{A brief note on Subcritical Spirals} 

If the spiral is subcritical,  $\tau$ has a fixed sign  (say positive).  Now 
$\cos\theta$ is represented by two different functions of $Z$, 
one for each sign of $Z^\bullet$,
$(+,-)$ and $(+,+)$ (cf. Eq.(\ref{eq:cosvsZ})). Their  
union forms a closed non-self intersecting loop on the $(Z,\cos \theta)$ plane.
A circuit of this cycle has a period $\Theta_0$.  
Such a cycle is represented by the upper loop in Figure \ref{Fig:Coscycle}c)
for $\gamma=1.2$ and $m=1.25$;\footnote{\sf 
The mirror image
with respect to the mid-plane, with $\tau<0$, represented  by the $(-,+)$ and $(-,-)$ graphs describes a second disconnected spiral.}
the mid plane is not crossed.  Now, however, $\theta$ exhibits a  maximum,
$\theta_{\sf Max}\le \pi/2$ and 
$\theta$ ranges between the minimum at $\theta_{\sf Min}$ occurring along the $(+,+)$ segment, and $\theta_{\sf Max}$ which occurs along $(+,-)$.
The spiral nutates between the two identically oriented cones defined by these angles. 
As $m$ is lowered from $\mathcal{C}_1$ towards $\mathcal{C}_0$ for a fixed $\gamma$, the cycles contract to a single point, as illustrated in Figure 6(c),
representing the merger of the two extremal cones  into one.
\\\\
In all subcritical trajectories, 
$Z<\gamma$ and $\tau$ has a fixed sign.
Eq.(\ref{eq:XparprimeZ}) then implies that $X_\|'>0$ and $X_\|$ increases monotonically along the spiral, no matter how deformed it may be. 
This remains true in the limiting, non-trivial asymptotically logarithmic geometries (cf. \ref{Limits}). The limiting behavior of $X_\|$ is examined in  \ref{asymptotrise} where the sub-linear power law controlling the approach to a planar logarithmic spiral is identified.


\section{ The distance to the apex increases monotonically}
\label{rhoprimebound}

\begin{figure}
\begin{center}
\includegraphics[height=5cm]{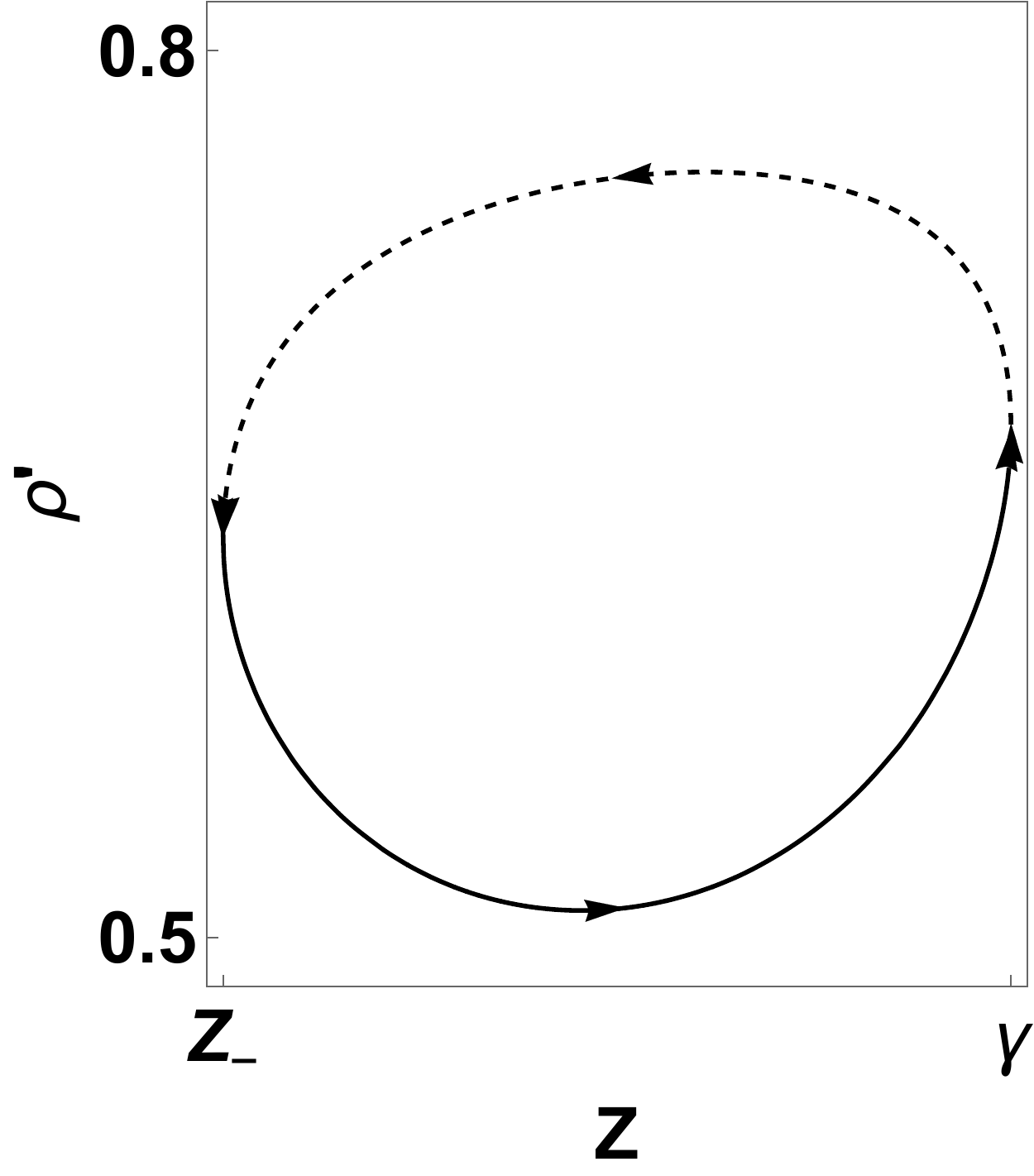}
\caption{\small \sf A supercritical $\rho'$ cycle ($\gamma=1.2$, $m=\sqrt{2}$). $\rho$ increases most rapidly outside the $\theta_\gamma$ cone, and most slowly within it close to the bounding  cone.}
\label{Fig:rhopcycle}
\end{center}
\end{figure}

Intuitively,  one would expect the distance $\rho$ to the apex  to increase monotonically with arc-length $s$. Yet it has been seen in section \ref{Xparosc}
that $X_\|$ oscillates in supercritical spirals. So it is worth confirming. As a payoff, a simple and useful identity for $\rho'$ is revealed.
\\\\
First rewrite Eq.(\ref{eq:kapparho}) in the form
\begin{equation}
\label{eq:rho2}
(m^2  + 1) \rho^2 = m^2  X_\|^2 + \frac{\gamma}{\kappa^2 Z} \,.
\end{equation}
Now
\begin{equation}
\label{eq:rho3}
(m^2  + 1)\rho \rho^\bullet =m^2  X_\| X_\|^\bullet- \frac{\gamma}{\kappa^2}\frac{{Z^\bullet}}{2Z^2}  +
 \frac{\gamma Z^{1/2}}{\kappa^2 } 
 = \frac{\gamma^2}{\kappa^2 Z^{1/2}} 
\,.
\end{equation}
The identities  (\ref{eq:XonM2}) and (\ref{eq:XparprimeZ}) for $X_\|$ and its 
derivative  have been used,  as well the quadrature (\ref{eq:Quadrature}) to eliminate $Z^\bullet$ in terms of $Z$ (modulo ${\sf sign}(Z^\bullet)$. Eqs.(\ref{eq:SigGamdef}) and
(\ref{eq:defsZgamma}) have also been used to eliminate $\kappa
^\bullet$ in favor of $Z$ and $\kappa$. The final  expression in Eq.(\ref{eq:rho3}) is manifestly positive confirming that $\rho$ increases monotonically everywhere. 
The simple identity for the dimensionless $\rho'$,
\begin{equation}
\label{eq:rhoprime2}
\rho\,'{}^2=\frac{\gamma^3 }{m^2  + 1} \,\frac{1 }{1+ \gamma W_\pm(Z)^2}
\end{equation}
follows ($W_\pm$ is defined in Eq.(\ref{eq:Wdef})).
Along a conical helix, $Q=0$, and $Z=\gamma^{-1/2}$, which reproduces the result $\rho' = 1/\sqrt{2}$ derived in  \ref{conehelix}.
\\\\
In \ref{rhoprimebounds}, it is shown that a 
sharp upper bound can be placed on $\rho\,'$,
bounding it strictly below $1$, the Euclidean limit.  
Sharp lower bounds, guaranteeing strict 
monotonicity, can also be established. These bounds are captured in the cyclic behavior of $\rho'$,
illustrated for $\gamma=1.2$ and $m=\sqrt{2}$ in Fig. \ref{Fig:rhopcycle}.
An immediate but nevertheless important consequence of the monotonicity of $\rho$ is that the spiral geometry never exhibits \textit{accidental}  self-intersections.   Nor do knotted self-similar spirals exist.
\\\\
To determine $\rho$ as a function of $\Theta$ one can  either integrate the $\rho'$ cycle given by (\ref{eq:rhoprime2}) 
or use the cyclical variable $\kappa \rho$ given by 
Eq.(\ref{eq:kapparho}), modulated by the falloff in $\kappa$,  Eq.(\ref{kapthet}),  following from the quadrature: for the first half-cycle,
\begin{equation} 
\label{eq:rhoTheta}
\rho  
 = \begin{cases}
        ( \kappa \rho)_- / \kappa , & 0\leq \Theta\leq \Theta_0/2\\
        (\kappa \rho )_+  / \kappa, & \Theta_0/2 \leq \Theta \leq \Theta_0\,;
         \end{cases}
 \end{equation}
to complete the determination of $\rho$ along the second half-cycle,  Eq.(\ref{eq:rhoTheta}) is modified with the appropriate continuation of Eq.(\ref{kapthet}) for $\kappa$. 

\section{ $X_\perp$ is not monotonic if $m$ is large}
\label{Xperpmonotonic}

\begin{figure}[htb]
\begin{center}
\includegraphics[height=5cm]{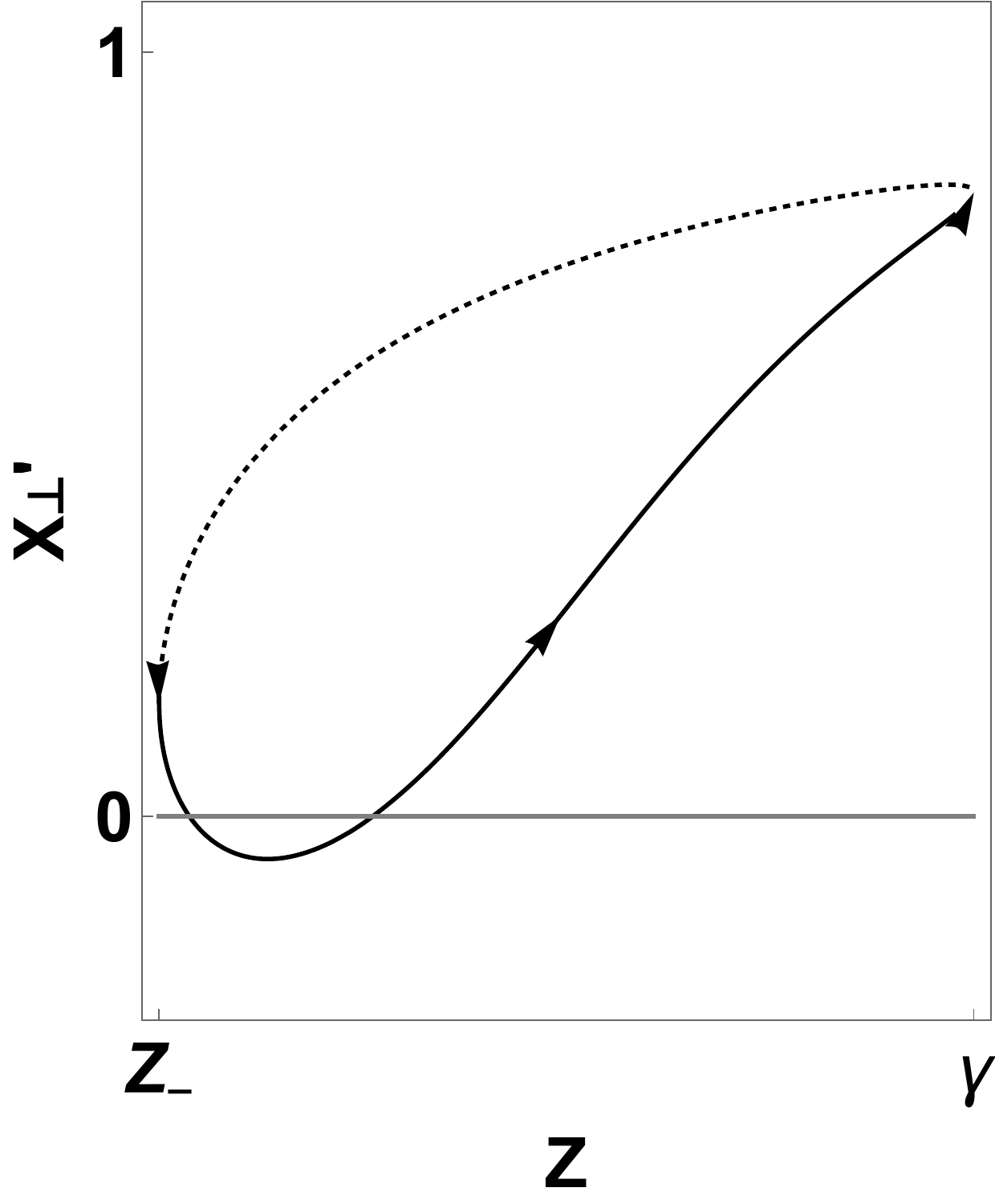}
\caption{\small \sf Supercritical $X_\perp'$ cycle ($\gamma=0.75$, $m=0.85$).   $X_\perp'$ is negative along connected segments of the $(+,+)$ and $(-,+)$ quarter cycles (or near the bounding cones). The length of these segments increases monotonically as $m$ is increased.}\label{Fig:XperpCycle}
\end{center}
\end{figure}

If $m$ is sufficiently large the perpendicular distance from the torque axis $X_\perp$ ceases to behave in a monotonic manner. But, unlike the oscillations in $X_\|$ which occur only in supercritical spirals,  in $X_\perp$ non-monotonic behavior is exhibited in both subcritical and supercritical spirals. Eq.(\ref{eq:Xperpprime}) of \ref{Xperppr0} indicates  that 
$X_\perp'$ forms a cycle of period $\Theta_0$ (it is independent of the sign of $\tau$). The supercritical cycle corresponding to the parameter values
 $\gamma=.75$, $m=.85$ is presented in Figure \ref{Fig:XperpCycle}. 
 
\begin{figure}
\begin{center}
\includegraphics[height=5cm]{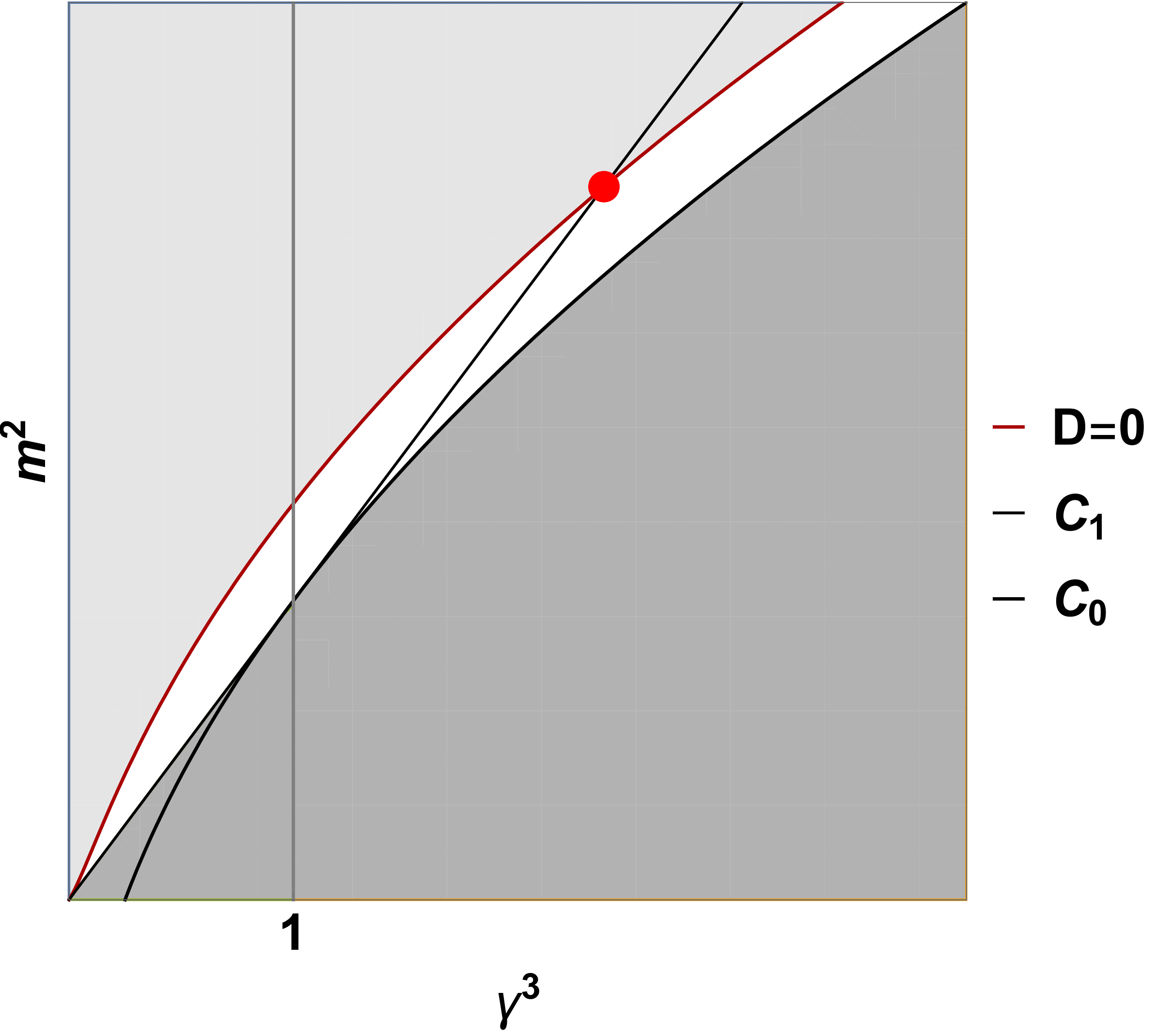}
\caption
{\small \sf  The contour $D=0$ partitions parameter space into two regions:  
$D<0$ where $X_\perp$ is monotonic everywhere  (white);  $D>0$
where $X_\perp$  is not monotonic in places (shaded light gray). The dark grey region is inaccessible.}\label{Fig:PhasesXperpp0}
\end{center}
\end{figure}

\vskip1pc
\noindent
Surprisingly, as Figure \ref{Fig:PhasesXperpp0} illustrates, monotonic behavior occurs only within a narrow band of values of $m$ for any fixed $\gamma$. 
The details are presented in  \ref{Xperppr0}.
If  $\gamma$ is fixed and $m$ is large enough, $X_\perp$ is not a monotonic function of $s$.

\section{The azimuthal advance and spiral precession} 
\label{phi}

\begin{figure}[htb]
\begin{center}
\subfigure[]{ 
\includegraphics[height=5cm]{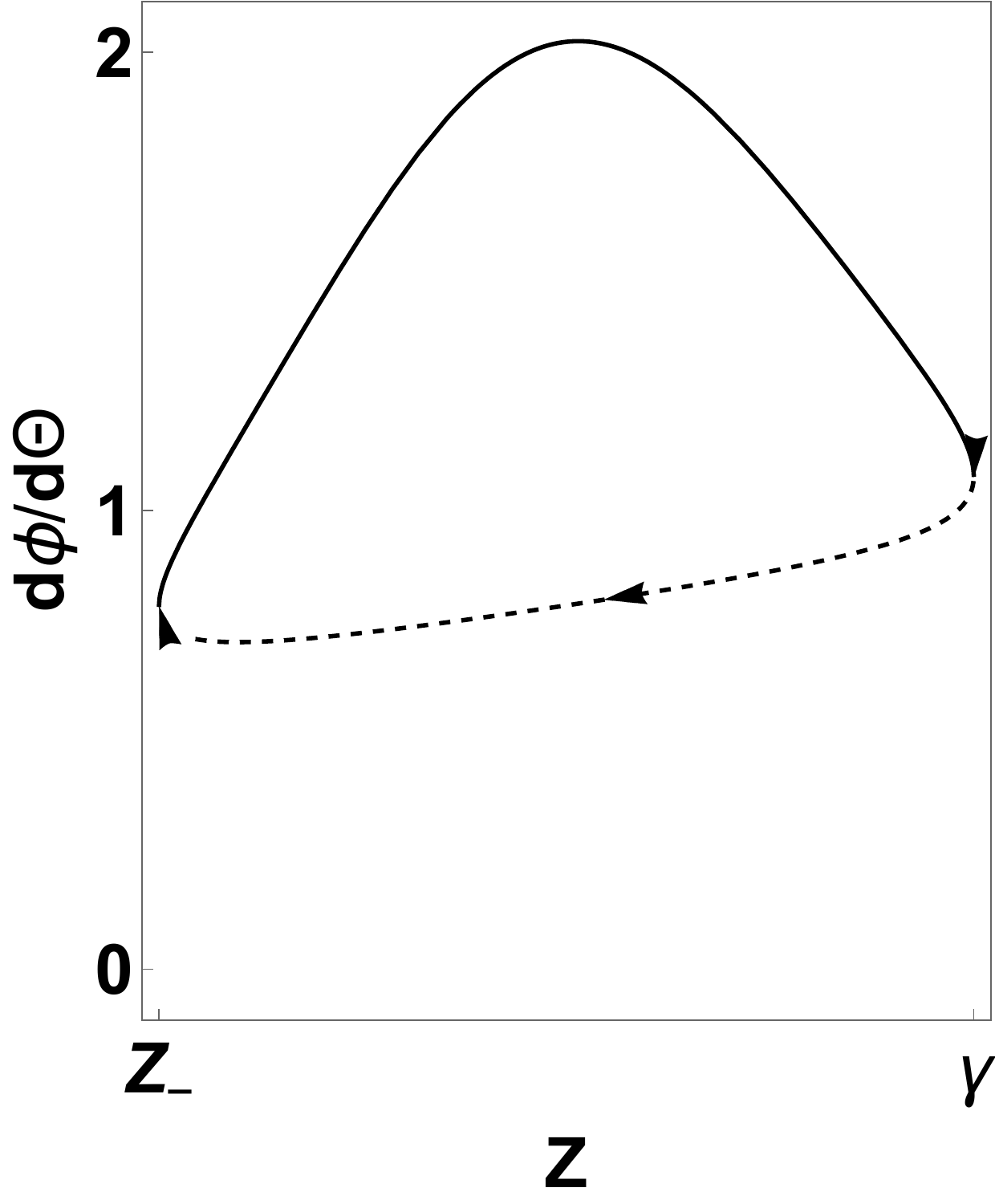}}
\hskip1cm
\subfigure[]{ \includegraphics[height=5cm]{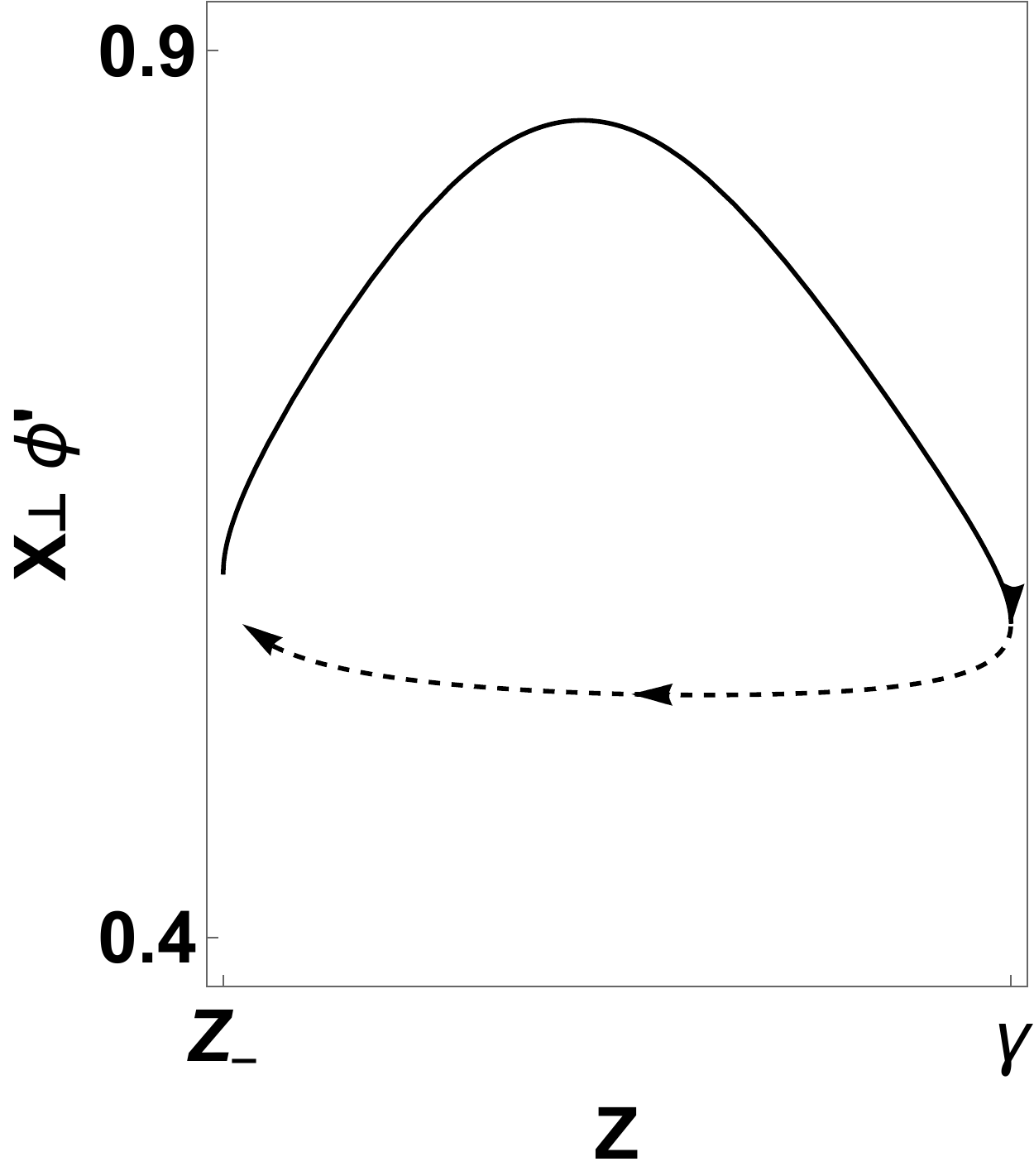}}
\caption{\small \sf (a) $\phi^\bullet$ cycle; (b) $X_\perp \phi'$ cycle (both for $\gamma=1.2$, $m=\sqrt{2}$).
$\phi$ evidently advances most rapidly along the $(+,+)$ and $(-,+)$ cycle segments 
close to the bounding cones.}
\label{Fig:phicycles}
\end{center}
\end{figure}

\begin{figure}[htb]
\begin{center}
\subfigure[]{\includegraphics[height=5cm]{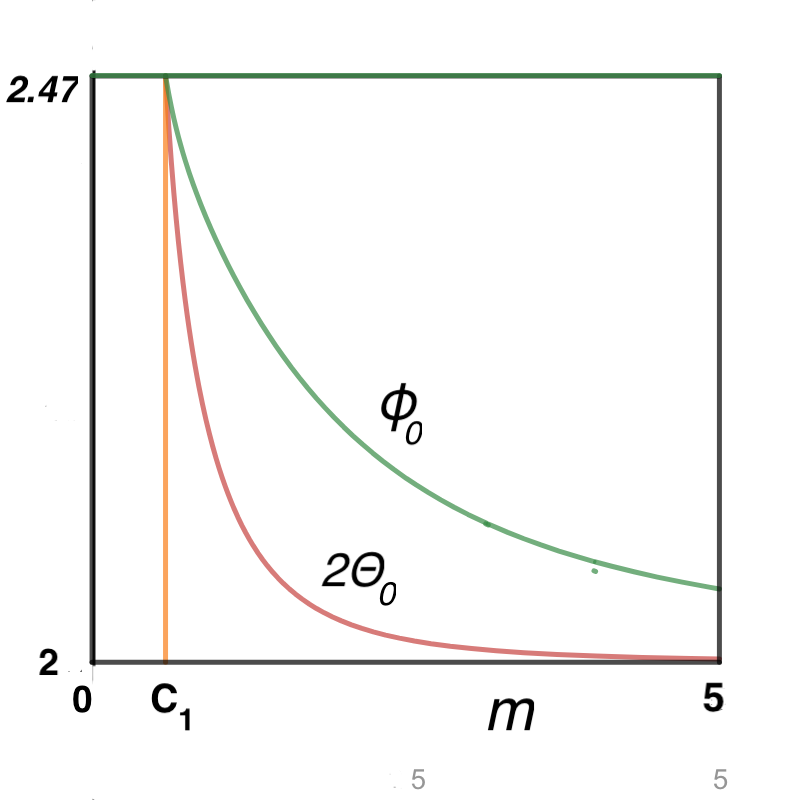}}
\hskip.5cm
\subfigure[]{\includegraphics[height=4cm]{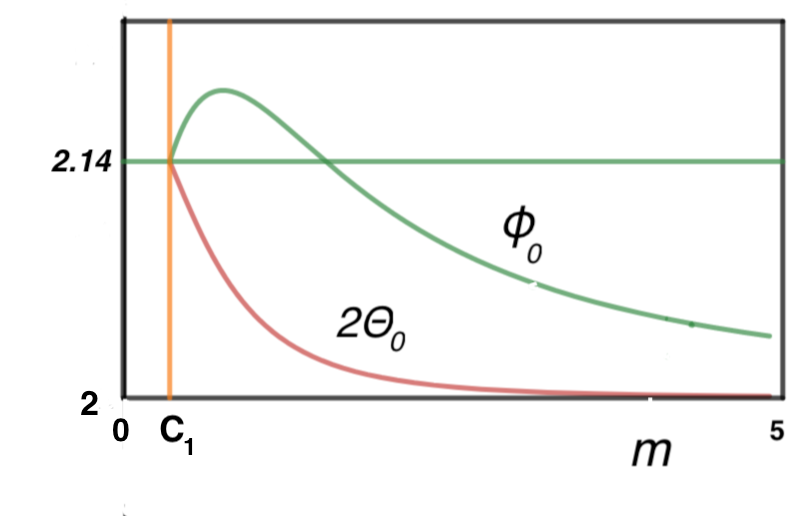}}
\caption{
\small \sf $\phi_0/\pi$ (green) and $2\Theta_0/\pi$ (red) vs. $m$  for $\gamma<1$ (a) $\gamma=0.7$ , $m\in[0.7^{3/2}, 5]$, (b)  $\gamma=0.5$, $m\in[ (0.5)^{3/2}, 4]$. 
Notice that $\phi_0\ge 2\Theta_0$, with equality along $\mathcal{C}_1$ where $m^2=\gamma^3$.  
Above $\mathcal{C}_1$,  $\Theta_0$ decreases monotonically to $2\pi$; $\phi_0$ does as well if $4\gamma^3>1$ (realized in (a)), but does not  below this value (b) (it increases initially before beginning its slow descent towards $2\pi$).}
\label{Fig:phiThetvsm}
\end{center}
\end{figure}

It remains to determine  the azimuthal progress of the spiral. One way to do this is the obvious one: integrate the arc length  identity 
$\rho'{}^2 + \rho^2 (\theta'{}^2 + \sin^2 \theta \phi'{}^2) =1$
using the expressions (\ref{eq:kapparho}) for $\rho$ and (\ref{eq:cosvsZ}) for $\theta$, or the equivalent expression for $X_\|$ and $X_\perp$ in cylindrical polars. 
However,  a  more direct  and transparent expression  for $\phi'$ is implied by  the conservation of the  special conformal current. 
\\\\
{\bf A vector orthogonal to the spiral position with respect to its apex}
\\\\
The projections of Eq.(\ref{eq:Gground2}) along $\mathbf{X}$ and orthogonal to it have been used to identify $\rho$ and $\cos\theta$. Its projection along $\mathbf{t}$ provides the identity
\begin{equation}
\label{eq:JdotX0}
\mathbf{X}\cdot (\mathbf{t}\times \mathbf{M}) +
  S\,(\mathbf{X}\cdot \mathbf{t}) =0\,;
 \end{equation}
as a consequence, the vector $\mathbf{J}$, defined by
\begin{equation}
\label{eq:Jdef}
\mathbf{J}=
\mathbf{t}\times \mathbf{M}
 +  S\,\mathbf{t} =  [\mathbf{X}\times \mathbf{M}
+ S\,\mathbf{X}]' \,,
\end{equation} 
is orthogonal to $\mathbf{X}$ at every point  
along tension-free spiral trajectories.  
\footnote{\sf The vector $\mathbf{J}$ itself is not conserved.  However, its projection along $\mathbf{t}$.
Using the expression (\ref{eq:Mdef}) for $\mathbf{M}$, it is possible to cast   
$\mathbf{J}=H_1\, \mathbf{N} - H_2 \,\mathbf{B} -  S\,\mathbf{t}$.  The identity $\mathbf{X}\cdot \mathbf{J}=0$ can then be read as a homogeneous constraint on the projections of $\mathbf{X}$ onto the Frenet frame, which is of interest in its own right.}
This identity $ \mathbf{X}\cdot \mathbf{J}=0$ determines $\phi'$ in terms of $\rho$, $\rho'$ and  $\theta$.
To see this  write Eq.(\ref{eq:JdotX0}) in the equivalent form,
\begin{equation}
\label{eq:JdotX0rewrite}
 \mathbf{X}\cdot\mathbf{J}=
- S\,{|\mathbf{X}|^2}'/2  + \mathbf{t}\cdot \mathbf{X}\times {\mathbf{M}} =0
\,.
\end{equation}
Expressed in terms of  the spherical polar coordinates adapted to the torque vector, Eq.(\ref{eq:JdotX0rewrite}) reads
\begin{equation}
\label{eq:rhopphip}
\rho' =m  \rho\, \sin^2 \theta \, \phi'\,.
\end{equation}
Because $\kappa \rho$, $\kappa X_\perp$ and $\rho'$ form periodic cycles, it is now evident that both $\phi^\bullet$ as well as the orbital velocity $X_\perp \phi' $ about $\mathbf{M}$ do also.\footnote{\sf
To derive  Eq.(\ref{eq:rhopphip}), 
decompose $\mathbf{t}$ with respect to the associated  orthonormal frame : $\mathbf{t}= \rho' \hat{\mathbf{X}}+ \rho \phi' \sin\theta \, \hat{\bm \phi}  
+ \rho \theta' \hat{\bm \theta}$, so that
\begin{equation}
 \mathbf{t} \times \hat{\mathbf{X}} =
-\rho \sin\theta \phi'\, \hat{\bm \theta}  + \rho \theta'\, \hat{\bm \phi}  \,;
\end{equation}
the identity (\ref{eq:rhopphip}) follows
upon substitution into Eq.(\ref{eq:JdotX0rewrite}).}
A significant consequence of  Eq.(\ref{eq:rhopphip}) is that $\phi'>0$: $\phi$ thus increases monotonically with $s$. There are no spiral reversals. Representative 
$\phi^\bullet$ and  $X_\perp \,\phi'$ cycles are illustrated in Figure \ref{Fig:phicycles}.
\\\\
Integrating Eq.(\ref{eq:rhopphip})  for each of the two signs of $Z^\bullet$ ($\phi$ does not depend on ${\sf sign}(\tau)$), 
\begin{eqnarray}
\phi_\pm 
        &=& \frac{\gamma}{m}\, \int d\Theta \, 
        \frac{Z^{1/2}}{ 1 - \gamma m^{-2}  
W_\pm(Z)^2} \,\,,
\label{eq:phiTheta}
\end{eqnarray}   
where  Eqs.(\ref{eq:kapparho}) and (\ref{eq:rhoprime2}), as well as the identity (\ref{eq:cosvsZ}) 
are used.
The integrand in Eq.(\ref{eq:phiTheta}) is a function of $Z$. Using the
quadrature, $\phi$ can now be expressed as a function of $\Theta$. 
If $\phi(0)=0$, the  advance of the azimuthal angle along a complete supercritical cycle is given by
\begin{equation} 
\phi 
 =\begin{cases}
       \phi_-(\Theta) \,,&   0\leq \Theta\leq \Theta_0/2\\
       \phi(\Theta_0/2) + \phi_+(\Theta)  \,,& \Theta_0/2\leq \Theta \leq \Theta_0 \\
         \phi(\Theta_0) + \phi_-(\Theta)  \,,& \Theta_0\leq \Theta \leq 3\Theta_0/2 \\
          \phi(3\Theta_0/2) + \phi_+(\Theta)  \,,& 3\Theta_0/2\leq \Theta \leq 2\Theta_0
  \,,  
          \end{cases}
 \end{equation}
 where $\phi_\pm$ is given by Eq.(\ref{eq:phiTheta}) with the  initializations indicated on the right.
 On completing the  cycle, the spiral has undergone a rotation about the torque axis by  the  value
\begin{equation}
\phi_0 = \frac{2\gamma}{m}\,\int_0^{\Theta_0} d\Theta\,  Z^{1/2}\,\left(  
        \frac{1}{ 1 - \gamma m^{-2}  
W_-(Z)^2} +\frac{1}{ 1 - \gamma m^{-2}  
W_+(Z)^2}\right)\,.
\label{eq:phi0def}
\end{equation}
In general,  
 Eq.(\ref{eq:phiTheta})  implies that
\begin{equation}
\label{eq:phiadvance}
\phi(\Theta + \Theta_0)= \phi(\Theta) + \phi_0/2\,;
\end{equation} 
$\phi_0=\phi(\Theta_0)$ is
generally not proportional to $\Theta_0$. It is clear from Eq.(\ref{eq:phiTheta}) that the local precession is not uniform within a 
cycle.  The total precession, captured by Eq.(\ref{eq:phiadvance}),  however, is identical in  consecutive cycles.  
 \\\\
 Upper and lower bounds 
on $W_\pm$ can be used to bound $\phi$ as a function of $\Theta$. 
 It is simple to see that $
\phi_0 \ge  2\Theta_0$.
But $\phi_0$ never significantly exceeds $2\Theta_0$.  The observed  pattern of cycle nutation and precession can be attributed to this \textit{coincidence}.
This bound is saturated along 
$\mathcal{C}_1$ ($m^2=\gamma^3$) when $\gamma<1$.  For now
$Z\approx \gamma$ and $W_\pm = 0$, so that $\phi_0 = 2\Theta_0$,
an unexpected coincidence.  Moreover $\Theta_0=1/ \sqrt{1-\gamma^3}$, a 
value completely determined by the curvature of the potential at $Z=\gamma$, given by Eq.(\ref{eq:linsupergamless10}). 
\\\\
In Figure \ref{Fig:phiThetvsm},  both $\phi_0$ and $2\Theta_0$ are plotted as functions of $m$ for two different fixed values of $\gamma<1$.  
\\\\
If $\gamma>1$, and $m^2\approx \gamma^3$ either above it or below, $\Theta_0$ diverges
(this is because the curvature of the potential at $Z=\gamma$ is negative). The integrands appearing on the right in Eq.(\ref{eq:phi0def}), nevertheless, remain finite along the cycle;  thus $\phi_0$ also diverges while the ratio $\phi_0/(2\Theta_0)$ remains finite.  

\subsection{An Integral Identity}

\begin{figure}[htb]
 \begin{center}
\includegraphics[height=4cm]{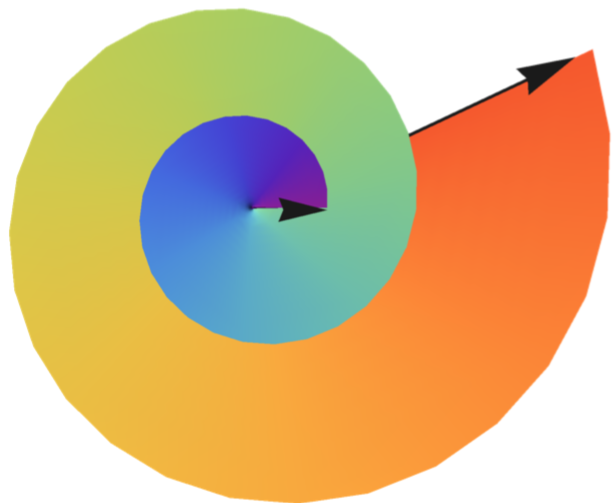}
\caption{\small \sf    
The projected area  for two consecutive 
cycles in  a supercritical deformation of
a logarithmic spiral [schematic]. The precession per cycle $2.125\pi$ exceeds $2\pi$.
As the spiral develops, $X_\perp$ sweeps out  a
multiply-covered region. Its progress has been color-coded using the visible spectrum. This area is 
proportional to  $\rho^2$. }
\label{Fig:Identity}
\end{center}
\end{figure}

An intriguing consequence of the identity  Eq.(\ref{eq:JdotX0rewrite}) is a  
connection between the distance from the apex to the area swept out by $X_\perp$ on the mid-plane 
along the trajectory from the apex. 
Integrating Eq.(\ref{eq:JdotX0rewrite}) implies the identity  
\begin{equation}
\label{eq:X2history}
\rho^2  = 2 m\, \int d\mathbf{X}\cdot \mathbf{X}\times \hat{\mathbf{M}}
 = 2 m\, \hat{\mathbf{M}} \cdot \int d\mathbf{X}\times \mathbf{X} = 2m \, A_\perp\,,
\end{equation}
where $A_\perp$ is the area of the multi-covered region on the mid-plane swept-out by $X_\perp$. The  proportionality is given by $m$.  This identity is represented schematically in Figure \ref{Fig:Identity}.
It should be mentioned that Eq.(\ref{eq:X2history})  completely characterizes a planar logarithmic spiral \cite{Paper1}.   

\section{Scaling and Self-Similarity}
\label{SelfSim}

\begin{figure}[htb]
\begin{center}
\subfigure[]{\includegraphics[scale=0.15]{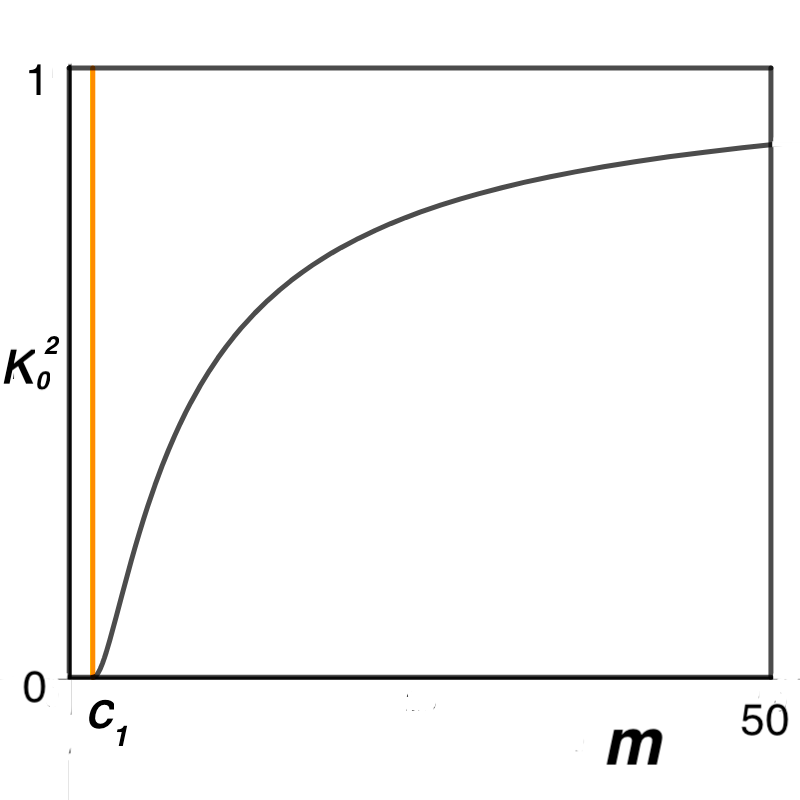}}
\hskip.5cm
\subfigure[]{\includegraphics[scale=0.15]{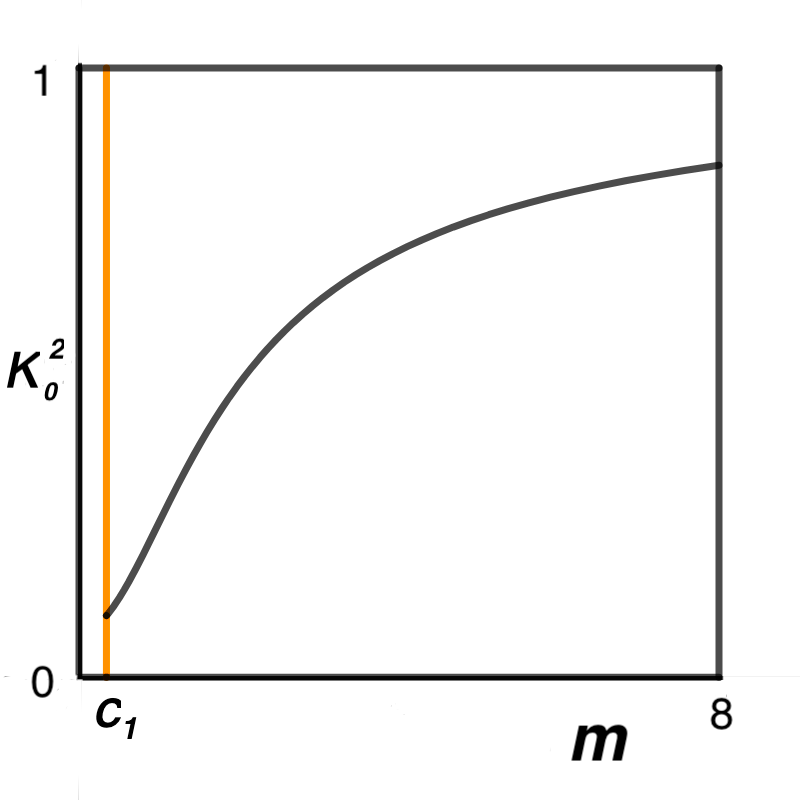}}
\caption
{\small \sf  (a) $\kappa_0^2$ vs. $m$ for $\gamma=\sqrt{2}$. Above $\gamma=0.979$ ($m=0.9683$), $\kappa_0$ rises monotonically from $0$ on $\mathcal{C}_1$, coasting slowly towards the asymptotic value $1$ at large $m$. The dilation factor ($\kappa_0^{-2}$) is large unless $m$ is large, as can be appreciated by noting the range of $m$ in the figure. As $m$ increases $\kappa_0\to  1$ so that the scaling on successive cycles is small. 
The vanishing of $\kappa_0$ along $\mathcal{C}_1$ for $\gamma>1$ reflects the fact that $\Theta_0$ diverges there so that the exponential defining $\kappa_0$ does too; $Z$ executes one round trip  in the well over the full spiral trajectory.  The behavior just below $\gamma=1$ is anomalous in that  even before $\gamma=1$ is reached the dilation diverges on $\mathcal{C}_1$.  
Below the value $\gamma=0.979$, as illustrated in Figure \ref{Fig:kappavsm} (b) for  $\gamma=0.5$, $\kappa_0$ does not vanish on $\mathcal{C}_1$. If $\gamma$ is lowered towards zero, the value of $\kappa_0$ on $\mathcal{C}_1$ tends towards its maximum value, $\kappa_0=1$. }\label{Fig:kappavsm}
\end{center}
\end{figure}

In section \ref{Coscycle}, it was seen that the cyclic variable $\cos \theta$  is periodic with period $2\Theta_0$ in supercritical spirals:
$\cos \theta (\Theta+ \Theta_0) = - \cos \theta(\Theta)$.
\\\\
The variables  $\kappa \rho$, $\kappa X_\perp$ and $\kappa X_\|$, captured by Eqs.(\ref{eq:kapparho}), (\ref{eq:XonM2}) and (\ref{eq:Xperp}),  are also cyclic variables. The first two, independent of 
${sf sign}\, \tau$, have  a period $\Theta_0$; the latter has a period $2\Theta_0$.
As a consequence, the dimensional variables $\rho$, $X_\perp$ and $X_\|$ are periodic functions of $\Theta$   
modulated by the growth rate $\kappa^{-1}$: over a half-period 
\begin{equation}
\label{eq:ratios}
\rho(\Theta + \Theta_0)/\rho(\Theta)= X_\perp(\Theta + \Theta_0)/X_\perp(\Theta) =-X_\|(\Theta + \Theta_0)/X_\|(\Theta)=\kappa_0^{-1} \,,
\end{equation} 
where 
the constant $\kappa_0$ is given by
 Eq.(\ref{kapthet0}).   
 The second half of the cycle is dilated by a factor of $\kappa_0$ with respect to the first.  
Like the precession, the growth rate  is not uniform within the cycle itself.  
When $m^2\approx \gamma^3$, $\kappa_0$ is small (significantly less than one), 
and the  spiral undergo significant inflation within a cycle.  If $m$ is large, however, $\kappa_0\approx 1$ and the scaling per cycle becomes modest. This behavior is illustrated  in Figure \ref{Fig:kappavsm} where $\kappa_0^2$ is plotted as a function of $m$.  

\vskip1pc\noindent 
The dimensionless variables 
$\rho'$, $X_\perp'$ and $X_\|'$ also form cycles (a happy accident of conformal symmetry) with 
periods identical to their counterparts   $\kappa \rho$, $\kappa X_\|$ and $\kappa X_\perp$.
\\\\
Whereas the zeros of an oscillating  function are preserved under modulation, its extrema generally are not.  However,  the extrema of $X_\|$  do occur periodically in $\Theta$, occurring whenever $\Theta=n\Theta_0$. Had $X_\|'$ not formed a cycle, this would not be expected. 
Eq.(\ref{eq:ratios}) implies that the  ratio of the magnitudes of successive extrema of $X_\|$ is given by $1/\kappa_0$.
\\\\
The second half-cycle is evidently identical to the first half, 
reflected in the mid-plane (switching $\tau$ and $-\tau$), 
rotated by $\phi_0/2$, and dilated by a factor of 
$\kappa_0^{-1}$. 
Modulo this additional symmetry, the irreducible spiral unit is described by a half-cycle.
\\\\
The existence of the internal structure captured by cycles is a significant feature distinguishing the three-dimensional self-similar spirals described here from their featureless logarithmic prototype. Logarithmic spirals are invariant under scaling composed with an appropriate compensating  rotation (if scaled appropriately, the spiral maps onto itself without rotation).  As such,  any one point on a logarithmic spiral is \textit{equivalent} to any other point along it. In their spatial counterparts, the internal structure quantizes the possible scaling so that the invariant subgroup is discrete. The spiral is invariant under scaling by the discrete factor $\kappa_0^{-1}$, composed with a reflection in the mid-plane androtation by $\phi_0/2$. 

\section{Translating cycles into spatial trajectories}
\label{TracingTrajectories}

\begin{figure}[htb]
 \begin{center}
\subfigure[$(\gamma,m)= (0.5,0.4)$]{\includegraphics[height=4cm]{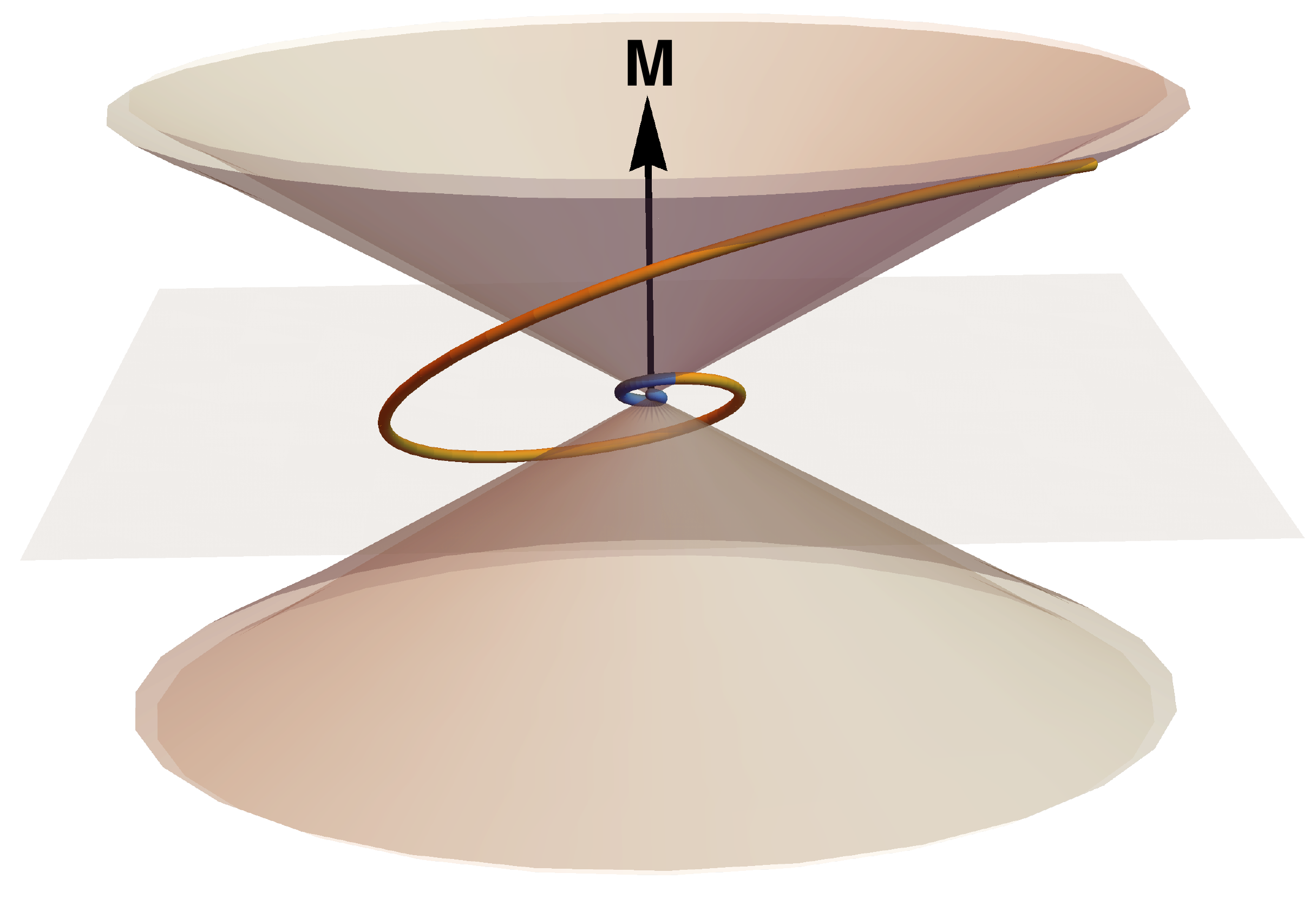}}
\hskip.5cm
\subfigure[$(0.75,0.85)$ {$1^{\sf st}$ half-cycle}]{
\includegraphics[height=5cm]{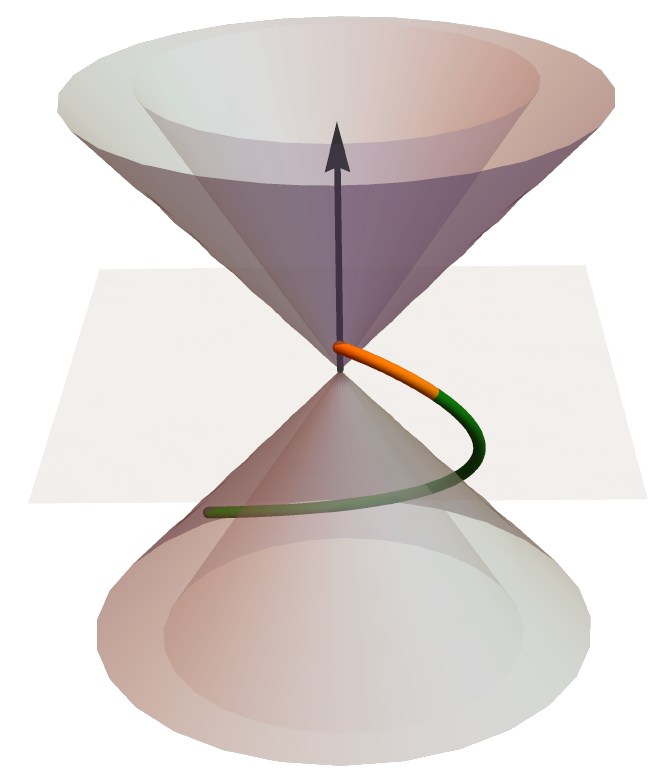}}
\\
\subfigure[$(\sqrt{2},4)$]{\includegraphics[height=6cm]{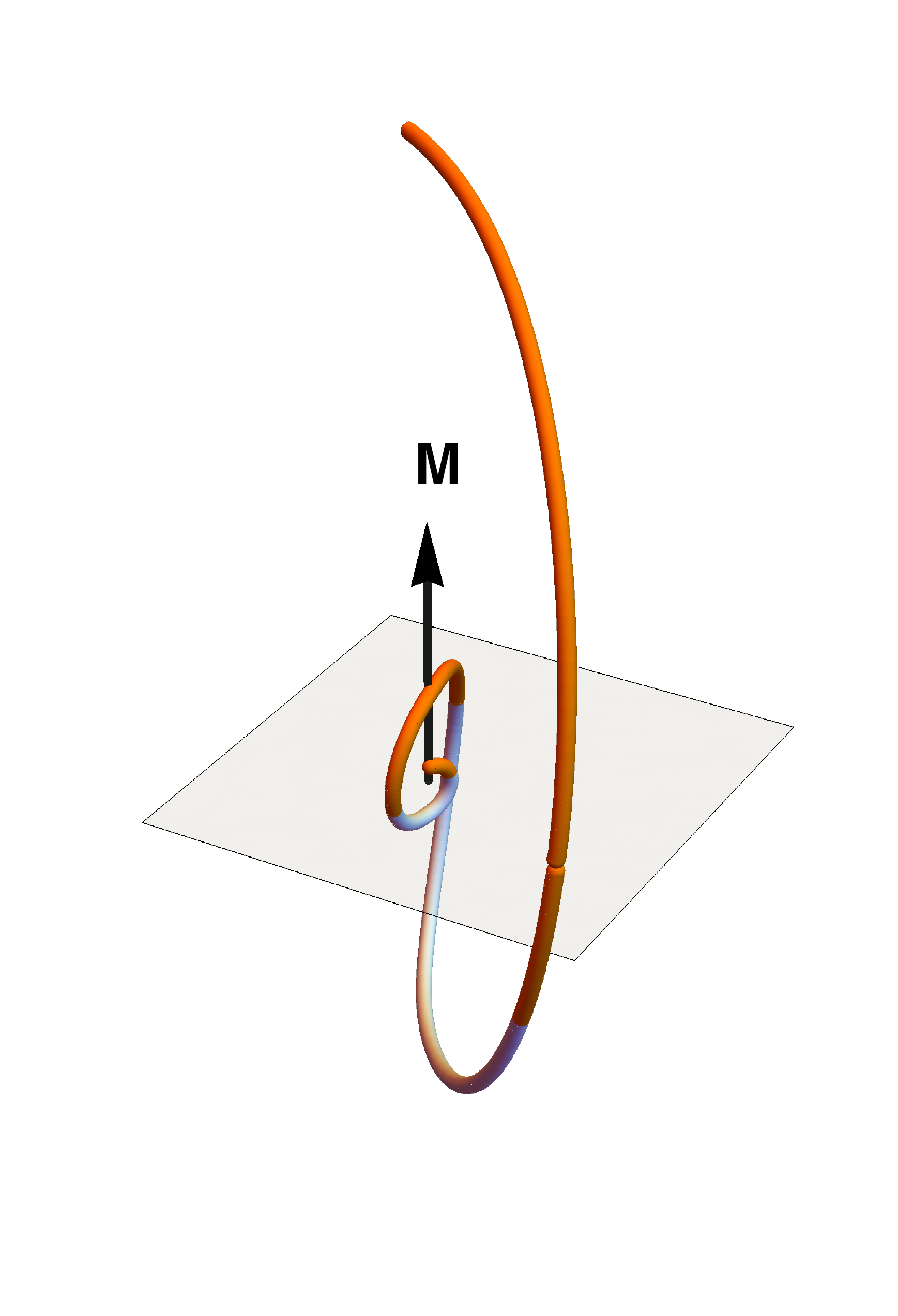}}
\hskip2.5cm \subfigure[$(0.75,0.85)$ {$2^{\sf nd}$ half-cycle}]{\includegraphics[height=5cm]
{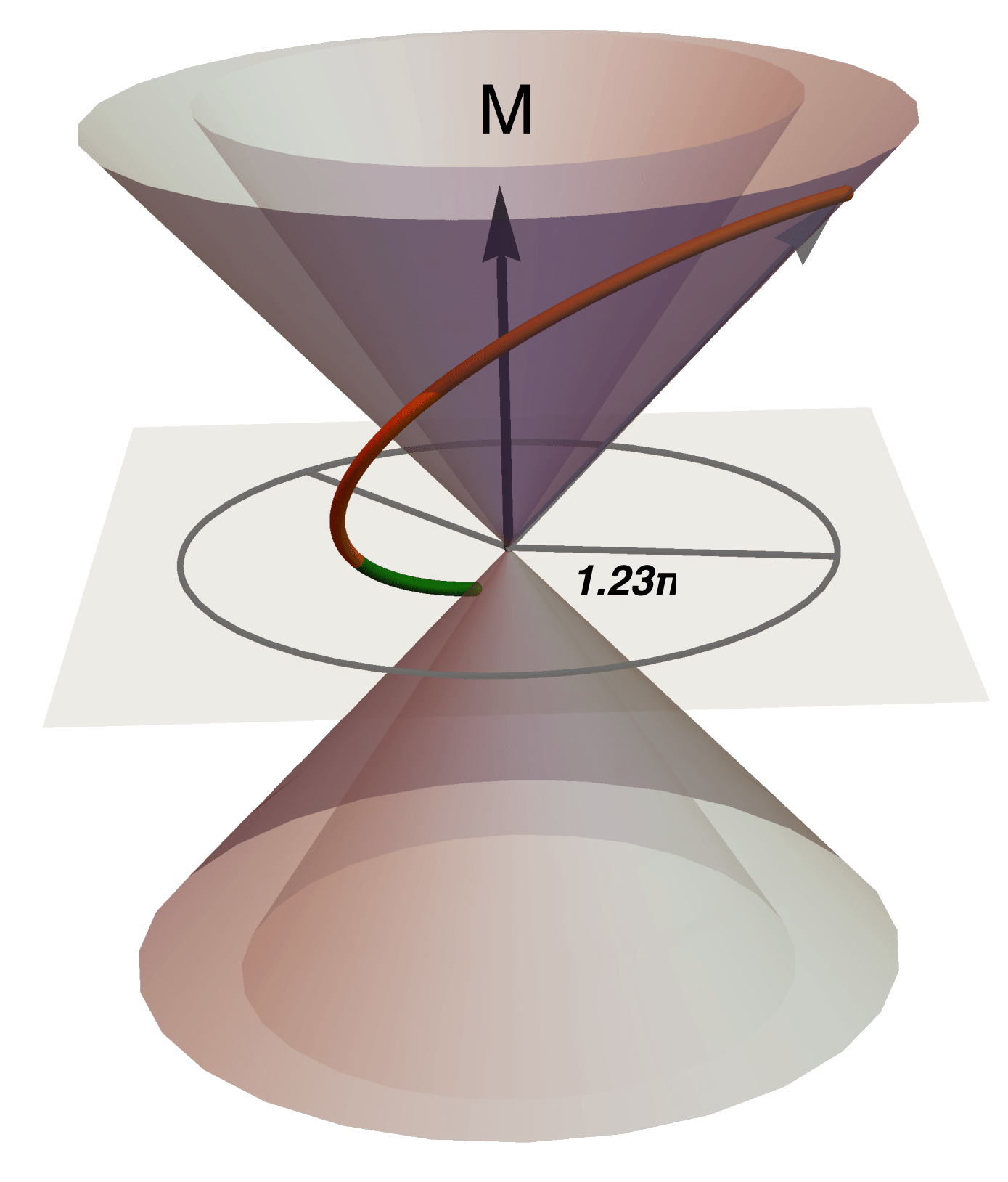}}
\caption{\small \sf   (a)  
Two consecutive cycles (blue, orange)
the diminutive first cycle compared to the second is explained by the relatively large scaling factor 
for a complete cycle of $\kappa_0^{-2}=9.94$); (b,d) together represent one complete cycle when $(\gamma,m)=(0.75,0.85)$; 
(intervals below the mid-plane are colored green, those above it are colored orange). Panels (b) and (d) are presented in register, with (b) scaled by a factor of $6.18$ with respect to (d); (c) Two cycle (light blue below mid plane; orange above). The torque axis is indicated by the vertical black arrows. 
All three trajectories advance anti-clockwise with respect to $\mathbf{M}$.
The torsion in all three cases changes sign mid-cycle, twisting one way in one half-cycle and then in the next. In (b) the angle rotated about $\mathbf{M}$ in a half-cycle,
$\phi_0/2= 221^0$ is indicated.
}
\label{Fig:Traject}
\end{center}
\end{figure}

One is finally in a position to provide a graphical representation of the spiral structure that has been described.  Three representative supercritical spirals  are illustrated in Figure \ref{Fig:Traject}, corresponding to the parameter values 
 (a) $\gamma=0.5$, $m=0.4$ (two cycles);
(b)  $\gamma=0.75$, $m=0.85$ (a half-cycle is represented in panel (b)); 
(c) $\gamma=\sqrt{2},m=4$ (two cycles).
Each example highlights a specific feature of the supercritical geometry. 

 \begin{figure}[htb]
\begin{center}
\subfigure[]{\includegraphics[height=5cm]{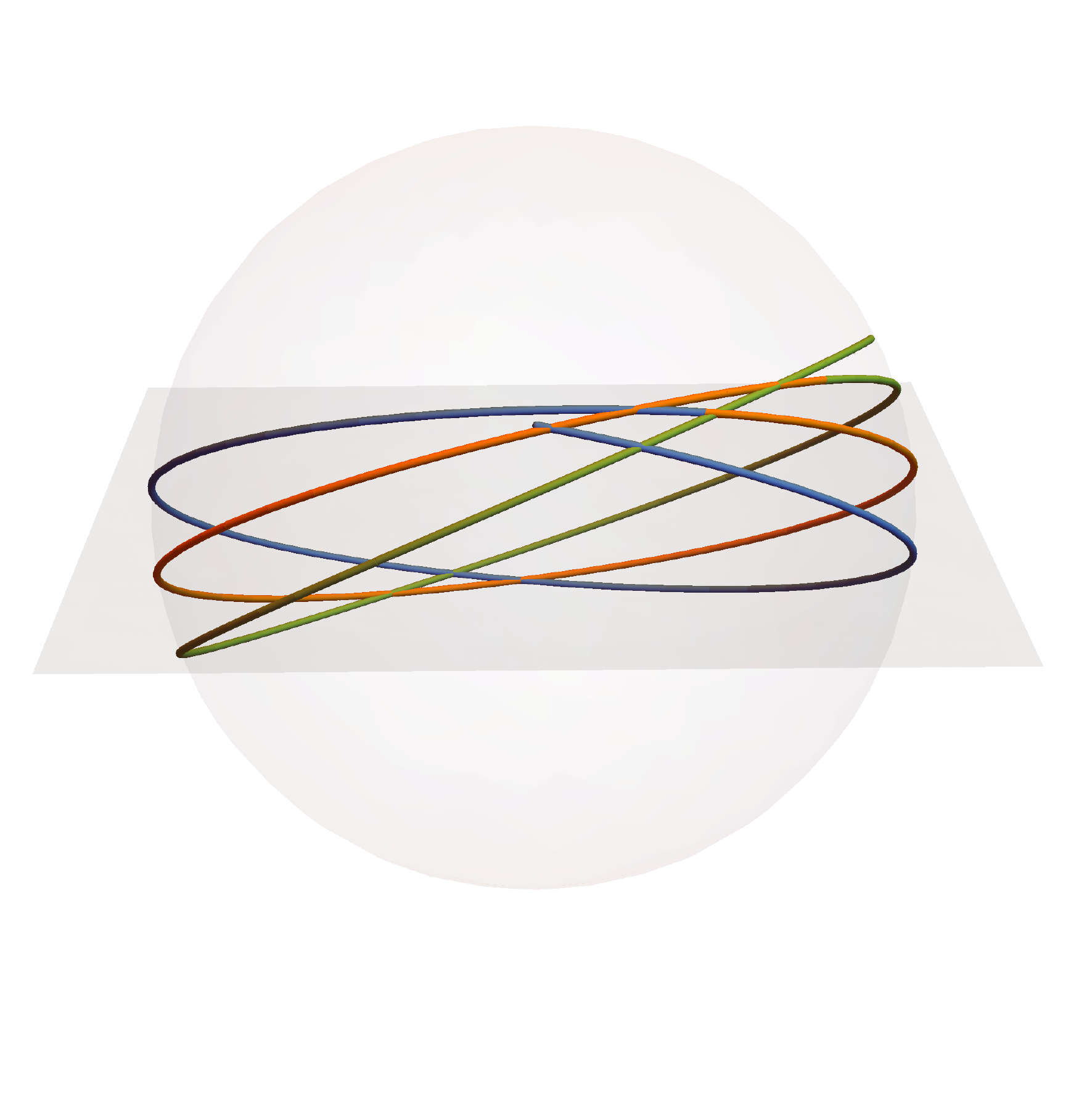}}
\hskip.5cm
\subfigure[]{\includegraphics[height=5cm]{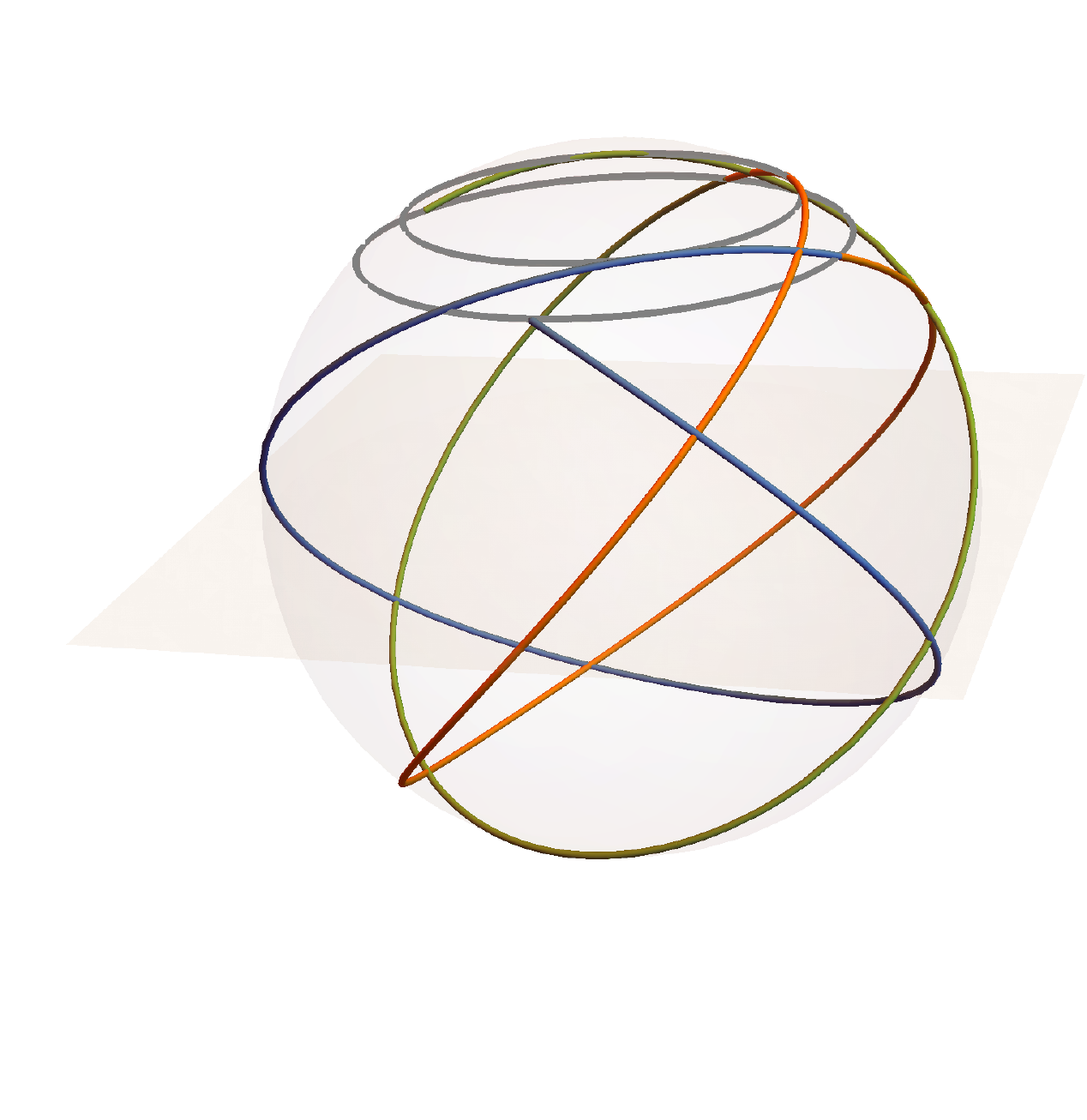}}
\hskip.5cm
\subfigure[]{\includegraphics[height=5cm]{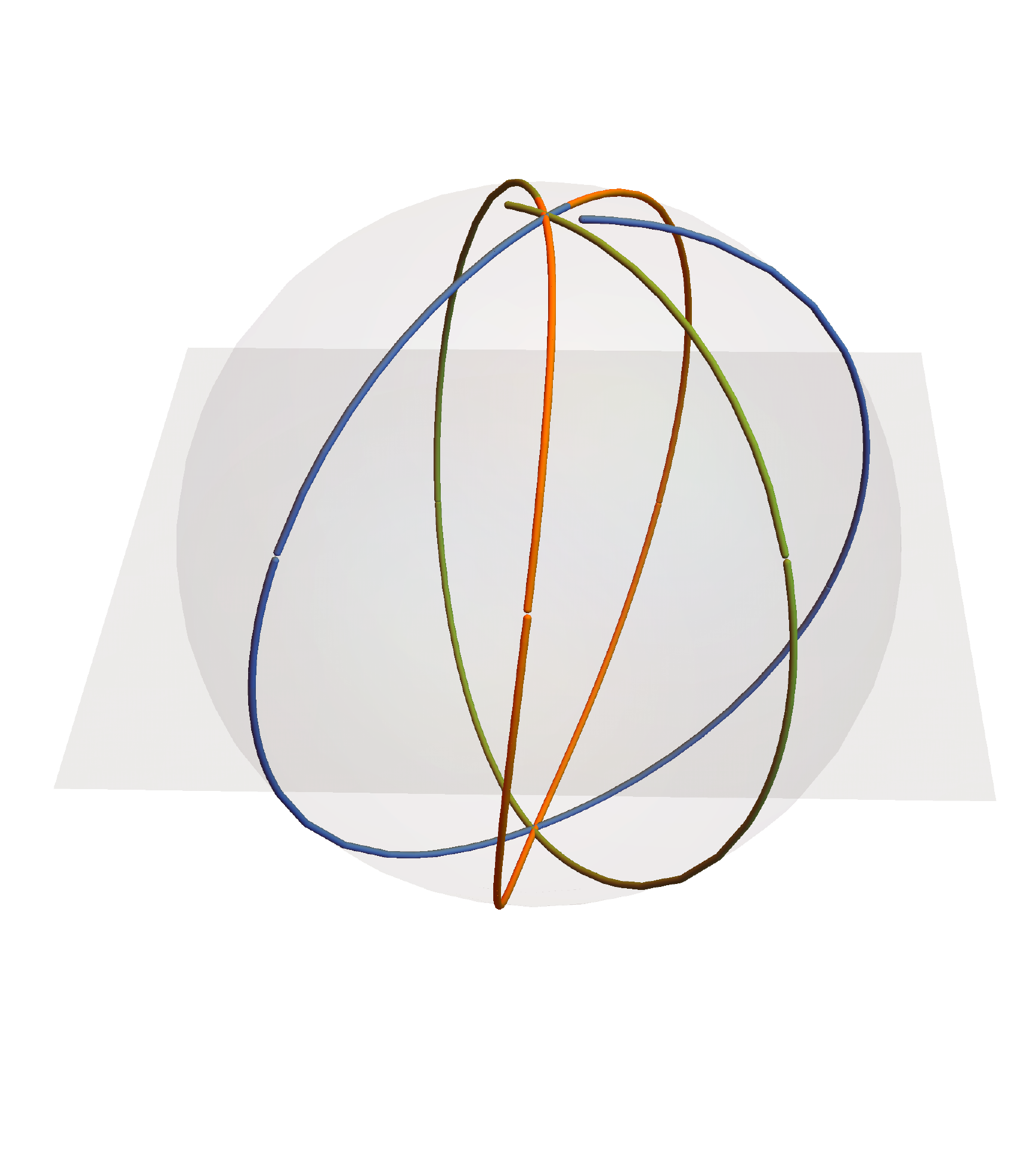}}
\caption{\small \sf Three  consecutive cycles (blue, yellow, green) projected onto a sphere in each of the three examples (a)-(c):  the cycles are bounded between the polar angles $\theta_{\sf Min}$ and $\pi- \theta_{\sf Min}$, the range increases as one proceeds from case (a) to (c) (with increasing values of $m$).
Consecutive cycles precess about the $\mathbf{M}$ axis. The precession angle is not itself sensitively dependent  on  $m$. In 
(a) $\phi_0/(2\pi) =1.07$;  (b) $\phi_0/(2\pi)= 1.23$; (c) $\phi_0/(2\pi)= 1.12$; in all three  $\phi_0 > 2\pi$, but otherwise the values are comparable. The different patterns exhibited in the three is, in large part, attributable to the decreasing value  
of $\theta_{\sf Min}$ as one moves from left to right as the bounding cones close. Whereas the individual cycles of type  (a) and (c) may appear superficially similar, it is clear here that the patterns of their precession are very different. In a logarithmic spiral the projection is represented by a horizontal  equatorial curve.
In a small deformation about a logarithmic spiral the projection will undergo small oscillations about this curve (cf. (a)). The higher value of $\phi_0$ in the intermediate case (b) is responsible for the  rapid azimuthal sweep of radial directions. In this case, the circles of constant $\theta_{\sf Min}$ and $\theta_\gamma$ are traced explicitly in the northern hemisphere to facilitate the \textit{visualization} of the spherical projection.}\label{Fig:Sphere}
\end{center}
\end{figure}

\begin{figure}[htb]
\begin{center}
\includegraphics[height= 4cm]{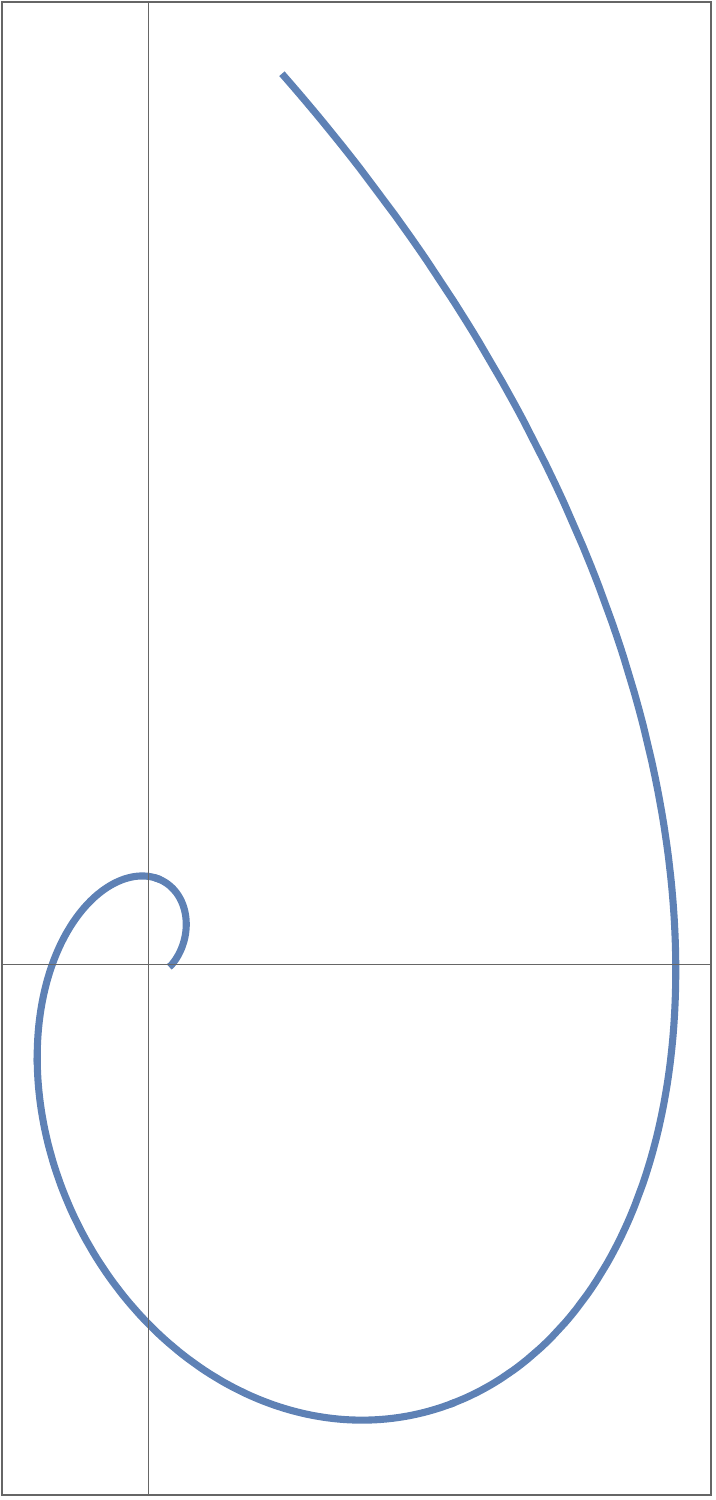}
\hskip1cm
\includegraphics[height= 4cm]{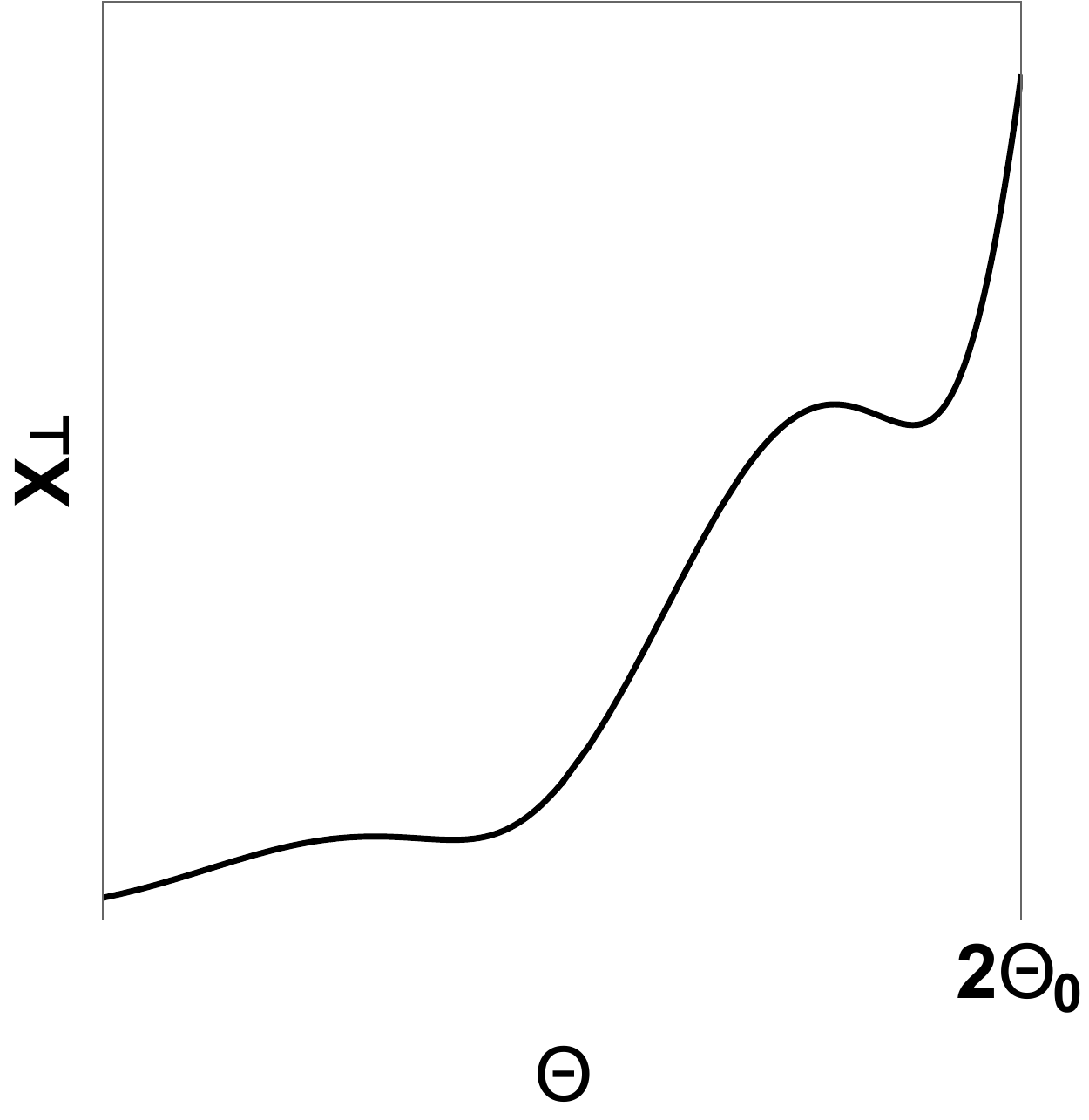}
\caption{\small \sf  (L) The projection of the cycle onto the mid-plane is superficially log spiral-like in example (b), but  (R) if one examines $X_\perp$ as a function of $\Theta$ it displays significant departures from monotonicity. }
\label{Fig:Xperp}
\end{center}
\end{figure}
\vskip1pc\noindent 
To interpret these figures, first focus on the invariant 
cones with polar angles $\theta_{\sf Min}$ and $\theta_\gamma$, discussed in section \ref{Coscycle}); the 
former bounds the trajectory away from the torque axis, the latter locates  
where the projection along this axis turns.  
Both  (a) and (b) represent modest deformations of a logarithmic spiral; $(c)$ does not. 
\\\\
The cumulative effect of precession is best appreciated by following 
the spiral trajectory over a number of cycles. 
This information is captured by the projection of the growing spiral onto a unit sphere centered at its apex,
illustrated in Figure \ref{Fig:Sphere}  for three consecutive cycles. The projection mods out the scaling by the exponential factor $\kappa^{-1}$ which obscures the process.
Even though $\theta_{\sf Min}$ is never strictly vanishing except in the limit of large $m$, so long as differs from $\pi/2$, the supercritical trajectory will eventually intersect every plane in three-dimensional space as it expands due to its precession about the $\mathbf{M}$ axis.
\\\\
In  example (a) 
$\theta_{\sf Min}=58.22^0$, $\theta_\gamma=60.32^0$; even though the angles themselves are large, they differ only by $2^0$, so that two cones are just about resolved (cf. Figure \ref{Fig:Traject}a). 
With the value $\gamma=0.5$, the logarithmic spiral with this value of $\gamma$, lying on the mid-plane, has $m=0.35$, in contrast to $m=0.4$ here.  Even though $\theta_{\sf Min}\uparrow \pi/2$ smoothly as 
$m\downarrow 0.35$ in this region of parameter space (cf. \ref{LimitSuper}), the cosine is sensitive to an increase in the value of $m$.  Despite this the spiral trajectory along a cycle remains approximately planar, even if this plane is not 
orthogonal to $\mathbf{M}$.  If  one were one provided the direction of $\mathbf{M}$,
one would be forgiven for interpreting the trajectory (a) as a logarithmic spiral.  The key distinction between the two is that this new spiral \textit{plane} will itself precess in successive cycles as illustrated in Figure \ref{Fig:Sphere}a.  
\\\\
In example
(b,d) $\theta_{\sf Min}=32.38^0$, $\theta_\gamma=42.67^0$; the increased torque has resolved the two cones.
In Figure \ref{Fig:Sphere},
the spiral trajectory along a single cycle is split over two panels, scaled but in register: (b) and (d) represent the first and second half-cycles respectively.  
The second half-cycle is magnified by a factor $\kappa_0^{-1}=6.18$ with respect to the first. Panel (b)  has been scaled with respect to panel (d) by this factor.  The overall expansion factor is always large when the cones are well-resolved ($\kappa_0^{-2}=38.15$). With this rescaling, 
panels (b) and (d) are identical modulo reflection in the mid-plane, and rotation by $1.226\pi$, the azimuthal angle rotated in a half-cycle. The asymmetry between entry and exit of the cone $\theta_\gamma$ is evident  ($X_\|$ achieves its local extrema on exit in a growing spiral).
\\\\
Example (c)  represents a spiral that is not related to a logarithmic spiral by a continuous increase of $m$ for fixed $\gamma$ with $\gamma>1$ (there is, of course, a continuous path connecting it to a  logarithmic spiral with $\gamma<1$ if $\gamma$ is let vary);  the behavior of this spiral differs from that of the previous examples in several respects.
The two conical polar angles are very small with
 $\theta_{\sf Min} <5^0$, $\theta_\gamma=6.43^0$.
 For this reason,  the cones have not been represented in Figure \ref{Fig:Traject}c. Even with the  chosen modest value $m=4$,  the two cones have essentially collapsed onto the axis (cf. Figure 18b).  
In Figure \ref{Fig:Traject}c,  the trajectory is followed along two consecutive cycles. 
Surprisingly the trajectory along a single cycle does not appear, on first sight, to differ significantly in a qualitative way from that displayed in (a). The two lie approximately within a plane. The difference is that  this plane is rotated significantly towards the vertical in (iii).  This is then reflected in the patterns of precession of the nutating spirals in the two which are very different even if the angles of procession themselves are similar (cf. Figure \ref{Fig:Sphere}).  
\\\\
In examples (a) and (b), the projection $X_\perp$ orthogonal to $\mathbf{M}$  resembles a logarithmic spiral on the mid-plane; in example (b), however, $X_\perp$ is not even monotonic as revealed in Figure \ref{Fig:Xperp}. In (c), where the spiral tilt is much more pronounced, the orthogonal projection looks nothing like a logarithmic spiral. Yet, 
over a single cycle, all three spirals resemble perturbations of a tilted logarithmic spiral.  What distinguishes them is the precession about the torque axis.

\section{Conclusions}

In this paper, 
three-dimensional analogues of planar logarithmic 
spirals have been constructed. This answers a question of interest  both from an intrinsic mathematical point of view as well as across the natural sciences where self-similar spiral patterns are wont to show up and need explaining. 
\\\\
This construction relies on the calculus of variations. The first step is to identify an energy functional exhibiting the maximum symmetry admitting self-similar equilibrium states. In reference \cite{Paper2}, this energy was identified as the conformal arc length along space curves (\ref{eq:Hdef}). This is a functional third-order in derivatives.  While it is not the unique scale invariant energy one can write down at this order,  it \textit{is} the unique conformal invariant.  As such, it  comes with an additional conservation law associated with the extra symmetry. The naive energy, $\int ds\, |\kappa'|^{1/2}$, may be simpler to write-down, but it is not a conformal invariant of space curves, so is missing this conservation law:  one is deceived by appearances. 
\\\\
Not all stationary states of scale invariant energies are themselves scale invariant. The critical missing criterion is that the Noether tension associated with translational invariance must vanish. Otherwise, a length scale is introduced spoiling the symmetry.  In the planar analog of this problem, 
the unique one-parameter family of  planar tension-free states of the conformal arc-length are found to be logarithmic spirals \cite{Paper1}.  
Scale invariance alone is sufficient to make this identification and logarithmic spirals are completely characterized by the constant scaling rate $S$.  
Unfortunately, 
this coincidence is a happy accident of the planar problem. There is also no spatial analogue of the local characterization of tension-free curves in terms of the constant angle that the tangent makes with the radial direction from the apex.  There is much more.
\\\\
Whereas  the magnitude $M$ of the torque  is completely determined by the scaling rate in planar tension-free states, in space the two parameters are independent.  
Scale invariance fixes  the ratio of torsion to curvature ratio locally in terms of $\Sigma=-\kappa'/\kappa^2$; 
the conservation of torque establishes the axis about which the spiral trajectory develops while its magnitude provides a quadrature for $\Sigma$. 
The quadrature indicates a repeating periodic dimensionless internal structure. To trace the trajectory,  the vanishing special conformal current $\mathbf{G}$ was shown to play 
a crucial role in the construction of a polar chart adapted to the  torque direction with its origin located at the spiral apex. The directness of this construction was not anticipated: after all it played no role in the--over-determined--planar reduction of the problem \cite{Paper1}. The intriguing identity
(\ref{eq:X2history}),  a direct consequence of the existence of the special conformal current, was stumbled upon, without forewarning.   It is clear that  we still lack a deeper understanding of why the different elements of the 
construction fall perfectly into place: a mechanical interpretation of $\mathbf{G}$, analogous to that of the other currents, remains elusive.
\\\\
In a planar logarithmic spiral, the two parameters $M$ and $S$ are 
constrained by the relationship $4MS=1$. Stepping up a dimension,  the qualitative behavior  of tension-free spirals is found to depend sensitively on whether $4MS>1$ or $4MS<1$.  
\\\\
In general, the spiral is characterized by an irreducible unit or cycle, describing its nutation between two fixed coaxial circular cones aligned  along the torque axis.  The nutating pattern precesses about this axis as the spiral expands.  
In supercritical spirals with $4MS>1$, the two cones are identical but 
oppositely oriented; the spiral nutates between these two cones, the region within the cones is excluded. 
As it nutates, the projection of the trajectory along the torque axis 
will oscillate, turning on a second pair of cones. At these turning points, the torsion changes sign. 
Thus the spiral twists one way on the way up and in the other on the way down.  
\\\\
The bounding cones of subcritical spirals are nested, oriented in the same direction;  the projection along the torque axis increases monotonically as the spiral nutates between these two cones. The 
torsion has a fixed sign. There is no twisting and untwisting as the spiral precesses and expands.
\\\\
The behavior as $4MS\to1$ in parameter space is not generally continuous.  Continuous supercritical deformations of a  planar logarithmic spiral     
by raising $M,$ keeping $S$ fixed, requires 
$2S>1$.   One  would thus expect spiral formations that can be modeled as small non-planar deviations of a logarithmic spiral to sit in this region of parameter space. There are no corresponding subcritical deformations. 
\\\\
If $2S<1$,  a small change in $M$---either up or down---results in a discontinuous change in the spiral morphology.\footnote{\sf In the introduction, it was suggested that the trajectory followed by the circumnutating tip of a growing tendril in a climbing plant follows  a supercritical template. It would appear, in this context, that these trajectories are not approximated by small deformations of a planar spiral.}
The limiting spiral geometry which corresponds to the limit 
$4MS\downarrow 1$, deviates significantly---towards its center---from the planar logarithmic template.
Curiously the subcritical limit $4MS\uparrow 1$, in the regime 
$2S<1$ is identical to its supercritical counterpart.\footnote{\sf 
It is tempting to speculate that this coincidence is related  to the absence of almost planar (as distinct from planar) spiral morphologies in mollusk shells.}  
\\\\
Matching a self-similar spiral to a template drawn from the minimum model examined here 
involves just two parameters. Any 
two independent measures of the geometry should be sufficient to do this, for example, the angle of nutation and the rate of precession, both captured in Figure \ref{Fig:Sphere}.  Any additional measure (say the expansion rate) had better be consistent with the first two if the minimal template is an accurate one. 
\\\\
Energies constructed from higher-order conformal invariants or conformally-invariant constraints 
will possess self-similar spirals exhibiting additional structure as equilibrium states.
As an example of the former,  a conformally invariant bending energy can be constructed using the conformal curvature, defined by  (cf. \cite{Sharpe1994,Musso1994})
\begin{equation}
\mathcal{K}= -\mu (\partial_s^2+ \kappa^2/2) \mu  +  (\mu')^2/2\,,
\end{equation} 
where $\mu$ is given by (\ref{eq:mudef}). In particular, the conformally invariant energy quadratic in $\mathcal{K}$, 
$H=\int ds \mu^{-1} \,\mathcal{K}^2 $, 
would be expected to exhibit self-similar spirals decorated with additional internal structure. While the details have yet to be worked out in three-dimensions,  
the planar reduction of this energy has been shown to
exhibit non-logarithmic spirals as equilibrium self-similar periodic structures. 
It is also possible to impose conformally invariant constraints on the simple conformal arc-length. The simplest example of this kind adds a twist to the problem, placing a constraint on the total torsion (a conformal invariant modulo $2\pi$), the direct analog of a problem of a physically relevant constraint on Euler-Elastica \cite{IveySinger} in the modeling of semi-flexible polymers.  

\section*{\bf Acknowledgments}

I have benefitted from discussions with Gregorio Manrique and Denjoe O' Connor. Gregorio's assistance navigating Mathematica is very much appreciated.   The hospitality of the Dublin Institute for Advanced Studies is acknowledged.  I thank Niloufar Abtahi for a careful reading of the manuscript. Partial support was received from CONACyT grant no. 180901.

\begin{appendix}

\setcounter{equation}{0}
\renewcommand{\thesection}{Appendix \Alph{section}}
\renewcommand{\thesubsection}{A. \arabic{subsection}}
\renewcommand{\theequation}{A.\arabic{equation}}

\section{Limiting spirals: $m^2\to \gamma^3$}
\label{Limits}

Along $\mathcal{C}_1$ in parameter space (cf. Figure \ref{Fig:Phases}), with $m$ and $\gamma$ tuned so that $m^2 =\gamma^3$ or $4MS=1$,
there is always an equilibrium with $Z=\gamma$ with $\tau=0$. These are planar logarithmic spirals. 
The behavior of trajectories in the neighborhood of $\mathcal{C}_1$, however, depends sensitively on the value of $\gamma$. This behavior will be examined in this appendix.
\\\\
{\bf Supercritical deformations, $m^2 >\gamma^3$, $\gamma<1$}
\\\\
Let $m^2 = \gamma^3 + \delta m^2$, where $\delta=\delta m^2/\gamma^3\ll 1$
Potentials representing this regime are not included in Figure 2, where $\gamma=2$.  A representative sequence of  potentials is displayed for $\gamma=.75$ in Figure 16, where they are labeled $(2)$, $(2)_{0+\delta}$, and $(2)_{0}$, corresponding respectively to a finite $\delta m^2$, $\delta\ll1$, and $\delta=0$.
When $\delta$ is small, the potential well will be small and its curvature at $Z=\gamma$  is positive. Using 
Eq.(\ref{eq:roots}), the lower turning point is approximated by  $Z_- \approx \gamma - 2 \gamma^4 \delta /(1- \gamma^3)$. The well between $Z_-$ and $\gamma$ is now both narrow and shallow and the motion within it describes small oscillations about a logarithmic spiral. The  period is finite and remains finite in the limit. Further details are provided in \ref{LimitSuper}. These are the only small deformations of logarithmic spirals. The behavior in the neighborhood of $\mathcal{C}_1$, when $\gamma>1$ is very different.
\\\\
{\bf Asymptotically Subcritical Logarithmic Helical Spirals: $m^2 \uparrow \gamma^3$; $\gamma>1$}
\\\\
First look at the  limit $m^2 \uparrow \gamma^3$, in the subcritical region. Now
$Z_+ \uparrow \gamma$, whereas 
$Z_-= 1/\gamma^2 < \gamma$.  In the limit the width of the accessible well, described by the potential (\ref{eq:Pdef}), remains finite unless $\gamma \approx1$.  More significantly, 
$Z=\gamma$ is now a double root of the potential, with negative curvature. Such a potential was labelled $(1)_\infty$ in Figure \ref{Potential1}.  The period of oscillation $\Theta_0$ diverges in the limit. 
As $Z\to\gamma$, $\tau\to0$ and $\kappa \to 2S/s$, so that  deformed helical spiral fans out asymptotically into a planar logarithmic spiral. 
Because the well is finite,  large deviations away from this asymptotic plane persist towards the center of the spiral.  The details are presented in \ref{asymptotrise}. Small subcritical deformations of logarithmic spirals do not occur as tension-free equilibrium states. 

\begin{figure}[htb]
\begin{center}
  \includegraphics[height=5cm]
{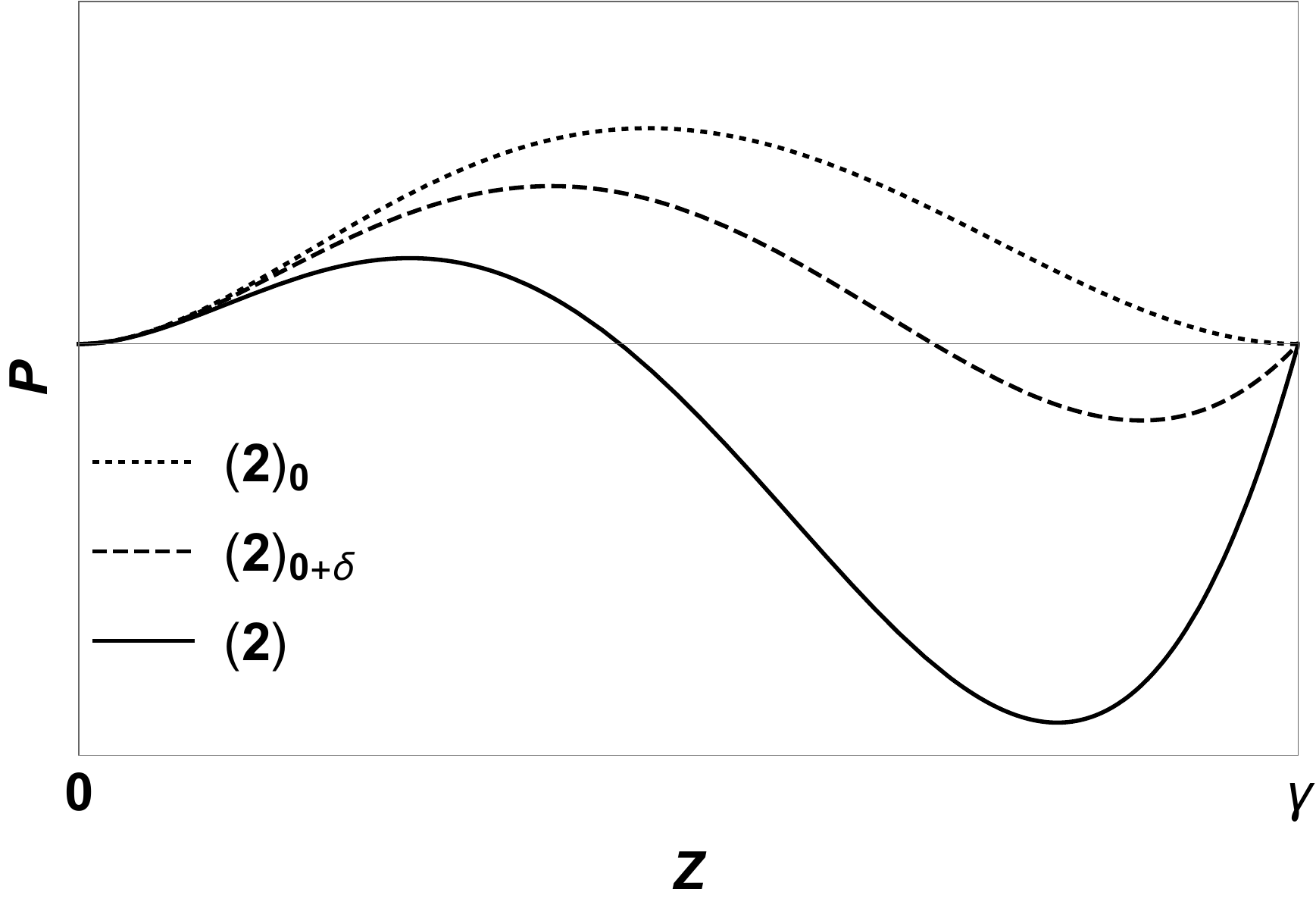}
\caption{\small \sf The potential for fixed $\gamma=.75$,  and $m=0.75^{3/2}, 0.85^{3/2}$ and  $1.3^{3/2}$.
Continuous supercritical deformations of logarithmic spirals require $\gamma<1$.  }
\end{center}\label{Potential2}
\end{figure}
\vskip1pc\noindent
{\bf Asymptotically Supercritical Logarithmic Spirals: $m^2 \downarrow \gamma^3$, $\gamma>1$}
\\\\
Because $Z_+>\gamma$ in the approach to the limit, 
both $\kappa$ and $\tau$ undergo large oscillations about the logarithmic value (and the sign of $\tau$ alternates) while the period of oscillation diverges. Thus $\tau$ does not get to change sign. 
Though the supercritical and subcritical approaches to the limit are very different (when $\gamma>1$),  
implying very different behavior, 
the two limiting potentials are identical ($(1)_\infty$ in Figure \ref{Potential1}); as a consequence, the limiting trajectories are also.  Supercritical deformations of logarithmic spirals are discontinuous if $\gamma>1$.  

\setcounter{equation}{0}
\renewcommand{\thesection}{Appendix \Alph{section}}
\renewcommand{\thesubsection}{B. \arabic{subsection}}
\renewcommand{\theequation}{B.\arabic{equation}}

\section{Extrema of $\tau/\kappa$}
\label{tau/kappabound}
In  section (\ref{torsionbound}) It was shown that, whenever $Z=Z_{\Gamma\,{\sf max}}=2\gamma/3$ is accessible, the extrema of $\tau/\kappa$ occur there. To determine if $
$ is accessible, first note that if $\gamma^3 \le 9/4$, then 
$Z_{\Gamma\,{\sf max}} \ge Z_-$, whenever  
\begin{equation}
\label{equalZs}
 m^2 - 2 \ge  2/3\,(\gamma^3-9/4)\,, 
\end{equation}
with 
equality holding along the red straight line with respect to the variables $(\gamma^3,m^2)$ in parameter space (cf. the red curve $\mathcal{C}_\Gamma$ in Figure 17). There is a small window of both subcritical and supercritical trajectories in which the maximum ratio occurs at $Z_-$.  
If $\gamma^3 > 9/4$, $Z_{\Gamma\,{\sf max}} \ge Z_-$ along all admissible trajectories. 
\\\\
On the other hand, if $\gamma^3 > 9/4$, $Z_{\Gamma\,{\sf max}} \le  Z_+$  whenever the inequality  (\ref{equalZs}) holds, whereas if $\gamma^3\le 9/4$, it holds for all admissible trajectories.  There is no upper constraint in supercritical trajectories.
\\\\ 
At the point $\gamma^3=9/4$, $m^2= 2$, $\mathcal{C}_\Gamma$ and $\mathcal{C}_0$ touch tangentially  in parameter space. The two roots of the potential coincide 
along $\mathcal{C}_0$.  Thus the conical helix with the largest ratio $\tau/\kappa$ occurs when $\gamma^3=9/4$ and $m^2= 2$. Now $\Gamma_{{\sf max}}= \sqrt{1- (2S)^2}= 1/\sqrt{3}$. This particular helix also saturates the bound (\ref{eq:Gammabound}). 
\begin{figure}[htb]
\begin{center}
\includegraphics[height=5cm]{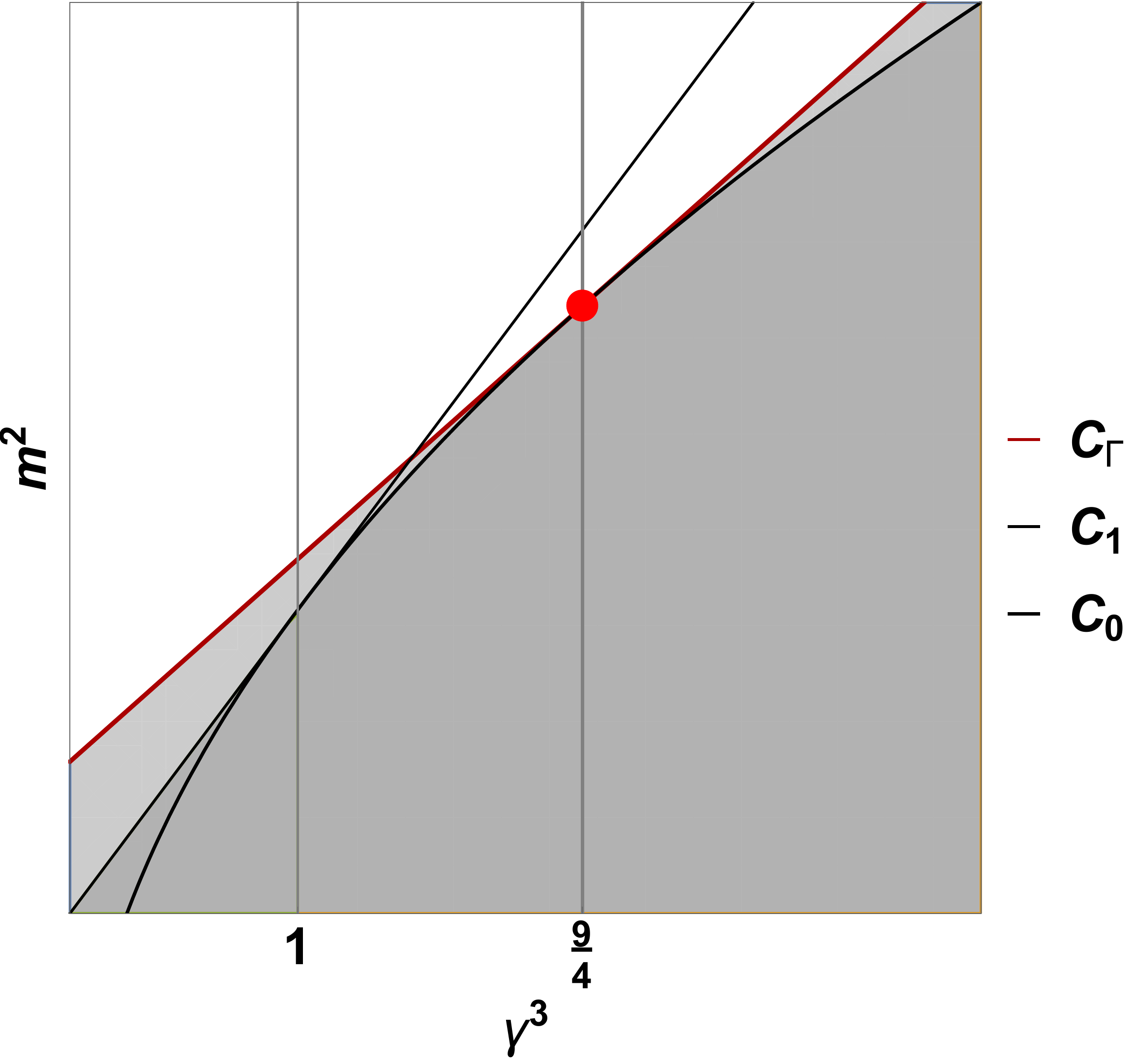}
\caption
{\small \sf Above $\mathcal{C}_\Gamma$ in parameter space, $\Gamma$ assumes its maximum $2\gamma^3/(3\sqrt{3})$ within the interval, $(Z_-, {\sf Min}(Z_+,\gamma)$.
If $\gamma^3<9/4$, $Z_\Gamma<Z_-$, so that the maximum is assumed at $Z_-$; 
If $\gamma^3>9/4$, $Z_\Gamma>Z_+$,  so that the maximum is assumed at $Z_+$.
These two regions  where the maxima of $\Gamma$ occur at the turning points of the potential
are represented by the light gray region lying below $\mathcal{C}_\Gamma$. The darker gray region is inaccessible. }
\end{center}\end{figure}

\setcounter{equation}{0}
\renewcommand{\thesection}{Appendix \Alph{section}}
\renewcommand{\thesubsection}{C. \arabic{subsection}}
\renewcommand{\theequation}{C.\arabic{equation}}

\section{Tension-free conical helices and their perturbations}
 \label{conehelix}
 
It is instructive  to examine 
 conical helices in the light of the conservation laws.  
They arise when the two roots of the quadratic 
appearing in the potential (\ref{eq:Pdef}) coincide.  As seen
below Eq.(\ref{eq:calC0}),  
this occurs when $Z=\gamma^{-1/2}$ and $\gamma\ge1$ (or
$\Sigma=2S$ and $2S\le 1$, where $\Sigma$ is defined in Eq.(\ref{eq:SigGamdef})). This integrates  to yield
\begin{equation}
\label{kappashelix}
\kappa s=(2S)^{-1}\,.
\end{equation}
Eq.(\ref{eq:GamSig0}) now implies that the ratio $\tau/\kappa$ is also constant;  
$\tau/\kappa= \sqrt{1-(2S)^2}$.  Using  Eq.(\ref{kappashelix}),  this implies   
$\tau s =\sqrt{1-(2S)^2} /(2S)$.
\\\\
To construct the spiral trajectory, note that $Q$, defined by Eq.(\ref{eq:Qdef}), vanishes so that 
Eq.(\ref{eq:kapparho}) reduces to 
\begin{equation}
(1 + m^2) \kappa^2 \rho^2 =  [ \gamma^2 (\gamma -Z_0)   + \gamma Z_0^{-1}]\,,
\end{equation}
where $Z_0=\gamma^{-1/2}$. 
Using the identity 
(\ref{eq:calC0}),  and Eq.(\ref{kappashelix}), the distance to the apex is found to be $
\rho= \, s/\sqrt{2}$.
Unlike the curvature or torsion, 
$\rho$ does not depend on the value of $S$. 
This behavior is also quite different from that of a planar 
logarithmic spiral, with $\rho=\cos\alpha\, s$, depending on $S$ through the angle $\alpha$, the tangent makes with the radial direction, $2 S  = \tan\alpha$.  This is consistent with the 
observation that there are no small subcritical deformations of logarithmic spirals discussed in \ref{Limits}, and expanded on in \ref{asymptotrise}. 
If $m$ is raised keeping $S$ fixed, the spiral will nutate and $\rho$ grows sublinearly with $s$. How 
it does this will depend on both $S$ and $M$.
\\\\
These tension-free helices lie on a cone with a constant opening angle. 
Of course, not all self-similar spirals on cones are realized by projection of tension-free equilibrium states.
The cosine of the opening angle,  determined by Eq.(\ref{eq:cosvsZ}), is given by \footnote{\sf its sine has a very simple expression in terms of $m$: $
\sin \theta_0 = {1}/{m}$.} 
  \begin{equation}
  \label{costhetacone}
  \cos\theta_0
    = \sqrt{ \frac{  1-(2S)^{2}}
{1 - 2S^2}}\,;
\end{equation}
The projection of the position vector along the torque $X_\|=\rho \cos\theta_0$ is completely determined by $\theta_0$, $X_\| (s)= s \cos\theta_0 /\sqrt{2}$.
When $2S=1$, $\theta_0=\pi/2$ so that the cone splays open into a plane and the conical helix splays out into a planar logarithmic spiral.  
 \\\\
Note that cones with a fixed $\theta_0$ are preserved under conformal inversions centered on the apex.  One can further confirm that, modulo the reversal of torsion, the spiral on this cone is as well.  Conical helices corresponding to distinct value of $S$ (or $\theta_0$), are conformally inequivalent.
\\\\
 In a conical helix, $\rho'{}^2 + \rho^2 \sin^2 \theta_0  \phi'{}^2 =1$, with $\rho=s/\sqrt{2}$. It follows that\footnote{
 \sf Notice that Eq.(\ref{philns})  is consistent with Eq.(\ref{eq:rhopphip})  with $\rho'$ and $\theta$  constant.}
 \begin{equation}
 \label{philns}
 \phi= \ln s /\sin\theta_0 = m\, \ln s\,.
 \end{equation}
An interesting corollary is that the conical pitch (or the angle $\Phi$  that the tangent makes with circles of constant $\rho$) is constant, given by $\tan \Phi= \rho'/(X_\perp \phi')=1$, or $\Phi=\pi/4$ independent of 
 $\theta_0$. This should not be conflated with the angle that the tangent makes with the torque axis,  introduced in 
 Eq.(\ref{eq:XparptdotM}),
which is also constant, but $\theta_0$ dependent, given  by $\cos \Psi= X_\|'= \cos \theta_0/\sqrt{2}$.
\\\\
To summarize, there is a one-to-one correspondence  between cones and the tension-free spirals they host. But key properties---notably the distance--arc length relationship and the pitch---turn out to be  independent even of this cone.
\\\\
{\bf Period of small oscillations about conical helices}
 \\\\
 The curvature of the potential (\ref{eq:Pdef}) at $Z=\gamma^{-1/2}$ determines the period of small oscillations about a conical helix. The harmonic approximation reads
\begin{equation}
\label{Quadraturehelix}
 Z^{\bullet\bullet} +
 4 (1 - \gamma^{-3/2}) (Z - \gamma^{-1/2}) \approx 0
 \,,
\end{equation} 
so that this  period  is given by $\Theta_0=2\pi/\omega$, where $\omega = 2 \sqrt{1 - \gamma^{-3/2}}=2 \sqrt{1- 4S^2}$. One is now in a position to express $\phi$ as a function of $\Theta$. The relationship
 $\kappa s= (2S)^{-1}$ implies, upon integration, that 
 $\Theta= \ln s\, /2S$.  Together with the identity  Eq.(\ref{philns}), it follows that 
 \begin{equation}
 \label{phiTheta0}
\phi= \sqrt{2} \sqrt{1 - 2S^2}\, \Theta \,,
\end{equation}
so $\phi$ is proportional to $\Theta$ in conical spirals; the two diverge with a ratio of $1$ as $2S\to 1$ (or $\gamma\to1$).  Importantly, $\phi$ always exceeds $\Theta$; this property is seen to hold in all tension-free states and, as discussed in sections \ref{phi} and \ref{TracingTrajectories},  establishes the pattern of procession of the spiral cycle about the torque axis. 
In one period $\Theta_0$, $\phi$ rotates by the angle
 $ \phi_0 = \sqrt{2} \pi \sqrt{(1 - 2S^2)/(1-4S^2)}$, increasing monotonically with $S$.  Below $S=1/\sqrt{6}$ (or $\theta_0<\pi/4$), $\phi_0 < 2\pi$; above it, $\phi_0 >2\pi$ (diverging as $2S\to1$ because $\Theta_0$ does). $\phi_0=2\pi$ occurs where $\tau/\kappa$ is maximized.  This behavior contrasts with that exhibited by supercritical spirals, in which $\phi_0$ always exceeds $2\pi$.
 \\\\
 One is now in a position to determine the decline in curvature $\kappa_0^{-1}$ given by Eq.(\ref{kapthet}), over  a period $\Theta_0$ 
 One determines  
 \begin{equation}
\kappa_0 = e^{- \gamma^{-3/2}\Theta_0}
=  \exp\left(- \pi \sqrt{\frac{4S^2}{1- 4S^2}}\right) =  \exp\left(- \sqrt{2} \pi \tan \theta_0\right)\,,
\end{equation}
where the identity $Z_0=\gamma^{-1/2}=(2S)^{2/3}$ has been used in the first expression,  the expression for $\Theta_0$ given below Eq.(\ref{Quadraturehelix}) in the second, and the trigonometric identity (\ref{costhetacone}) in the third.
As $2S\to1$, $\theta$ opens and $\Theta_0$ diverges, so that $\kappa_0\to 0$.

\setcounter{equation}{0}
\renewcommand{\thesection}{Appendix \Alph{section}}
\renewcommand{\thesubsection}{D. \arabic{subsection}}
\renewcommand{\theequation}{D.\arabic{equation}}

\section{The bounding cones}
\label{BoundingCones}

Using Eq.(\ref{eq:cosvsZ}) with
$W_\pm$ defined by Eq.(\ref{eq:Wdef}), it is evident that the opening angle ($\theta_{\sf Min}$) of the bounding cone is always located where $dW_+/dZ=0$;  the maximum in a subcritical cycle $\theta_{\sf Max}$ occurs when $dW_-/dZ=0$.
\\\\
One can show that $d W_\pm/dZ =0$ when 
\begin{equation}
[\gamma^6 + 4 \gamma^3 - (m^2+1)^2 ] Z^2 + \gamma [(m^2+1) (m^2+1 -\gamma^3) - 4 \gamma^3] Z+ \gamma^5=0\,.  
\label{eq:Quadratic}
\end{equation}
To see this, first note that 
\begin{eqnarray}
W_\pm^\bullet &=& 
 \big[ (m^2 +1)/\gamma^2 -2 Z\big] Z(\gamma -Z)^{1/2}  \pm 
[ \gamma - 2Z]  Z^{1/2} Q  \,.\label{Wprime}
\end{eqnarray}
Thus, at an extremum,  $Q$ is given by
\begin{equation}
Q=\mp \left(\frac{ (m^2 +1)/\gamma^2 - 2Z} {\gamma - 2Z}\right) \, Z^{1/2}  (\gamma -Z)^{1/2}  \,,
\end{equation}
or, equivalently,
\begin{equation}  
Z(\gamma-Z )\,\left(\frac{m^2+1 }{\gamma^2} - 2  Z \right)^2 + \left(Z^2 - \frac{m^2+1 }{\gamma^2}\, Z + \frac{1}{\gamma}\right) (\gamma-2Z)^2 =0\,,
\label{P40}
\end{equation}
which reproduces Eq.(\ref{eq:Quadratic}). 
\\\\
The two roots of the quadratic are given by 
\begin{equation}
\label{eq:rootsZtheta}
Z_{{\sf \theta}\,1,2 }/\gamma =\frac{- [(m^2+1) (m^2+1 -\gamma^3) - 4 \gamma^3]\pm
\sqrt{- (1 - \gamma^3 + 
    m^2)^2 (4 \gamma^3 - (1 + m^2)^2)}
}{2[\gamma^6 + 4 \gamma^3 - (m^2+1)^2] }\,.
\end{equation}
The opening angle $\theta_{\sf Min}$ occurs at the smaller root 
 $Z_{\theta\, {\sf Min}}$. The functional dependence of $Z_{\theta\, {\sf Min}}$ and $\cos (\theta_{\sf Min})$ on $m$, for a fixed value of $\gamma$, is indicated
by the solid black curves in Figures 18a and b. 

\begin{figure}[htp]
\begin{center}
\subfigure[]{
\includegraphics[scale=0.40]{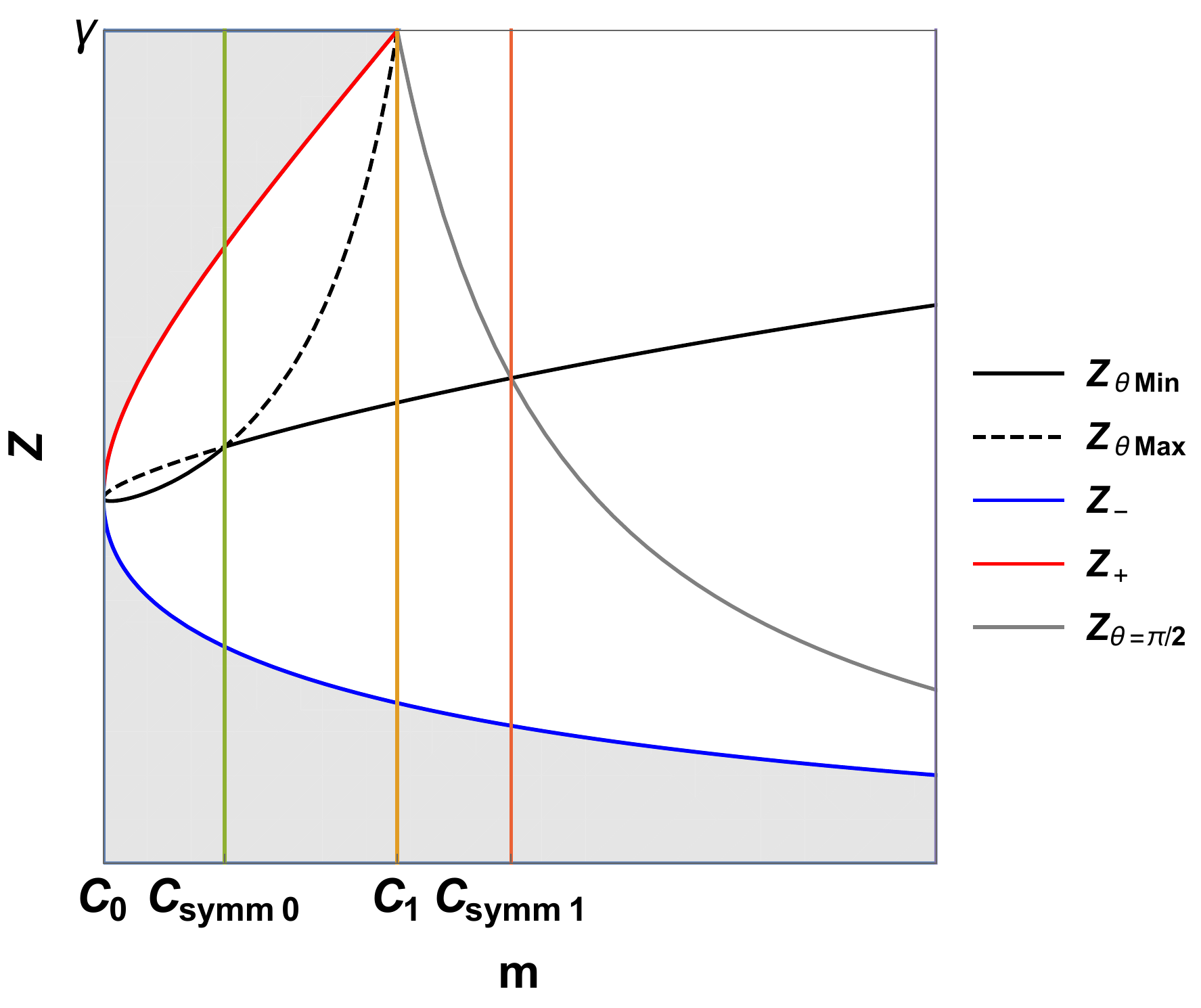}}
\hskip.25cm
\subfigure[]{\includegraphics[scale=0.42]{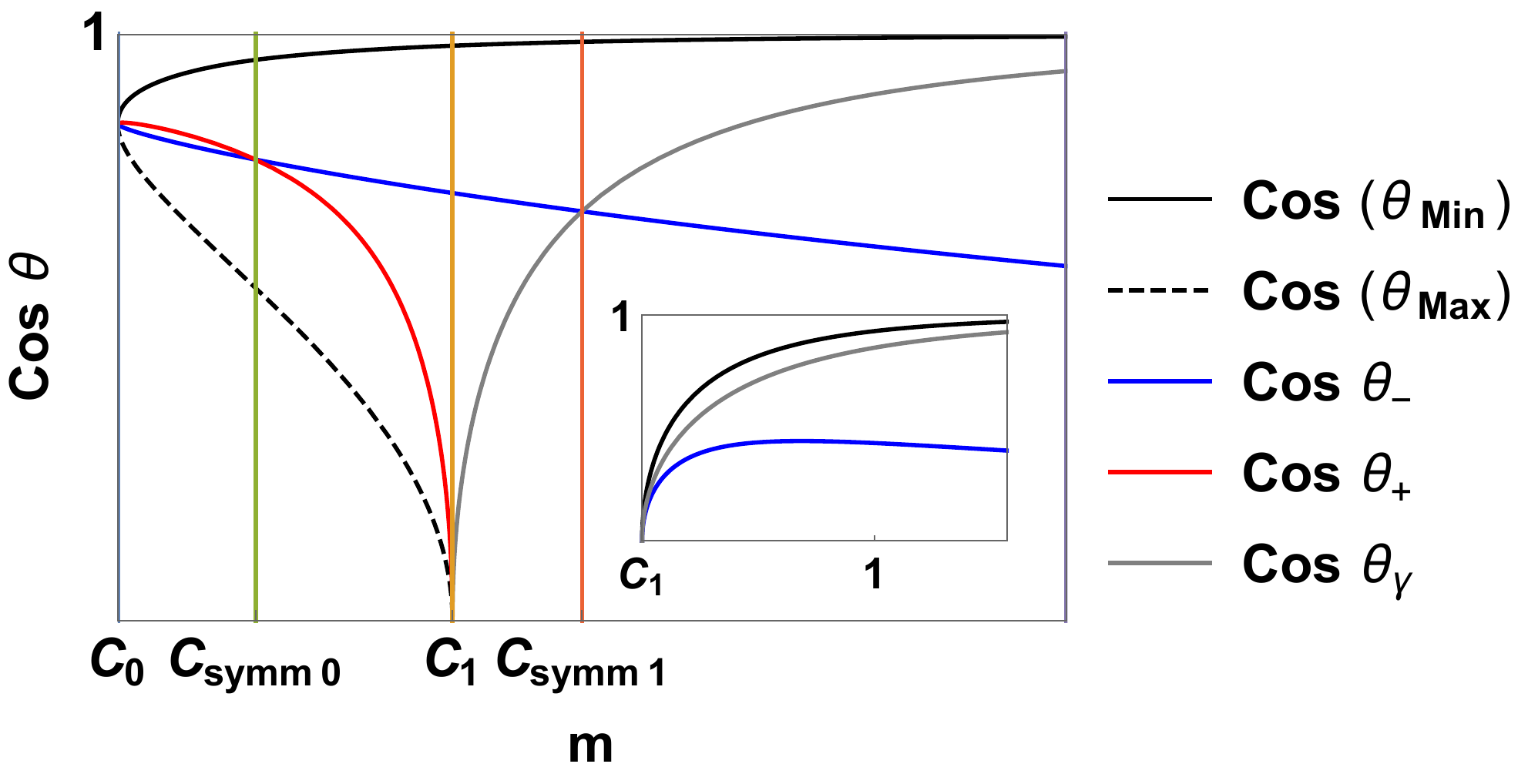}}
\caption{\small \sf  (a) Locating $Z$ where $\theta=\theta_{\sf Min}$ and (below  $\mathcal{C}_1$) where $\theta=\theta_{\sf Max}$ as functions of $m$ for $\gamma=\sqrt{3}$); (b) The corresponding apex angle cosines of the bounding cones; the insert reveals the very different behavior of $\theta$ when $\gamma<1$ (here  $\gamma=0.75$). The values of the turning points 
$Z_-$ and ${\sf Min}[Z_+,\gamma]$ and the corresponding cosines at these points are included (in blue and red/gray) to facilitate reading the figures.}
\end{center}\label{Fig:Cosvsm}
\end{figure}

\vskip1pc\noindent
In subcritical spirals  the two roots  
lie within the interval $[Z_-,Z_+]$; the larger root identifies the opening angle of the exterior bounding cone, $\theta_{\sf Max}$.  
The dependence of $Z_{\theta\, {\sf Max}}$ vs. $m$ and $\cos (\theta_{\sf Max})$ vs. $m$ are indicated by the dashed black curves in Figure 18.  The minimum occurs along the $(+,+)$ cycle segment, 
the maximum along $(+,-)$. The  corresponding values  of $Z$ coincide,  or $Z_{\theta\, {\sf Min}}= Z_{\theta\, {\sf Max}}$, when the 
discriminant on the right in Eq.(\ref{eq:rootsZtheta}) vanishes.  
This occurs either trivially  if $(m^2+1)^2=4\gamma^3$, or, non-trivially if $m^2=\gamma^3-1$.
The former corresponds to the collapse of the Cosine cycle to a single point along 
 the interface $\mathcal{C}_0$ in parameter space, so that $\theta_{\sf Max}=\theta_{\sf Min}$ and  the  two bounding cones coincide.  The non-trivial possibility, with $\theta_{\sf Max}\ne \theta_{\sf Min}$ is consistent with the lower bound on $m$ only when $\gamma^3\ge 4$.  The corresponding  locus in  parameter space  is indicated by the curve $\mathcal{C}_{\sf S\,0}$ in Figure 18. 
The  $\cos\theta$ cycle exhibits a left-right symmetry where this occurs.  

 \begin{figure}[htb]
\begin{center}
\includegraphics[scale=0.30]{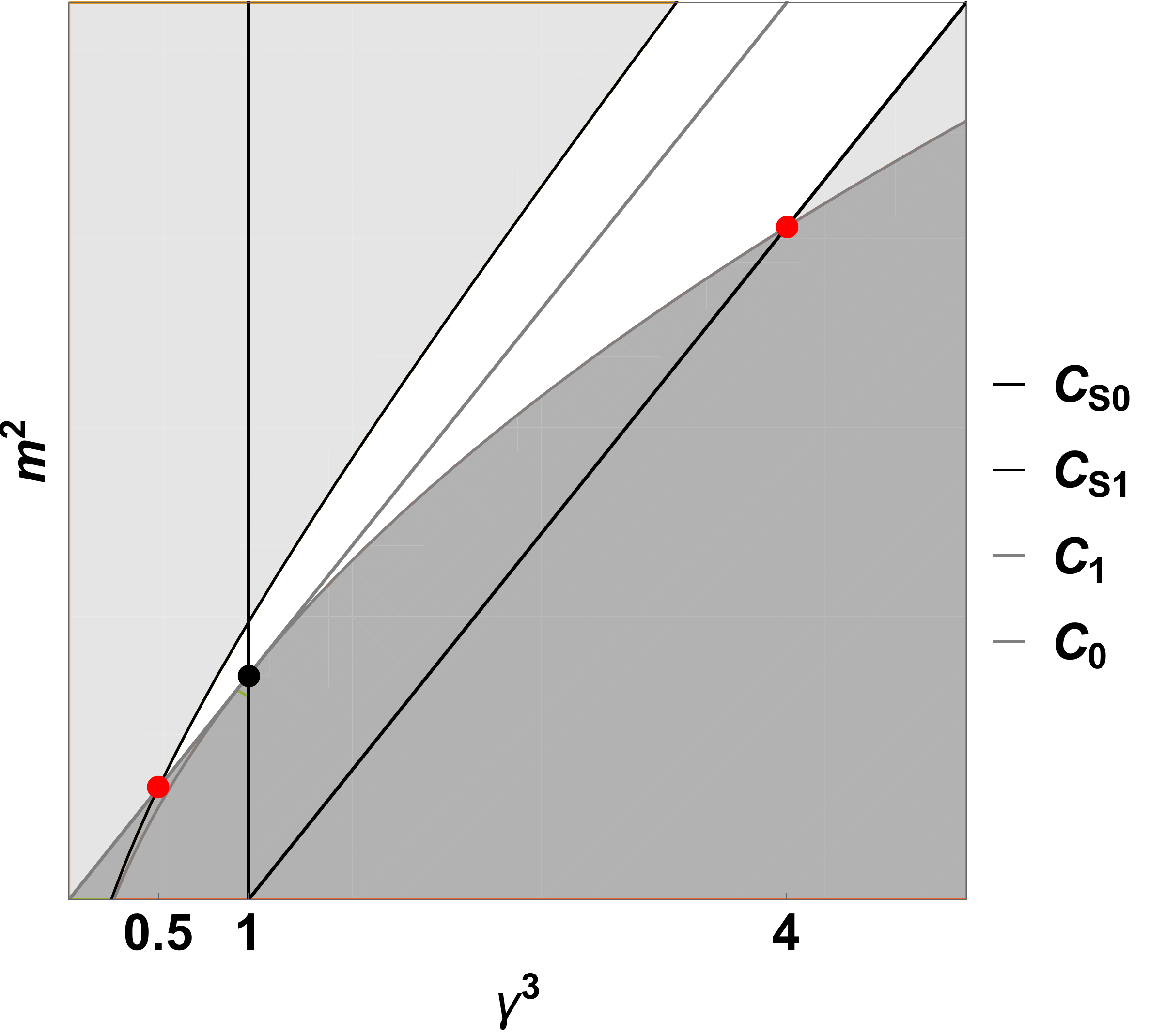}
\caption{\small \sf  $\mathcal{C}_{\sf S\,0}$ ($\mathcal{C}_{\sf S\,1}$) locates symmetric subcritical (supercritical) cosine cycles  in parameter space.  Below $\gamma^3=4$, all non-conical subcritical cycles are asymmetric; below $\gamma^3=0.5$, all cycles are asymmetric. 
} 
\end{center}\label{PhaseSymmCos}
\end{figure}

\vskip1pc\noindent In supercritical spirals the larger root, given by (\ref{eq:rootsZtheta}),
lies outside the interval $[Z_-,\gamma]$.  Even though it lies outside the well it still conveys information concerning behavior within it. In particular, it diverges when the coefficient $a=(m^2+1)^2-4\gamma^3 -\gamma^6$ of $Z^2$ in (\ref{eq:Quadratic})  vanishes,  and the quadratic reduces to a linear equation.  At this value of $m$, $Z_{\theta\, {\sf Min}}=Z_{\theta=\pi/2}$,  the value of $Z$ where mid plane crossing occurs (cf. section \ref{mid}), the former assumed along the $(+,+)$ and $(-,+)$ cycle segments with $Z^\bullet>0$, the latter along the $(+,-)$ and $(-,-)$ segment with $Z^\bullet<0$.
At coincidence, the $\cos\theta$ cycle is again  left-right symmetric about this value of $Z$.
 The corresponding  locus  in  parameter space  is indicated by the curve $\mathcal{C}_{\sf S1}$ in Figure 18. This curve lies above $\mathcal{C}_1$ in the admissible region, intersecting it when $\gamma^3=1/2$.  There are no symmetric supercritical cycles if $\gamma^3<1/2$.  
\\\\
In Figure 18b, observe that $\theta_{\sf Min}$  is continuous as a function of $m$ across $\mathcal{C}_1$;
while $\theta_{\sf Max} \uparrow \pi/2$ as $m^2\uparrow \gamma^3$,
and $\theta_\gamma\uparrow  \pi/2$ as $m^2\downarrow \gamma^3$.
These results are consistent with the  common limiting asymptotically planar logarithmic spiral form as $\mathcal{C
}_1$ is approached from above and below when $\gamma>1$.. Above  $\mathcal{C}_1$, 
$\theta_\gamma$ decreases monotonically to zero from $\pi/2$.  The position of $Z_{\theta=\pi/2}$, locating mid-plane crossing (discussed in section \ref{Coscycle}), is indicated by the descending gray curve in Figure 18a. It originates  at $Z=\gamma$ along $\mathcal{C}_1$ and is always bounded below by $Z_-$. 
\\\\
As $m$ is increased further, as illustrated in Figure 18, all cycles tend to a qualitatively identical form with $\theta_{\sf Min}$ and $\theta_\gamma \to 0$ as the two cones close onto the pole;  $\theta_{Z_-}$ however decreases monotonically to $\pi/2$, while the range of $Z$ approaches its maximum extent, $[0,\gamma]$. This asymmetry in the 
corresponding cycles in illustrated in Figure \ref{Morecycles}a, as well as the left-most cycle in \ref{Morecycles}b.
\\\\
The insert in Figure 18b, describing $\gamma<1$ supercritical cycles, indicates that $\theta_{\sf Min}=\pi/2$ along  $\mathcal{C}_1$ so that cycles vanish there.  The vanishing point represents a logarithmic spiral. 
Further details of the distinction between
limiting supercritical spirals with $\gamma<1$ and $\gamma>1$ as reflected in their respective bounding cones is  examined in \ref{LimitSuper}.
  
\begin{figure}[htb]
\begin{center}
\subfigure[]{
\includegraphics[height=4cm]{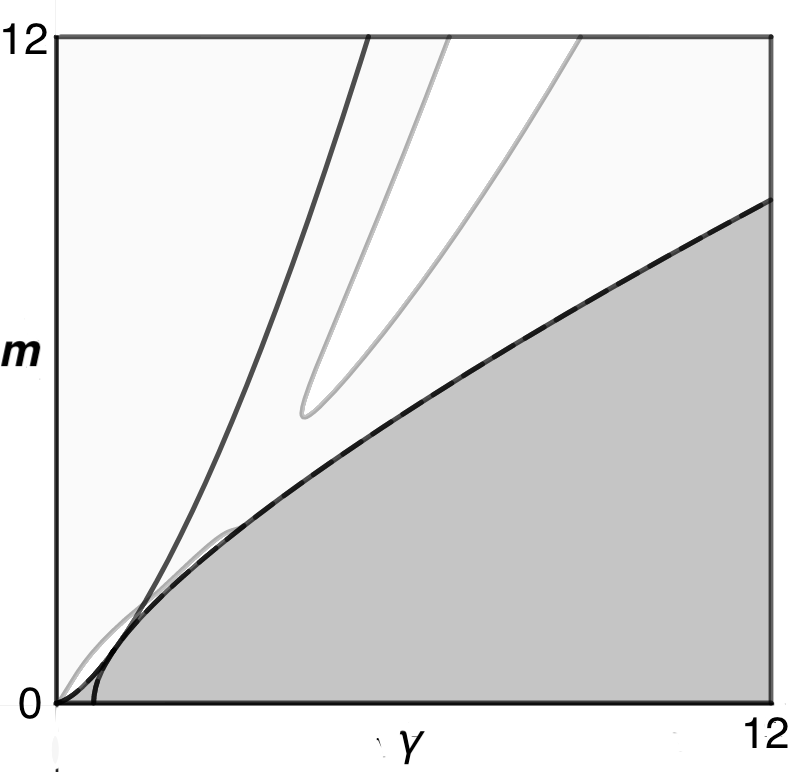}}
\hskip.5cm
\subfigure[]{\includegraphics[height=4cm]{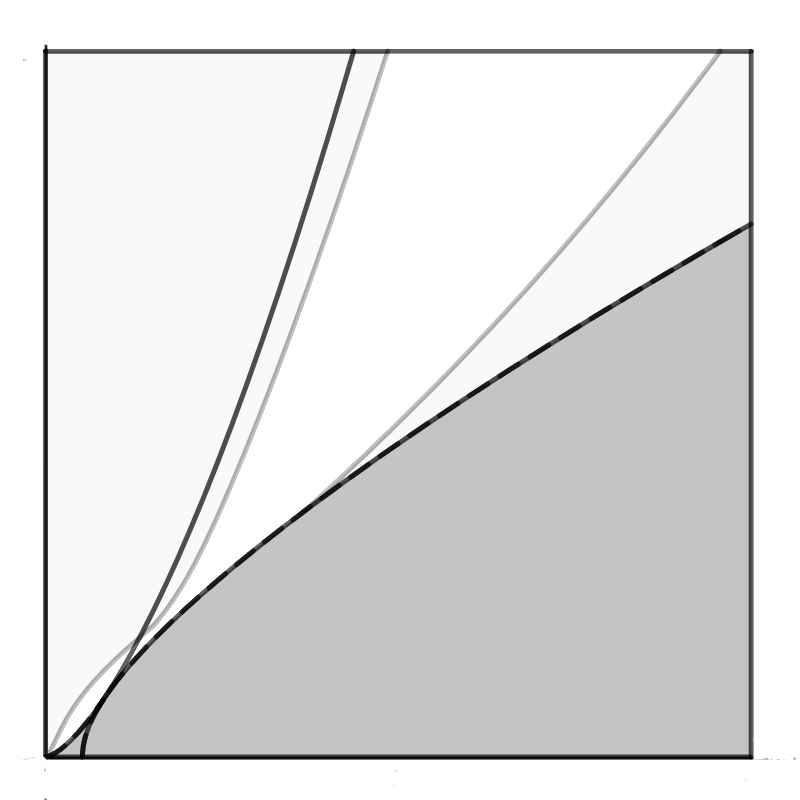}}\\
\subfigure[]{\includegraphics[height=4cm]{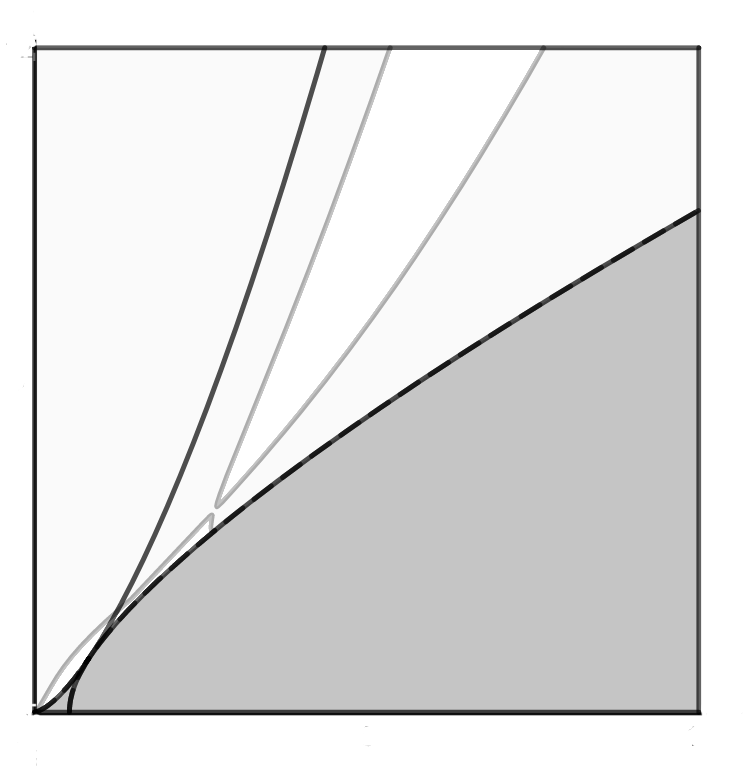}}
\hskip.5cm
\subfigure[]{\includegraphics[height=4cm]{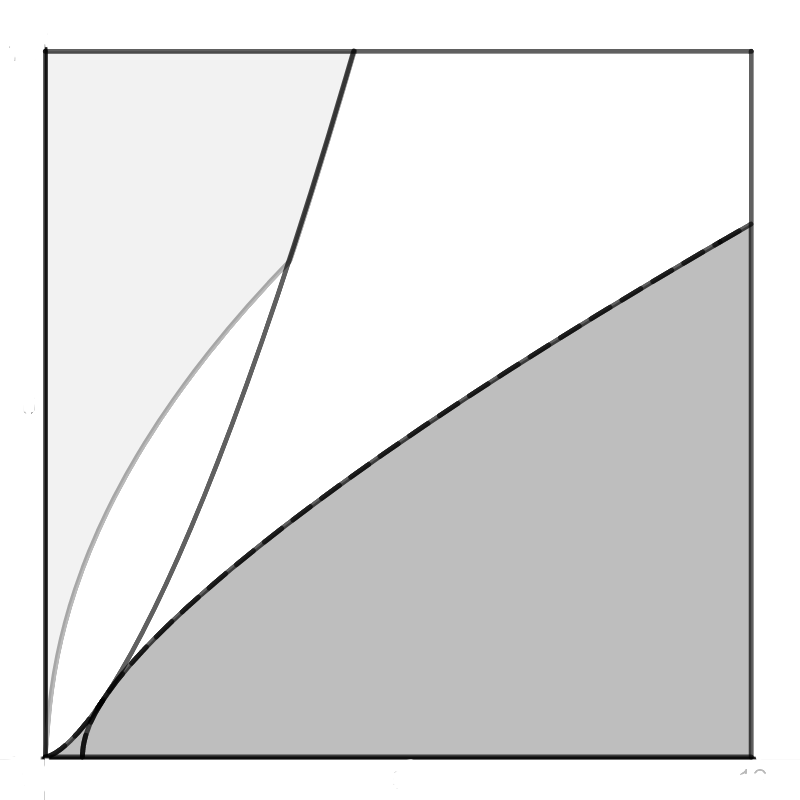}}
\caption{\small \sf  $(\gamma,m)$  parameter space with contours of constant 
$\cos (\theta_{\sf Min})=0.95,0.9529,0.97$ and $0.9999$ (a-d), on same scale. 
The white (light gray) region indicates where 
$\cos (\theta_{\sf Min})$ is lower (greater) than the indicated value; the dark gray region is forbidden. 
The curves $\mathcal{C}_0$ and $\mathcal{C}_1$ are traced in black. Notice that $\cos (\theta_{\sf Min})\ge 0.95$ over a large expanse of parameter space, both subcritical and supercritical. Indeed, all supercritical spirals outside a narrow bend terminating just above $\gamma=1$ possess a vanishingly small opening angle; witness what occurs when the  cosine is raised to  $0.9999$. It is also evident that these contours 
reveal a number of intriguing unanticipated complex patterns worthy of closer examination but beyond the scope of this paper.} 
\end{center}\label{PhasesCosmax}
\end{figure}

\vskip1pc\noindent{\bf Cone closure}
\\\\
Just how rapidly $\theta_{\sf Min}\to 0$ as $m$ increases can be best appreciated by examining a contour of constant $\theta_{\sf Min}$ in parameter space. In Figure 20
these contours are represented  for a sequence of very small conical opening angles. 
Panel (a) in this figure indicates that
spiral trajectories describing \textit{small} deviations of a planar logarithmic spiral (say even $\theta_{\sf Min}=\pi/3$) are confined within a very
narrow sliver of parameter space extending from the origin to the point  $(1,1)$. 
\\\\
Figure \ref{Fig:Sphere} in section \ref{TracingTrajectories} illustrates clearly  how the closure of the limiting cones on the spiral trajectory facilitates its passage though its three-dimensional environment.

\setcounter{equation}{0}
\renewcommand{\thesection}{Appendix \Alph{section}}
\renewcommand{\thesubsection}{E. \arabic{subsection}}
\renewcommand{\theequation}{E.\arabic{equation}}

\section {\bf Cone opening in limiting supercritical spirals}
\label{LimitSuper}

In the neighborhood of the interface $\mathcal{C}_1$ in parameter space the behavior of  $\theta_{\sf Min}$ in supercritical spirals in the regime $\gamma<1$ contrasts significantly with that in the regime $\gamma>1$.
Let $m^2=\gamma^3 + \delta m^2$, where $\delta=\delta m^2/\gamma^3$ is small (cf. \ref{Limits}). In the case $\gamma<1$, as shown in \ref{Limits},
$Z_- \approx \gamma - 2 \gamma^4 \delta /(1- \gamma^3)$.  In the quadratic approximation, the quadrature describing the small well lying
between $Z_-$ and $\gamma$ is given by
\begin{equation}
\label{eq:linsupergamless10}
\frac{1}{4}{{Z^\bullet}}^2 - (1-\gamma^3) (\gamma- Z)\left(Z-\gamma + \frac{2\gamma\, \delta m^2}{1- \gamma^3} \right)=0\,.
\end{equation}
The motion in this well is harmonic in $\Theta$,
\begin{equation}
\gamma-Z=  \frac{\gamma^4\, \delta}{1- \gamma^3} \,\left( 1- \cos \omega \Theta \right) =0\,,
\label{eq:linsupergamless1}
\end{equation}
where $\omega = 2 \sqrt{1-\gamma^3}$ is the curvature of the potential at $Z=\gamma$ in this limit. The period of oscillations in the vanishing small well remains finite (just as it does in deformed conical helices) so long as $\gamma\ne 1$. In this limit, it is possible to approximate $\cos\theta$, given by Eq.(\ref{eq:cosvsZ}) to first order in $\delta$:
\begin{equation}
\label{eq:coslimit}
\cos\theta \approx 
\frac{1+ \gamma^3}{1-\gamma^3} \, \gamma^3 \,( 1- \cos \omega \Theta) \, \delta \,,
\end{equation}
with a maximum linear in $\delta m^2$. 
 Notice that perturbation theory breaks down as the critical point
$\gamma=1=m$ is approached. Away from this point, Eq.(\ref{eq:coslimit}) describes small \textit{symmetric} oscillations about a logarithmic spiral.\footnote{\sf  
Supercritical templates in this regime would appear to track the out-of-plane trajectories of the arms of spiral galaxies.} 
\\\\
If, on the other hand, $\gamma>1$,  then $Z_-= 1/\gamma^2$, while $Z_+\downarrow \gamma$. and there are no small \textit{symmetrical} excursions about logarithmic spirals.  
The minimum conical angle $\theta_{\sf Min}$ is bounded away from $\pi/2$ 
for each value of  
$\gamma$, consistent with  the continuity of  $\theta_{\sf Min}$  across $\mathcal{C}_1$, as illustrated in Figure 18b.  

\begin{figure}[htb]
\begin{center}
\includegraphics[height=3cm]{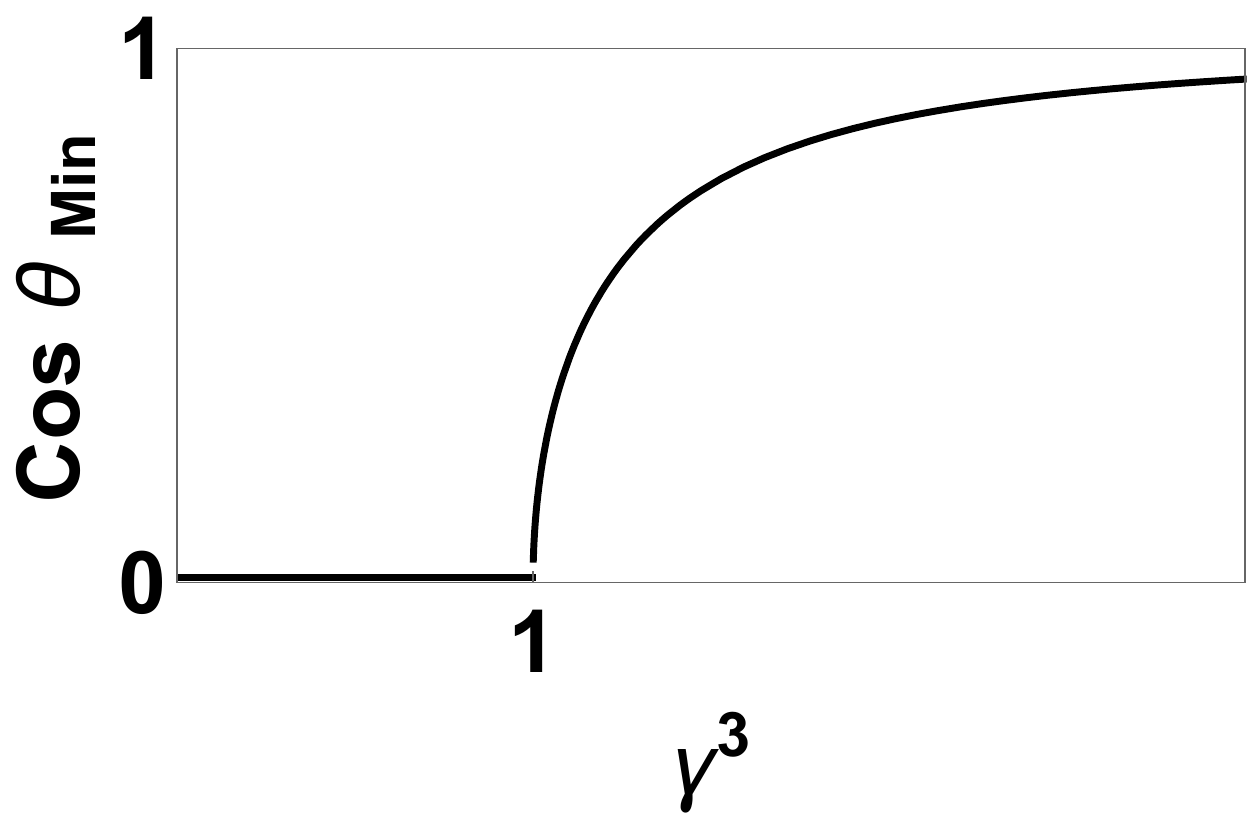}
\caption{\small \sf $\cos (\theta_{\sf Min})$ vs. $\gamma$ along the boundary $\mathcal{C}_1$. The 
details in the immediate vicinity of $\gamma=1$  require a more careful (non-perturbative) treatment than the one sketched here. }
\label{Fig:Cvsgam}
\end{center}
\end{figure}
\vskip1pc\noindent 
A qualitative plot of $\theta_{\sf Min}$ vs. $\gamma$ along 
the boundary $\mathcal{C}_1$ is provided in Figure \ref{Fig:Cvsgam}. At lowest order, $Z_+=\gamma$.  The minimum $\theta_{\sf Min}$ occurs where $Z$ satisfies Eq.(\ref{eq:Quadratic}), with $m^2=\gamma^3$, which reduces to 
$Z_{\sf Min}={\gamma^4}/(2\gamma^3-1)$, 
and $\cos (\theta_{\sf Min}) =  \sqrt{1 -\gamma^{-6}}$.

\setcounter{equation}{0}
\renewcommand{\thesection}{Appendix \Alph{section}}
\renewcommand{\thesubsection}{F. \arabic{subsection}}
\renewcommand{\theequation}{F.\arabic{equation}}

\section{ Sharp upper and lower bounds on $\rho'$}
\label{rhoprimebounds}

It is straightforward to confirm that $\rho' \le 1$ which, in turn, implies $\rho \le s$.
\\\\
{\bf Supercritical vs. Subcritical spirals}
\\\\
In a supercritical spiral, Eq.(\ref{eq:rhoprime2}) implies that $\rho'$ is maximized when $W_-=- Q  + Z^{1/2}  (\gamma -Z)^{1/2}=0$.  As discussed in  section \ref{mid}, this  is satisfied when the trajectory crosses the mid plane, or $\cos\theta=0$, so that 
$Z=  \gamma/ [m^2 +1 - \gamma^3]$ (cf. Eq.(\ref{eq:ztheta0})). 
\\\\
In a subcritical spiral, there is a  
non-vanishing minimum value of $\cos^2\theta$ which places a sharp upper  bound on $\rho'$.   This  
occurs where $W_-^\bullet=0$,  which reduces to the quadratic appearing in Eq.(\ref{eq:Quadratic}). 
\\\\
In summary
\begin{equation}
\label{rhoprimesuper}
\rho'^2 \le 
\begin{cases} 
  \frac{\gamma^3 }{m^2  + 1} \le \frac{\gamma^3 }{\gamma^3  + 1}
 \,, \quad m^2\ge \gamma^3\,;\\
\frac{\gamma^3 }{m^2  + 1} \,\frac{1 }{1+ \gamma W_-(Z_0)^2} \le 
\frac{m^2 }{m^2  + 1} \,,\quad m^2\le \gamma^3\,.
\end{cases}
\end{equation}
It is possible to also place sharp lower bounds on $\rho'$ by determining the maxima of $|W_+|$.  A weaker lower bound on $\rho'$,
$ \rho'^2\ge \gamma^4/ [(m^2  + 1)(m^2+ \gamma)]>0$, follows from the
 trigonometric bound implied by the identity (\ref{eq:cosvsZ}), $|W_+|^2 \le m^2/\gamma$.

\setcounter{equation}{0}
\renewcommand{\thesection}{Appendix \Alph{section}}
\renewcommand{\thesubsection}{G. \arabic{subsection}}
\renewcommand{\theequation}{G.\arabic{equation}}

\section{ Where $X_\perp'$ is negative}
\label{Xperppr0}

\begin{figure}[htb]
 \begin{center}
\includegraphics[height=5cm]{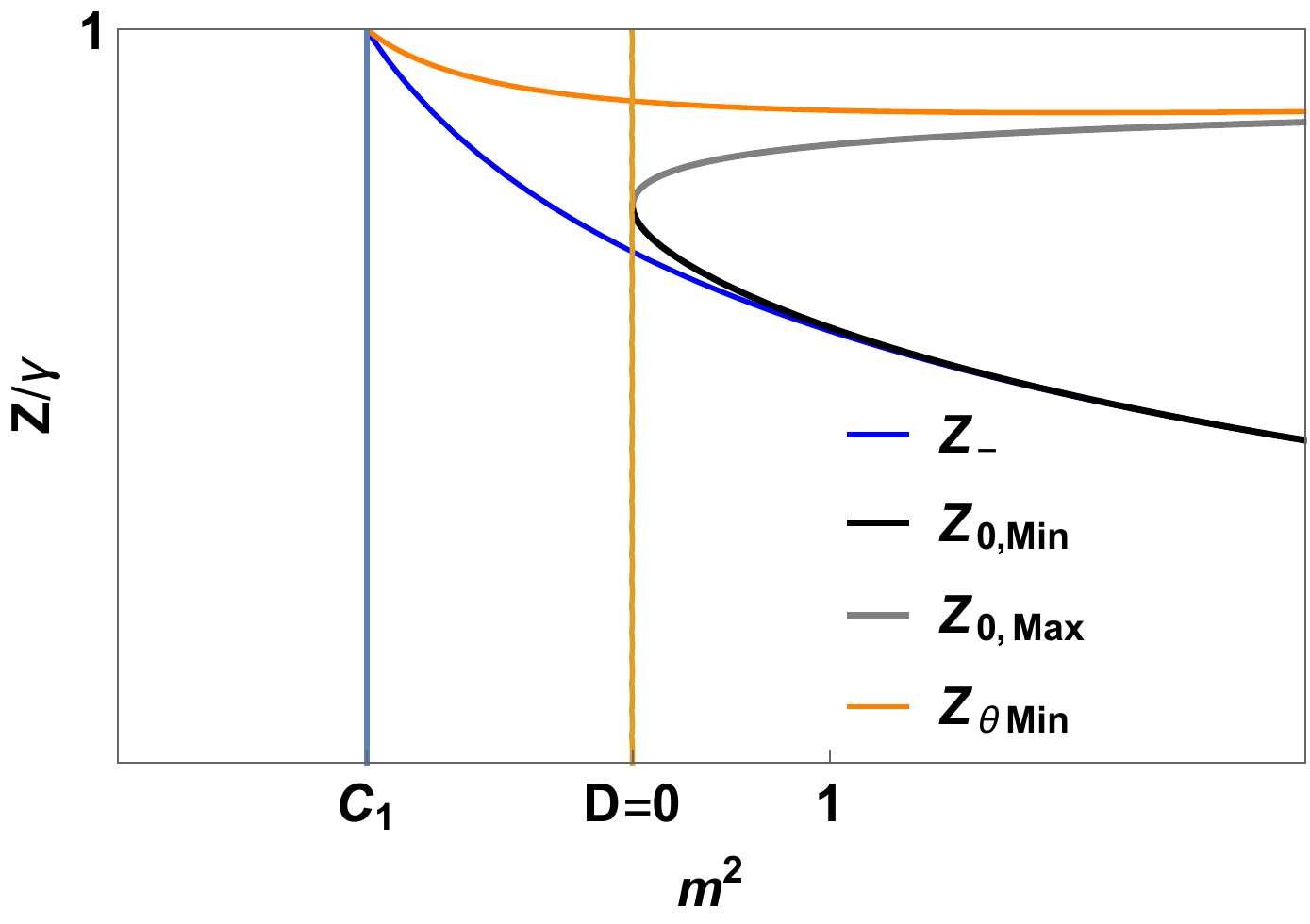}
\caption 
{\small \sf $Z_{0\,{\sf Min}}$ (black) and  $Z_{0\,{\sf Max}}$ (gray) as functions of $m$ for $\gamma=0.8$.
Thus $X'_\perp<0$ for all values of $Z$ within the band $[Z_{0\,{\sf Min}},Z_{0\,{\sf Max}}]$ along the $(+,+)$ and $(-,+)$ quarter cycles. Note that this band lies below $Z_{\theta\,{\sf Min}}$, the value of $Z$ where the polar angle achieves its minimum, and above $Z_-$. As $m$ increases, however, the band extends from $Z_-\to 0$ to $Z_{\theta\,{\sf Min}} \to \gamma$. In the limit $m\to\infty$, $X_\perp$ is negative everywhere along the $(+,+)$ and $(-,+)$ quarter cycles.}\label{Fig:ZXperpp0}
\end{center}
\end{figure}

The Pythagorean decomposition of $\rho^2$ in terms of $X_\|$ and $X_\perp$ implies 
$\rho\rho' = X_\perp X_\perp' + X_\| X_\|'$.
The identities  (\ref{eq:cosvsZ}), (\ref{eq:XparprimeZ}) and (\ref{eq:rhoprime2}) for the cosine, $X_\|'$ and $\rho'$ respectively, then imply
\begin{equation}(m^2+1)m^2 (\kappa X_\perp) X_\perp' = \frac{\gamma}{ Z} \, \Big[ - (m^2+1) \,  {\sf sign} (Z^\bullet) Q  (\gamma-Z)^{1/2} 
+  Z^{1/2}  \left((m^2+1) \, Z - \gamma \right) \Big]\,,
 \label{eq:Xperpprime}
 \end{equation}
 independent of the sign of $\tau$.
It is evident that $X_\perp'$ is not necessarily positive. If  $Z^\bullet<0$, however, 
Eq.(\ref{eq:Xperpprime}) implies it will be.  If $X_\perp'$ is non-positive anywhere,  it must occur along the 
 $(+,+)$ and $(-,+)$ cycle segments). It vanishes whenever
$(m^2 +1) \,   Q\, (\gamma-Z)^{1/2}
 = (m^2+1) \, Z - \gamma$, 
 or, equivalently, 
 \begin{equation}\frac{m^2+1}{\gamma^2} [ (m^2+1)^2  + (m^2-1)\gamma^3  ]Z^2 
- \frac{1}{\gamma} [(m^2+1)^3 + (m^2+1)^2 - \gamma^3] Z + (m^2+1)^2=0\,.
\label{eq:Xperpquad}
\end{equation}
This quadratic possesses two real roots whenever the discriminant, $D$,
is non-negative, or 
\begin{equation}
1 + 4 m^2 + 5 m^4 + 2 m^6 + 
  2 \sqrt{ m^2 (1+ m^2)^5}\ge \gamma^3/m^2\,.
  \label{eq:Dge0}
  \end{equation}
This is the region  in parameter space, shaded light gray,  above the curve $D=0$ in Figure \ref{Fig:PhasesXperpp0}.  One can now confirm the following:
 \\\\
(i) All supercritical deformations of logarithmic spirals with $\gamma<1$ are monotonic in $X_\perp$ if $m$ is not too large, behavior consistent with their interpretation as small deformations of logarithmic spirals. This pattern is followed above $\gamma=1$ up to a critical value $\gamma$ ($\gamma^3=2.38298$ or
$\gamma^{1/3} =(54 - 6 \sqrt{33})^{1/3} + (2 (9 + \sqrt{33}))^{1/3}/3^{2/3}$). 
This critical value is indicated by the red point in Figure \ref{Fig:PhasesXperpp0}.
Above this value,  all supercritical spirals exhibit oscillations in $X_\perp$. 
\\
(ii) Below the same critical value, $X_\perp$ is monotonic in all subcritical spirals; above it monotonic behavior occurs only in ever smaller deformations of conical helical spirals. 
\\\\
When $D>0$,  the two real roots of  the quadratic (\ref{eq:Xperpquad}) are given by  
 \begin{equation}
Z_{0\,{\sf Min,Max}}/\gamma = \frac{2 - \gamma^3 + m^2 (5  + 4 m^2 + m^4) \pm \sqrt{
     \gamma^6 - 
      2 \gamma^3 m^2 (1 + 2 m^2) (m^2+1)^2 + m^4( 
        m^2+1)^4}}{2 (1 - \gamma^3 + 3 m^2 ( m^2 + 1) +
      \gamma^3 m^4 + m^6)} \,.
      \label{eq:Z0pm}
   \end{equation}
 It is easy to confirm that they both lie within the accessible region $[Z_-,{\sf Min}(Z_+,\gamma)]$. 
 Their behavior for fixed $\gamma=0.8$ as a function of $m$ is displayed in Figure \ref{Fig:ZXperpp0}.
\\\\
The identity $\sin \theta X_\perp' = \rho' -  \cos \theta  X_\|'$  reveals that $X_\perp'=0$ never 
occurs on the mid-plane. 
In addition, the extrema of $\theta$ and 
$X_\perp$ never coincide (unless asymptotically in $m$). 
This is because the extrema of $
\cos\theta={ X_\|}/{\rho}$ occur when 
$\cos\theta={ X_\|'}/{\rho'}\ne 0$. This feature is evident in Figure \ref{Fig:ZXperpp0}.
\\\\
Even though it is not generally monotonic, 
$X_\perp$ never returns to zero if  $\gamma$ or $m$ is finite.  On the other hand,  as shown in 
section \ref{Coscycle}) and quantified in \ref{BoundingCones},
even with modestly large values of $m$  the bounding cones close down with $\theta_{\sf Min}\approx 0$ and with it $X_{\perp\,{\sf Min}}\approx 0$, so that the spiral trajectory visits the polar neighborhoods in every cycle.

\setcounter{equation}{0}
\renewcommand{\thesection}{Appendix \Alph{section}}
\renewcommand{\thesubsection}{H. \arabic{subsection}}
\renewcommand{\theequation}{H.\arabic{equation}}

\section{Asymptotic rise of $X_\|$ as $m^2 \uparrow \gamma^3$}
\label{asymptotrise}

As $m^2 \uparrow \gamma^3$, the linear helical ascent along the torque axis of the deformed helices slows down,
growing sub-linearly with $s$ (compared to the linear growth in a conical helix).  
The increasingly deformed helix degenerates as $s\to \infty$ into a planar logarithmic spiral. It is possible to determine the exponent associated with this critical slowdown.  
\\\\
If $Z\to \gamma$ in Eq.(\ref{eq:XonM2}), then 
\begin{equation}
\label{Xpargammaf}
X_\| \to 
S\, \left( \frac{1}{\kappa}\right)\, \Big[(1 -\gamma^{-3})^{1/2} + 1\Big](\gamma - Z)^{1/2}
\,.
\end{equation}
The potential is now expanded about the double root at
$Z=\gamma$ in Eq.(\ref{eq:Quadrature}): one has 
\begin{equation}{Z^\bullet}^2/2 + 2\gamma^3  (1- \gamma^{-3}) (\gamma- Z)^2=0 \,, \end{equation}
so that 
\begin{equation}
\gamma - Z \approx \\\\ A e^{- \Omega \Theta}\,,
\end{equation} 
where
$\Omega^2= 4 \gamma^3  (1- \gamma^{-3})$. This in turn implies that
$\kappa \approx c e^{-\gamma^{3/2}\Theta}$, so that
$s\approx c^{-1} \gamma^{- 3/2}\,e^{\gamma^{3/2}\Theta}$ consistent with 
the asymptotic behavior of the trajectory as a planar logarithmic spiral, $\kappa s \approx \gamma^{-3/2} = (2S)^2 $.  Now
$\gamma - Z \approx A s^{- \gamma^{-3/2}\Omega} = 
A s^{-2 \sqrt{1- \gamma^{-3}}}$. Substituting into Eq.(\ref{Xpargammaf}), this gives
\begin{equation}
\label{XonMgamma}
X_\| \to 
 [(1 -\gamma^{-3})^{1/2}+  1] s^{1 - (1 -\gamma^{-3})^{1/2}}
\,,
\end{equation}
diverging sub-linearly with $s$. In the limit $\gamma\to \infty$, the limiting value is finite.

\end{appendix}

\end{document}